\documentclass[12pt]{article}
\usepackage{graphicx}      
\usepackage{setspace}      
\usepackage{enumerate}     
\usepackage{amssymb,amsthm,amsmath,amsfonts,amsbsy}
\usepackage{mathrsfs,syntonly,graphicx,color}
\usepackage{setspace,geometry}
\usepackage{indentfirst}
\usepackage{bm}
\usepackage{float}
\usepackage{tabularx}      
\usepackage{tikz}
\usepackage{ctable}        
\usepackage{multirow}      
\usepackage{comment}       
\usepackage{caption}
\usepackage[colorlinks, urlcolor=blue, linkcolor=blue,citecolor=blue]{hyperref} 
\usepackage{subcaption} 
\usepackage{soul} 
\usepackage{lineno}
\usepackage{lscape} 
\usepackage{chngcntr} 
\usepackage[labelfont=bf]{caption} 
\usepackage[flushleft]{threeparttable} 
\usepackage{authblk}
\usepackage{diagbox}
\usepackage{dcolumn}

\usepackage[round]{natbib}
\usepackage{multibib}
\newcites{app}{Appendix References}

\title{Agree to Disagree: Measuring Hidden Dissent in FOMC Meetings}
\author{Kwok Ping Tsang and Zichao Yang\thanks{Tsang: Department of Economics, Virginia Tech,
		Pamplin Hall, Blacksburg, VA 24061; E-mail: byront@vt.edu. Yang: Wenlan School of Business, Zhongnan University of
		Economics and Law, Wuhan, 430073, China; E-mail: yang\_zichao@outlook.com. We would like to thank seminar participants at ZUEL, HUST and CUM. We also thank Lei Lu, Andre Silva, Dick Startz, Xiaojin Sun, Ming Yi, Cloud Yip, and an anonymous referee for helpful comments.}}
\date{\today}    

\begin{document}
	
\maketitle

\begin{abstract}
	
\noindent Using FOMC transcripts and customized deep learning models, we quantify ``hidden dissent'', or disagreement in the FOMC that is unobserved in formal votes. We find hidden dissent to be prevalent and systematically driven by macroeconomic conditions like inflation and unemployment. It strongly correlates with divergent member projections (SEP) and measures of policy sub-optimality, reflecting heterogeneity among members in policy preferences. Furthermore, we show that the financial markets respond to the hidden dissent implied in FOMC minutes.

	~
	
	\noindent \textbf{JEL Codes}: E52, E58, C55.
	
	\noindent \textbf{Keywords}: Natural language processing, disagreement, monetary policy, FOMC.
\end{abstract}
\clearpage
\onehalfspacing
\section{Introduction}

\noindent Statements, minutes, speeches, and other texts released by the Federal Reserve have been scrutinized extensively using techniques ranging from simple word clouds to advanced machine learning models. However, the transcripts of Federal Open Market Committee (FOMC) meetings—word-for-word records of members' views—have been less studied. This is partly due to the five-year delay in their release, but also because of the difficulty in analyzing the vast volume of text produced by over a hundred past FOMC members.

This paper begins by evaluating each member’s position during meetings, using data (i.e., NO votes) to identify language that indicates strong disagreement with policy actions. The total percentage of NO votes from 1976 to 2018 is $6.93\%$. In \autoref{vote_stat}, we show the annual counts of YES and NO votes, which reveal minimal variation over time.

While NO votes are rare, they provide enough data to train a deep learning model. Using members' words, our model accurately predicts whether a member voted to dissent, allowing us to assess hidden dissent that may not have led to a NO vote. We demonstrate that \autoref{vote_stat} creates a misleading impression of consensus among FOMC members, and show that the level of hidden dissent measured by our model aligns with economic conditions more consistently than the voting outcomes. 

Consider two examples that illustrate our main idea. In the FOMC meeting on May 19, 1998, President Minehan of the Federal Reserve Bank of Boston raised concerns about inflation and proposed a less accommodating policy, and she said:

\emph{``Mr. Chairman, at our last meeting I expressed concern that, even in the absence of clear indications that inflation was rising, both the strength of the domestic economy and the frothiness in financial markets required some policy tightening to reduce the risk that even more tightening might be needed later...''}

Other members, like Presidents Jordan, Poole and Broaddus shared Minehan's concerns. Even though Broaddus was not invited to vote in that meeting, he commented that:

\emph{``Mr. Chairman, if it were my choice, I would still prefer to make a small move today for many of the reasons that Cathy Minehan and Bill Poole have adduced here...So, to paraphrase Governor Kelley, I would say that if we sit still, I hope we will at least sit uptight. [Laughter]''}

Eventually the Committee seeks to maintain the federal funds rate unchanged at an average of around 5-1/2 percent, and President Minehan, despite her concerns, voted YES.

Another example comes from the FOMC meeting on April 24, 2012. Governor Raskin voted for the final decision (no action), but the governor also raised concerns:

\emph{``Thank you, Mr. Chairman. I, too, support alternative B, but not without a vague sense that the actions taken today, i.e., no action, could be eventually interpreted by markets as being contractionary, an interpretation that could eventually effectively make them contractionary...Analogously, if our statement today sends a signal that we are content with the current path of the economy, this could work like contractionary policy—the last thing we need right now in my view. I hope that is not the message the public takes from our meeting today.''}

These examples suggest that some members voiced disagreements that were not reflected in their votes. In this paper, we introduce a hidden dissent measure based on a customized deep learning model, ranging from 0 (strong support) to 1 (substantial opposition). We also aggregate these individual measures to derive the level of hidden dissent for each FOMC meeting.

To validate the relevance of our measure, we regress it on two sets of explanatory variables at both individual and meeting levels. The first set includes current or projected targets under the dual mandate: inflation and unemployment rates. The second set comprises member characteristics like age, gender, political ideology, and education. Our results show that hidden dissent is strongly correlated with economic conditions, increasing when inflation rises or unemployment falls. This correlation suggests that members have more reservations when there seems to be a need for ``cooling down'' the economy, and members are more in unison when the economy is in a downturn. Among characteristics, we find that only education background has consistent explanatory power.

We also present evidence showing that hidden dissent is driven by members’ differing policy preferences and economic projections. First, using data from the Summary of Economic Projections (SEP), we show that disagreements over the trajectory of monetary policy and projections for macroeconomic variables have roughly equal explanatory power. We also find evidence that the disagreement on monetary policy contains information beyond the disagreement on economic outlook, and this information is correlated with our measure of hidden dissent. Second, we observe a strong correlation between hidden dissent and the optimal policy perturbation (OPP) proposed by \cite{barnichon2023sufficient}. Since OPP can be driven by variations in loss functions or impulse responses, we again conclude that both forms of heterogeneity are important.

Finally, we investigate the real-world relevance of this hidden dissent by examining its impact on financial markets. While previous research has focused heavily on the sentiment conveyed in central bank communications (e.g., overall hawkishness or dovishness), we hypothesize that the degree of internal hidden disagreement represents a distinct and previously under-explored information channel. Using FOMC minutes for timeliness, we demonstrate that our hidden dissent measure provides information beyond simple sentiment, triggering significant and unique reactions in stock and bond markets. 

\subsection{Related papers}
\label{litreview}
Central bank communication is a well-researched field, with numerous studies exploring various aspects \citep{born2014central, hansen2016shocking, tobback2017between, hansen2019long, husted2020monetary, lunsford2020policy, leombroni2021central, malmendier2021making}. Among central banks, the Federal Reserve has received the most attention due to its detailed communication records. Researchers have studied statements, minutes, speeches, news articles, press conferences, testimonies, and even facial expressions to understand Federal Reserve communication \citep{handlan2020text, arismendi2021federal, shapiro2022measuring, gorodnichenko2023voice, curti2023let, alexopoulos2024more}. Recently, there has been growing interest in the less-explored FOMC meeting transcripts. \citet{romer2008fomc} pioneered this area by manually examining transcripts from three FOMC meetings to show the link between forecast differences and policy actions. However, the release of transcripts, containing over three million words, makes manual analysis challenging. Advances in natural language processing (NLP) now enable researchers to apply sophisticated machine learning methods to analyze this large volume of text \citep{jurafsky2023speech}. For example, \citet{hansen2018transparency} applied an LDA model to FOMC transcripts to study transparency policies, while \citet{cieslak2021policymakers} and \citet{shapiro2022taking} used Bag of Words models to measure members' sentiment. More recently, \citet{shah2023trillion} trained a BERT-based model to gauge monetary policy stance using FOMC speeches, meeting minutes, and press conference transcripts.

In this study, we develop a customized deep learning model leveraging transfer learning and the self-attention mechanism to analyze FOMC transcripts and uncover nuanced shifts in hidden dissent. We further aggregate these individual-level measures to construct a meeting-level hidden dissent score for each meeting. Additionally, we explore how this measure relates to economic conditions, members’ personal characteristics and the factors contributing to hidden dissent. Finally, we investigate how financial markets respond to this new hidden dissent measure.

Our research contributes to the literature characterizing FOMC members' policy preferences through various forms of communication. For instance, \citet{kahn2018understanding} examined the Compilation and Summary of Individual Economic Projections to study how hawks and doves in the FOMC differed in their views of appropriate monetary policy, while \citet{apel2022much} used minutes and transcripts to measure participants' ``hawkishness'' and found it improves forecasting of future policy. \citet{bertsch2022central} employed a BERT-based model to analyze speeches, focusing on non-dual mandate concerns that influenced policy. Similarly, \citet{bordo2023perceived} used a hawk-dove measure to study how members' personal characteristics shaped their preferences.

Our paper moves beyond traditional static classifications by using FOMC transcripts, which provide a more detailed record of members' views. By applying our deep learning model to the full content of transcripts, we construct a hidden dissent score at the individual-meeting level rather than the individual level, allowing members' policy preferences to fluctuate over time rather than categorically labeling them as hawks or doves. This method provides a more flexible and granular view compared to assigning fixed labels, capturing the nuances of policy deliberation. We also show that hidden dissent is closely related to current economic conditions, with little influence from personal characteristics.

In addition, our work connects to the literature examining explicit voting dissent as an information channel in monetary policy. Studies like \citet{riboni2014dissent} demonstrate that dissenting votes can predict future policy actions, while \citet{madeira2023origins} link dissent frequency to specific economic conditions, such as supply shocks that create policy trade-offs. However, a key limitation of focusing solely on recorded votes is their rarity. In contrast, our hidden dissent measure proves to be more prevalent and, as we demonstrate, shows a closer connection with the macroeconomic situation and tangible influences on financial markets, complementing the insights gained from studying explicit dissent.

Furthermore, we also apply our deep learning model to FOMC meeting minutes to derive a more timely assessment of hidden dissent. This approach takes advantage of our model's ability to capture nuanced dissent, offering a consistent, data-driven index that can be utilized soon after meetings, rather than waiting for the five-year lagged release of transcripts.

This paper is structured as follows. We lay out the various variables and data sources in \autoref{data}. Then \autoref{methodology} describes the customized deep learning model and the generation of hidden dissent scores. In \autoref{results} we discuss our empirical strategies and examine the economic meaning of the hidden dissent measure. And \autoref{conclusion} concludes.

\section{Data}
\label{data}

\noindent The data used in this study fall into four categories: (1) personal characteristics of FOMC members, (2) FOMC communication materials and voting outcomes, and (3) macroeconomic and financial market data.

\subsection{Personal characteristics}
\label{data_personal}

Personal characteristics for each FOMC member come from multiple sources, including the \href{https://www.federalreservehistory.org}{Federal Reserve History} website, members' Wikipedia entries and related news articles. The list of variables we collect is shown in \autoref{personal_info}. 

When categorizing the hometown region of each FOMC member, we follow the practice established by the \href{https://www.bls.gov/cex/csxgeography.htm}{Bureau of Labor Statistics} and divide the U.S. into four regions: Northeast (NE), Midwest (MW), South and West. We base the hometown classification on the location where a member was raised rather than their birthplace and we use ``OTH'' to represent hometowns outside of the U.S. Similarly, the school regions where members attained their highest degree are classified in the same fashion. 

To measure the political ideology of FOMC members, we employ two different methods due to the different appointment procedures for governors and presidents. Members of the Board of Governors are appointed by the President of the United States (POTUS) and confirmed by the U.S. Senate. However, the POTUS and Senate are not directly involved in picking presidents of the twelve regional Federal Reserve Banks. Based on the information on the \href{https://www.federalreserve.gov/faqs/how-is-a-federal-reserve-bank-president-selected.htm}{Board of Governors of the Federal Reserve} website, presidents are chosen by their own boards of directors and confirmed by the Board of Governors. Accordingly, we use the party affiliation of the incumbent POTUS at the time of each governor's appointment as a proxy for their political ideology. For the presidents of the regional Federal Reserve Banks, we proxy their ideology based on the party that won the most recent presidential election in the state where the respective bank is located. For example, President Jeffrey M. Lacker assumed the role of president at the Federal Reserve Bank of Richmond in 2004, so we use the presidential election result in Virginia in 2004; President Dennis P. Lockhart was appointed as the president of the Federal Reserve Bank of Atlanta in 2007, we use the presidential election result in Georgia in 2008 rather than 2004. The state level presidential election data after 1976 is downloaded from \href{https://electionlab.mit.edu/data}{MIT Election Data and Science Lab}, and earlier data is downloaded from the \href{https://history.house.gov/Institution/Election-Statistics/Election-Statistics/}{Election Statistics} published by the U.S. House of Representatives. 

We also  collect the significant events experienced by FOMC members in their early ages. Following \citet{bordo2023perceived}, we collect information related to three major events: the Great Depression (1929-1939), World War II (1939-1945), and the Great Inflation (1965-1982). If a FOMC member experienced one of these events before turning 21, we count the corresponding variable as one.\footnote {We choose to not include World War I (1914-1918) because only nine FOMC members experienced WWI in our sample, and this number is significantly smaller than that in other events. We also look at the active periods of these nine members and find that they only served on the FOMC from 1976 to 1986. In contrast,  members who experienced the Great Depression can be found from 1976 to 2008, members who experienced the Great Inflation can be found from 1982 to 2018, and members who experienced WWII were on the FOMC throughout the entire data sample. Furthermore, given that some macroeconomic data used in our later study is only available after 1986, we choose to focus on the listed three events. } Eventually, we collect the personal information of FOMC members who cast votes in meetings from 1976 to 2018, a selective description of the personal characteristics information can be found in \autoref{personal_info_summary}.

\subsection{Communication materials and voting outcomes}

How FOMC communicates its monetary policies changes over time, and \citet{timtodd2016} gives a detailed discussion on this topic. In this study, we download the available transcripts spanning from March 29, 1976 to the end of 2018, remove all the content from Fed staff, and only keep the transcripts related to the members who cast votes at the end of meetings. These transcripts are then disassembled and re-clustered at the meeting-person level. For example, suppose ten FOMC members attended a FOMC meeting, combining the transcript with the members' voting records resulted in the creation of ten labeled observations. And, to reduce the data processing time, we also remove irrelevant sentences like ``Thank you, Chair", ``Thank you, President", ``Good morning..." etc. To remove them, we follow the method in \citet{shapiro2022taking} and use the Oxford Dictionary of Economics and the Oxford Dictionary of Finance and Banking to filter out sentences that do not contain any economics-related terms. 

The FOMC meeting minutes were introduced in 1993 to provide the public with ``a timely summary of significant policy issues addressed by meeting participants.'' Initially, the minutes were released approximately three days after the subsequent Committee meeting. However, since December 2004, the release schedule has been moved up to three weeks following the meeting. Prior to 1993, the content of the current minutes was available in two separate documents: the \emph{Record of Policy Actions} and the \emph{Minutes of Actions}. In this study, we consider these two documents as substitutes for the current meeting minutes for the period before 1993. Corresponding documents are available on the \href{https://www.federalreserve.gov/monetarypolicy/fomc_historical.htm}{Board of Governors of the Federal Reserve} website, from which we compile meeting minutes data spanning from 1976 to 2024.

The voting records we have extracted are cross-checked with the FOMC Dissents data maintained by \citet{thornton2014making} to ensure accuracy. By pairing individual-level transcripts with their respective voting records, we process data from 370 FOMC meetings and a set of 3,950 labeled observations was assembled for the training of our NLP model.\footnote{We include conference calls as long as they contain voting records. If not, we discard the corresponding transcripts.} The distributions of meeting and observation counts per year are shown in \autoref{num_obs_meetings}.

\subsection{Macroeconomic and financial markets data}
\label{macro_variable}

Macroeconomic data mainly comes from the \href{https://www.philadelphiafed.org/surveys-and-data/real-time-data-research/philadelphia-data-set}{Tealbook} (formerly Greenbook) data set maintained by the Federal Reserve Bank of Philadelphia. Tealbook is prepared by the Board of Governors' staff in preparation for each regularly scheduled FOMC meeting. Tealbook contains in-depth analysis of the U.S. and international economy. Additionally, staff at the Board of Governors also prepare projections for the U.S. economy using an assumption about monetary policy. We use the unemployment and core CPI data from the Tealbook to measure the dual mandate, and the core CPI data only became available after the FOMC meeting on February 12, 1986. Meanwhile, all financial markets data unitized in this study is openly accessible on \href{https://finance.yahoo.com/}{Yahoo Finance}.

\subsection{Transformation of meeting-level variables}
\label{variable_trans}

Combining the macroeconomic variables with personal characteristics, we create a list of variables at the FOMC meeting level, and they are summarized in \autoref{meeting_info}.

A member’s experience is calculated based on the start date of their term and the date of each meeting.\footnote{For members like Governor Janet Yellen who served as the president of the Federal Reserve Bank of San Francisco first (2004-2010), then joined the Federal Reserve Board (2010-2018), we count her term started in 2004 as the president.} Similarly, a member's age at a certain FOMC meeting is calculated based on the birth date and the corresponding meeting date. To measure the diversity in a meeting, we employ the concept of Shannon entropy and calculate:
\begin{equation*}
H = -\sum_{i=1}^n p_i \log_b p_i
\end{equation*} 
where $n$ is the number of elements in the system, $p_i$ is the frequency of element $i$, and $b$ is the base of the logarithm.

In the case of quantifying the diversity of hometowns among meeting participants, we count the numbers of members whose hometown locates in the northeast (NE) region, the midwest (MW) region, the south region, the west region, and outside of the U.S. (OTH). It is then fed into the aforementioned entropy function with a base of five to standardize the entropy value. Shannon entropy reaches its maximum value when all elements have equal probability, implying that greater diversity in hometown distribution corresponds to a higher entropy value. Consequently, value 0 indicates minimal diversity of hometowns, while a value of 1 signifies the highest level of diversity on hometowns. The same approach is applied to calculate the school diversity. 

For the macroeconomic variables, unemployment and inflation, we use corresponding five data points: \emph{Variable\_B2}, \emph{Variable\_B1}, \emph{Variable\_F0}, \emph{Variable\_F1} and \emph{Variable\_F2} from the Tealbook to calculate both trend and standard deviation for these two variables. The trend value is determined by the slope obtained from a first-degree polynomial regression applied to the five data points. And the standard deviation is also calculated based on these five data points, which measures the uncertainty faced by FOMC members.

\section{Natural Language Processing}
\label{methodology}

\noindent With the continued evolution of deep learning, a multitude of NLP tasks are seeing improving outcomes \citep{kamath2019deep, lauriola2022introduction}. In this section, we present the architecture of our customized deep learning model, which is built on the self-attention mechanism. We implement the model on FOMC communication materials. For background information on the application of deep learning models in NLP, please refer to \autoref{appendix_ml_nlp_intro}. A more detailed discussion can be found in \citet{dell2024deep}.

\subsection{A customized deep learning model}
\label{model structure}

Our primary goal is to quantify the extent of disagreement expressed by FOMC members during meetings. To achieve this, we develop a deep learning model designed to analyze the textual content of members' statements. Since computers process numerical data rather than raw text, the first step involves converting the language used by members into a format the model can understand. We employ a state-of-the-art technique based on Sentence-BERT \citep{reimers2019sentence} to transform each sentence within a member's transcript into a numerical representation, known as a sentence embedding. This process captures the semantic meaning of the sentences. For technical consistency, we ensure all these numerical representations have a uniform size through the padding process.\footnote{Specifically, using Sentence-BERT, each sentence is converted into a 768-dimension vector. We then group these sentence vectors for each member's transcript in a meeting and pad them to a standard size of (256, 768) before feeding them into our model. This standardization aids the model training process.}

The core of our approach lies in comparing each member's statements against those of the FOMC chair. Given the Chair's consistent record of voting with the majority to implement the final policy decision, we use their transcript as a proxy for the meeting's ultimate policy stance or consensus view. Our customized model architecture, inspired by the self-attention mechanism prevalent in modern NLP \citep{vaswani2017attention}, is specifically designed for this comparison. Intuitively, the self-attention mechanism allows the model to weigh the importance of different parts of the text, focusing on the segments most relevant for gauging agreement or disagreement. The model processes the numerical representations of both the member's and the chair's transcripts, identifies the key differences in their expressed views, and uses these differences to predict the member's likely voting outcome (YES=0 or NO=1). \autoref{fomc_dl} provides a schematic overview of this architecture.

Training such a model requires careful consideration of the data and methodology. We optimize the model's configuration by fine-tuning its hyperparameters, such as the complexity of its internal layers and the learning rate, using the \emph{optuna} framework developed by \citet{akiba2019optuna} to find the best settings (details in \autoref{hyperparameter_search_space}). A key challenge is the inherent imbalance in the voting data: dissenting (NO) votes are rare compared to assenting (YES) votes.\footnote{After filtering out observations that contain fewer than five sentences, the final dataset contains 3,523 observations. Among these, only 244 (6.9\%) are linked with NO votes, compared to 3,279 with YES votes. This severe imbalance poses challenges not only for current model training, but also for further distinguishing the direction of dissent (e.g., dovish vs. hawkish). With so few NO votes, further subdividing them into directional categories would leave too few observations per class for reliable prediction. Accordingly, our analysis is constrained to measuring the magnitude of dissent. Nevertheless, we show that our proposed measure of hidden dissent is a potent, and previously unmeasured, channel of information that provides statistically significant and economically meaningful insights.} To prevent the model from simply learning to always predict YES, we employ an over-sampling technique, effectively showing the model an equal number of YES and NO vote examples during training. Furthermore, we incorporate standard deep learning practices like learning rate scheduling (gradually adjusting the learning rate during training) and early stopping (halting training when performance on unseen data stops improving) to enhance training efficiency and, crucially, to prevent overfitting—ensuring the model generalizes well to new, unseen transcripts rather than just memorizing the training data.

The optimally configured model achieves high accuracy in predicting actual votes on both the training data (85.88\%) and unseen test data (84.20\%). While the model is trained to predict a binary outcome (0 or 1), the underlying prediction is a probability score between 0 and 1, indicating the likelihood of a member dissenting (voting NO). We leverage this probability directly as our measure of \emph{hidden dissent}. We denote this score as $hd_{ij}$ for member $j$ in meeting $i$. A score close to 0 signifies strong agreement with the chair's position, while a score close to 1 indicates substantial disagreement, reflecting a high probability of the model predicting a NO vote. Summary statistics for these hidden dissent measures are reported in \autoref{disagree_measure}.This continuous measure allows us to capture nuances in members' positions that a simple binary vote cannot. Furthermore, in \autoref{results}, we demonstrate that $hd_{ij}$ provides unique information not captured by other FOMC communication channels.

\subsection{Hidden dissent inside FOMC meeting transcripts}
\label{transcripts_data}

With the hidden dissent scores ($hd_{ij}$) from our customized model, we can now study what characteristics contribute to FOMC members' voting behaviors. Moreover, we are able to describe the hidden dissent that cannot be fully captured by voting records ($v_{ij}=0$ or $1$).

\autoref{pred_distr} shows the boxplot of $hd_{ij}$ scores in each year. In the plot, the red box represents the interquartile range (Q1 to Q3), with the black line in the middle indicating the median (Q2). The whiskers extend to 1.5 times the interquartile range (Q3-Q1). Meanwhile, we compare the boxplot with the percentage of total NO votes, shown by the black dotted line, and they reveal a similar trend. For example, the high median score in 1980 corresponds with a significant proportion of total NO votes recorded in history, whereas lower median scores in recent years are associated with fewer NO votes.

Furthermore, \autoref{wordcloud_voting_prediciton} presents word clouds for different groups based on hidden dissent scores and FOMC meeting policy actions. The first two figures represent word clouds for members with $hd_{ij}$ scores below and above 0.5, respectively. The subsequent rows further refine these word clouds based on FOMC meeting policy actions (i.e., maintaining, increasing, or decreasing the federal funds rate).

In the plots, both the gray shading and word sizes represent word frequency in meeting transcripts. One pattern is obvious: inflation has a high priority in every FOMC meeting. From the first row we can see that no matter the members have low (i.e., align with the meeting decision) or high hidden dissent scores (i.e., deviate from the meeting decision), inflation remains focus in both groups. While words like "economy" and "risk" show relatively high priority in the low-score group, the high-score group's emphasis stays on inflation. This pattern persists when groups are further refined based on policy actions. When FOMC opts not to reduce the policy rate, the low-score group prioritizes inflation, followed by economy, growth, and risk, etc. On the other hand, the high-score group mainly focuses on inflation alone, followed by economy, growth, unemployment, etc. However, a deviation from this pattern arises in the periods when FOMC seeks to decrease the rate. In those meetings, inflation drops to the third most frequent word for the low-score group. Meanwhile, inflation still has the highest priority in the high-score group, but other words like growth and economy also become more important. The observation may suggest that the decision to reduce the rate usually is reached when FOMC has to pivot its priority from inflation to other goals, like economic growth or market stability. For a robustness check, we also show word clouds for different groups based on members' voting records (see \autoref{wordcloud_vote}), and yield similar results.

\subsection{Hidden dissent: hawkish or dovish?}

{Sentiment analysis that classifies policy preferences along a hawkish-dovish spectrum is a widely adopted and valuable method for studying the FOMC \citep{bordo2023perceived, malmendier2021making, istrefi2019fed}. We propose our hidden dissent score as a complement to this approach, designed to capture a different aspect of policy deliberation. While sentiment indices effectively measure the directional leaning of preferences, we find that disagreements are often multi-dimensional and cannot be fully captured by a binary classification. Our measure therefore quantifies the intensity of these nuanced disagreements.

The limitations of a simple hawk-dove classification can be illustrated by the voting record of President Jerry Jordan. In the October 6, 1992 meeting, President Jordan dissented in favor of a more accommodative policy, a seemingly ``dovish'' action. However, his reasoning was not a simple call for more loosening but a critique of the Fed's lack of credibility:

\emph{``I ask business people why they don't believe us when we say we're going to zero inflation...They just sort of shrug and say they don't believe it...I think the issue we have to attack is the credibility of our commitment; we as a Committee need to convey in the clearest way we possibly can that we are going toward zero inflation. That's what we should tell them about how to gauge our policy.''}

Less than two year later, in the March 22, 1994 meeting, President Jordan dissented again, this time in favor of a more aggressive tightening (a 50 basis point hike instead of the Committee's 25). This appears to be a classic ``hawkish'' dissent. His reasoning was again rooted in establishing a systematic policy to bolster credibility, arguing that the Committee was not moving decisively enough toward a neutral stance:

\emph{``And it's a useful idea to think that there is such a thing as a neutral policy stance. I feel very strongly that we are nowhere near a neutral stance and that we ought to be aggressive in moving toward it.''}

President Jordan was not simply a ``dove'' in 1992 and a ``hawk'' in 1994. Rather, he was a consistent advocate for a credible, systematic policy framework. He dissented when he believed the Committee's actions—whether by being too timid in tightening or too discretionary in easing—failed to achieve that primary objective. Our model captures this consistency, assigning high hidden dissent scores to President Jordan in both instances (0.82 in 1992 and 0.80 in 1994), reflecting his significant deviation from the Committee's discretionary approach, regardless of the policy direction.

Besides ``No" votes that are more than hawkish or dovish, we also have ``No" votes that do not have a directional leaning.   According to the dataset compiled by \cite{thornton2014making}, out of a total of 503 dissenting votes have been recorded between March 19, 1936 and September 17, 2025, 269 are categorized as being for a tighter policy and 172 for an easier policy. The remaining 62 dissents—over 12\% of the total—could not be classified in either category.\footnote{Restricting the data to our 1976–2018 sample period, 30 of the 281 recorded dissents are classified as neither ``tighter'' nor ``easier.''} One example of such a dissent is that of President Charles Plosser in the September 17, 2014, meeting. He objected explicitly to the wording and communication strategy of the policy statement, not its immediate policy direction:

\emph{``At the last meeting I dissented over forward guidance language in alternative B...I have to renew my objection to the language, which continues, in my mind, to ignore the significant progress the economy has made toward our goals. It is not just the last number that matters. It is the accumulated progress we have made over the past year. And both the statement language and the forward guidance, in particular, fail to acknowledge that.''}

Ultimately, our hidden dissent score complements existing hawk-dove sentiment indices by capturing these nuanced, multi-dimensional disagreements. The analysis in \autoref{section4_3} will further show that this measure provides additional information not captured by traditional sentiment indices.

\section{Results}
\label{results}

\noindent Since the hidden dissent scores derived from transcripts are inferred at the individual level, a natural starting point is to examine the relationship between these scores and macroeconomic variables as well as personal characteristics. To test whether hidden dissent is related to the dual mandate (i.e., maximum employment and stable prices), we use the unemployment rate and core CPI as the target macroeconomic variables in this study.\footnote{As mentioned in \autoref{macro_variable}, the core CPI data documented in the Tealbook (formerly Greenbook), is available starting from February 12, 1986. Therefore, the following regression analyses in this paper are exclusively based on data from this date forwards.} And personal characteristics are discussed in \autoref{data_personal}. Additionally, we average the scores at the committee level and  assess whether the hidden dissent level can explain deviations of the FOMC from the optimal policy and whether financial markets respond to this information.

After obtaining the hidden dissent score $hd_{ij}$ for FOMC member $j$ in meeting $i$, we measure the hidden dissent level within each FOMC meeting by averaging the hidden dissent scores ($HD_i = \sum_{j=1}^{N} hd_{ij}$), which captures the hidden dissent intensity toward the chair. Similarly, for each meeting, we define the average of final voting results (0 or 1) as $V_i$. 

\subsection{What drives the hidden dissent inside FOMC meetings?}
\label{section4_1}

To examine how members' opinion divergence is related to macroeconomic variables and personal characteristics, we run the following regressions:
\begin{equation}
	y_{ij} = \beta_0 + \bm{\beta_1} X_i^{Macro} + \bm{\beta_2} X_{ij}^{Char} + \varepsilon_{ij},
\end{equation}

\noindent where $y_{ij}=hd_{ij},v_{ij}$. Considering the unbalanced nature of the panel data and the conditions where $hd_{ij} \in (0, 1)$ and $v_{ij} \in \{0, 1\}$, we employ mixed-effects beta panel regression for $hd_{ij}$ and mixed-effects logistic panel regression for $v_{ij}$, respectively.\footnote{For these and all the following regressions, using OLS yields similar results and does not change our conclusions.} Both regressions are clustered at the individual level and the findings are presented in \autoref{personal_level_reg_1}, showing that certain macroeconomic variables and personal characteristics can help explain the hidden dissent. In particular, as inflation is trending up, members on average have more reservations about the FOMC decision. This observation resonates with the conclusion from \citep{ball1992does} that high inflation may raise inflation uncertainty since when inflation is high, the Federal Reserve is facing a dilemma: curbing inflation or a potential recession. Consistent with \cite{bordo2023perceived}, where the member went to school matters. Regarding other personal characteristics, the findings remain inconclusive, leading us to consider these factors as control variables. For a comprehensive presentation of the results, please refer to \autoref{appendix_personal_detailed}. Furthermore, if we regress on voting results, $v_{ij}$, instead, while results for education mostly remain, macroeconomic variables do not matter anymore. 

Our analysis thus far shows that, while hidden dissent is strongly related to the inflation mandate, its connection to unemployment is surprisingly modest. This is somewhat puzzling, given that unemployment is the other key mandate of the Federal Reserve. One potential explanation, advanced by \citet{malmendier2021making}, is that members' personal experiences generate heterogeneity in their sensitivity to macroeconomic variables.\footnote{We thank an anonymous referee for the suggestion to explore this channel.} To test this hypothesis, we conduct another analysis with interaction terms between macroeconomic variables and members' formative-year experiences, estimated using a fixed-effects model:

\begin{equation}
	y_{ij} = \beta_0 + \bm{\beta_1} X_i^{Macro} + \bm{\beta_2} X_i^{Macro} \cdot X_{j}^{Exp} + \tau_j + \varepsilon_{ij},
\end{equation}
where $X_{j}^{Exp}$ represents one of three events experienced by FOMC members in their early ages: the Great Depression, World War II, or the Great Inflation. $\tau_j$ is the fixed effect for member $j$.

The results, presented in \autoref{personal_level_reg_3}, strongly support this hypothesis. The modest aggregate effect of unemployment masks significant heterogeneity driven by members' early-age experiences. Specifically, during periods of high unemployment, members who, in their early age, lived through the Great Depression or WWII (two adjacent events) are significantly more likely to support the consensus accommodative policy. In contrast, members whose views were shaped by the Great Inflation exhibit significantly greater conservatism and are more likely to express dissent from policies of trading off unemployment for inflation.

This finding not only provides further support for the mechanism in \citet{malmendier2021making} but also demonstrates the value of our hidden dissent measure. When we repeat the analysis using official votes ($v_{ij}$) as the dependent variable (see \autoref{personal_level_reg_3_appendix}), these nuanced effects disappear entirely, which demonstrates again that the official voting record is too coarse to detect these subtle yet economically important differences in policy preferences.

Next, we conduct a similar regression  as equation (1) at the meeting level to examine the relationship between the hidden dissent level $HD_i$ for meeting $i$, macroeconomic variables, and the composition of the FOMC committee’s characteristics in meetings. The committee’s characteristics at the meeting level are summarized as follows: 

(i) Standard deviations of members' experience ($D_{experience}$), ages($D_{age}$), and the endowment per student of the school where the member received her highest degree ($D_{SchoolWealth}$);

(ii) Percentage of female members ($P_{gender}$), members with highest degrees in econ-related field ($P_{major}$), and members who have lived through significant historical events (e.g., Great Depression, Great Inflation, WWII) ($P_{event}$);

(iii) Entropy measures of diversity, including the regions of members’ hometowns ($E_{hometown}$), regions of the schools where members earned their highest degrees ($E_{school}$), and the political party of the president ($E_{POTUS}$) who appointed the governor.\footnote{For chairs from regional Federal Reserve Banks, the state-level presidential election results are used as a proxy for the appointing party.}

Considering that $HD_{i} \in (0, 1)$ and $V_{i} \in [0, 1)$, we employ beta regression to model $HD_{i}$ and fractional logistic regression for $V_{i}$, respectively. The results are reported in \autoref{voting_pred_reg1}.\footnote{Consistent with the individual-level analysis, the majority of these member characteristics are treated as control variables. A detailed report on these variables is available in \autoref{appendix_personal_detailed}.} In column (1),  the macroeconomic variables account for a significant portion of the variance in hidden dissent levels $HD_i$, with a pseudo $R^{2}$ of approximately 23\%. Both lower unemployment and higher inflation are associated with increased hidden dissent. Meanwhile, column (2) shows that educational factors continue to play a crucial role at the meeting level. Beyond the locations of educations institutions, a higher proportion of FOMC members holding economics-related degrees is linked to lower level of hidden dissent. Furthermore, columns (3) and (4) reveal that the dissent measured by voting results exhibit an insignificant correlation with the macroeconomic variables after controlling personal characteristics. Our results are consistent with \citet{thornton2014making} who report that the dissent in general is not correlated with either inflation or unemployment rate.

We also consider other variables that may be correlated with the hidden dissent measure. There is evidence that yield curve can predict recessions or output growth (see, among many examples, \cite{ang2006does}). In particular, an inverted yield curve (i.e., when the long rate is lower than the short) is often viewed as an ominous sign of an upcoming recession. FOMC members may take the yield curve into account when forming their view on the economy and monetary policy. In  \autoref{voting_pred_reg1_inverted_yield_curve}, we include the yield curve either as the spread between 10-year and 2-year Treasury bonds or a dummy variable for an inverted yield curve (i.e., when the 2-year yield is higher than the 10-year one). Both variables show an insignificant correlation with the hidden dissent level. However, when we regress the hidden dissent level solely on the yield curve variable, it becomes significant, suggesting that the macroeconomic variables already capture the relevant information in the yield curve.

Lastly, we examine whether the hidden dissent level is driven by uncertainty, as measured by either the CBOE Volatility (\href{https://www.cboe.com/tradable_products/vix/}{VIX}) index or the Economic Policy Uncertainty (\href{https://www.policyuncertainty.com/}{EPU}) index. As shown in \autoref{voting_pred_reg1_uncertainty_indices}, these indices have no explanatory power, which is consistent with the finding that aggregate uncertainty is driven by monetary policy and less so the other way round (\cite{bekaert2013risk}). 

The results in this section clearly confirm that the hidden dissent captured by our model is not merely noise. Compared to dissent revealed through voting outcomes in FOMC meetings, hidden dissent shows a stronger correlation with economic conditions and certain personal characteristics.

\subsection{Is hidden dissent about what will be or what should be?}
\label{section4_2}

Having established that hidden dissent is closely tied to economic conditions, we now focus on unraveling the specific sources of this hidden dissent. Specifically, we explore whether disagreements stem more from differences in members' views on the expected trajectory of the economy or from divergences in opinions about what the appropriate monetary policy should be. To answer this, we use data from the Summary of Economic Projections (SEP) to understand how these two types of disagreements contribute to hidden dissent within the FOMC.

First, we use the data from the Summary of Economic Projections (SEP) to assess the divergence in members' view on the direction of the economy. Starting from 2007 (2012 for the federal funds rate), individual responses in the SEP are made available four times a year. The SEP contains FOMC members' anonymous assessments on future monetary policy and projections for unemployment rate, GDP growth, PCE inflation, and core PCE inflation. 

In our analysis, we categorize SEP variables into two groups: projections of the federal funds rate, which represent what the member considers as the ``appropriate monetary policy'',  and projections of other economic indicators, which indicate the members' expectations for the economic outlook. To quantify disagreement embedded within these projections, we calculate the average absolute deviation from the median projection for each category, following the approach used by \cite{foerster2023evolution}. Due to the limited sample size and high correlations among the projections, we consolidate the SEP policy disagreement measures across different horizons (current year, next year, the following year, and long term) into up to two principal component, which explains $78.61\%$ of the variations. For the economic outlook-related disagreement measures (same horizons, except no long run values for inflation measures), we use up to three principal components to capture $82.76\%$ of the total variance.

We begin by regressing our hidden dissent measure on each category of SEP disagreement separately, with the results presented in \autoref{voting_pred_reg1_dot_plot}. A clear pattern emerges: disagreement over the appropriate path of monetary policy is a strong and statistically significant predictor of hidden dissent, alone explaining over $12\%$ of its variation (Column 1). In contrast, disagreement over the economic outlook is neither statistically nor economically significant, and its explanatory power is negligible in the longer sample (Column 3). This initial evidence suggests that our hidden dissent measure primarily captures heterogeneity in members' policy preferences rather than differing views of the economic outlooks.

A potential challenge to this interpretation, however, is that the two forms of disagreement are themselves moderately correlated. The correlation of $\text{Policy}_{PC1}$ and $\text{Economy}_{PC1}$ is $-0.5640$. A member's view on the appropriate policy path is naturally informed by their economic outlook. To isolate the independent contribution of each channel and test the robustness of our initial finding, we must therefore disentangle these effects.

To achieve this, we apply the Double Machine Learning (DML) method proposed by \citet{chernozhukov2017double}. DML combines machine learning techniques with econometric models to control for confounding variables, and it aims to isolate the causal effect of a treatment variable ($T$) on the variable of interest ($Y$). The DML process involves two stages: first, machine learning algorithms are used to predict $Y$ and $T$ separately based on confounder variables $X$, and the algorithm helps us to capture the complex relationships between $Y$ or $T$ and $X$.  Next, the residuals from these predictions are regressed against each other to estimate the causal effect. In our setting, each SEP disagreement takes the role of $T$ or $X$ alternately, with our transcript hidden dissent measure ($HD_i$) being $Y$.\footnote{SEP policy disagreement is characterized by four distinct measures, while SEP economic outlook disagreement encompasses fourteen measures. We treat the first principal component of one SEP disagreement as $T$, and utilize all measures from the other SEP disagreement as $X$.}  The Lasso model is employed in the first stage, followed by linear regression in the second stage. Hyperparameter tuning is conducted to select the optimal model parameters. To mitigate overfitting and reduce estimation bias, cross-fitting is used in our model.

The second-stage results are reported in columns (5) and (6) of \autoref{voting_pred_reg1_dot_plot}. The significant coefficient for $\text{Residualized Policy}_{PC1}$  implies that the disagreement on monetary policy significantly contributes to hidden dissent inside FOMC meetings, even after accounting for disagreement on the economy outlook. This suggests that policy disagreement encompasses additional factors related to our measure. One possibility is members having different loss functions, but it is also possible that members differ on variables not covered in the SEP. In contrast,  $\text{Residualized Economy}_{PC1}$ is not statistically significant, suggesting that disagreement about the economic outlook does not have an independent influence on our measure. 

Further evidence supporting this interpretation comes from the optimal policy perturbation (OPP) proposed by \citet{barnichon2023sufficient}.\footnote{The OPP dataset spans from 1980 to 2022 and has varying frequencies due to data availability constraints. We matched the OPP data with the hidden dissent derived from the FOMC meetings held immediately before the date of the OPP data.} Based on a loss function, the median projections, and impulse responses of inflation and unemployment rates to monetary policy shocks, OPP measures the deviation of monetary policy from its optimal level. A positive OPP means that monetary policy is looser than the optimal, while a negative OPP means it is too tight. \autoref{disagree_opp} shows the correlation between the hidden dissent inside FOMC meetings and OPP measures across three different monetary policy definitions: the federal funds rate, the shadow rate, and the slope of the yield curve through quantitative easing.\footnote{To improve the readability, the hidden dissent level in \autoref{disagree_opp} is smoothed with a rolling window size equals to twelve.} Except the Financial Crisis period, during which the nominal interest rate approached the zero lower bound, significant deviations of the OPP measure from zero are frequently associated with higher level of hidden dissent. 

To closely examine their relationship, in \autoref{voting_pred_reg3_OPP} we present regression results for these three different monetary policy definitions.  We use the hidden dissent level calculated from the FOMC meeting immediately preceding the generation of the OPP measure as the independent variable. The results show that the absolute value of the OPP (federal funds rate) is strongly correlated with our hidden dissent measure. The results for the other two OPP measures appear different, primarily due to the large negative OPP values during the Financial Crisis. Once those observations are dropped, the relationship between hidden dissent and deviations from optimal policy becomes consistent across all OPP measures.

Unlike the SEP data, the OPP does not allow us to distinguish between members' preferences and their projections. The OPP is derived from a reduced-form model and a loss function where the weights on inflation and unemployment gaps are equal, and we can at best conclude that both forms of heterogeneity contribute to hidden dissent. A more precise evaluation of their relative importance will require additional SEP data in the future.

\subsection{Does hidden dissent move the market?}
\label{section4_3}

In the preceding subsections, we explored various methods to offer insights into the internal divergences of opinion among FOMC members and their relationships with economic indicators. Now, we shift our focus to understanding whether and how this hidden dissent influences financial markets.

Numerous studies have explored the impact of central bank communication on financial markets \citep{gurkaynak2005actions,rosa2011words,schmeling2016does,rosa2016fedspeak,alexopoulos2024more,gordon2024effects}. However, due to the five-year embargo on FOMC meeting transcripts, the hidden dissent level within meetings cannot be promptly revealed to the public through transcripts, making it difficult to examine how financial markets react to the hidden dissent we measured above. 

To address this issue, we consider two approaches. First, we can use public speech data from FOMC members as a proxy for hidden dissent. However, results show that public speeches by FOMC members can be informative only if we have knowledge of what the chair will say in the subsequent meeting. Without that reference point, information in the speeches becomes elusive (see \autoref{appendix_speech_data}).

Our second approach involves FOMC minutes. Unlike transcripts, minutes are released just three weeks after the FOMC meeting, allowing the public to assess the hidden dissent in a more timely manner.\footnote{Before December 2004, the minutes were released three days after the Committee's subsequent meeting.} This shorter delay makes it possible to observe immediate financial market reactions to the hidden dissent detected from the minutes, whereas the five-year delay in the release of transcripts diminishes the likelihood of any market reaction. As such, meeting minutes, despite being less detailed than transcripts, serve as a crucial source for measuring hidden dissent within the meetings and its impact on financial markets. In this subsection, we propose using hidden dissent scores, $HD_i$, to predict the hidden dissent in minutes, $HD_i^{min}$. By utilizing scores derived on transcript data, we provide a consistent, data-driven approach for estimating hidden dissent within FOMC meetings and minimize the reliance on subjective domain expertise.

To accomplish this, we employ a MHSA-based deep learning model, a modified version of the model used earlier in this paper, to predict the hidden dissent from minutes. The model architecture, illustrated in \autoref{fomc_dl_minutes}, is designed to effectively handle the complex language in FOMC minutes, and accurately extract key features to predict the hidden dissent level revealed in corresponding transcript. The training set is constructed by extracting content related to committee members from the minutes and labeling it with $HD_i$,  the hidden dissent level detected from transcript $i$. Following the training procedures in Section 3, the optimal model configuration includes six MHSA modules (each with four heads), a dropout rate of 0.46, and an initial learning rate of $4.57 \times 10^{-5}$. This setup achieves an average MAE of $5.835 \times 10^{-2}$, an $R^2$ of $0.704$ on the training set, with  $5.838 \times 10^{-2}$ and $0.697$ on the test set, respectively, based on five-fold cross-validation. We then apply the trained model to post-2018 minutes to extend  the date to the present. As shown in \autoref{disagree_trans_min} the hidden dissent level in transcripts and minutes align closely, with a correlation of $0.848$. Meanwhile, $HD_i^{min}$ shows a clear increase in hidden dissent during the second half of 2022, as the FOMC implemented four consecutive 75-basis-point rate hikes followed by a 50-basis-point hike, pushing its benchmark interest rate to the highest level in 15 years.\footnote{ \href{https://www.cnbc.com/2022/12/14/fed-rate-decision-december-2022.html}{CNBC: Fed raises interest rates half a point to highest level in 15 years}}

While the overall sentiment (hawkish/dovish tone) of FOMC communications is known to influence markets, we propose in this section that the hidden dissent provides a separate, crucial information channel reflecting policy uncertainty.\footnote{A potential concern with using generated regressors in downstream tasks, as highlighted by \citet{battaglia2024inference}, is that measurement error can bias coefficient estimates. The authors show this bias is governed by the parameter $\kappa = \sqrt{n} \times \mathbb{E}[1/C_i]$, where a small $\kappa$ indicates that the bias is likely to be minimal. In our application, the average text length ($C_i$) is large, leading to very small $\kappa$ values: approximately 0.011 for our transcript-based model and 0.006 for our minutes-based model. These values are substantially smaller than the 0.44 benchmark in the CEO application re-analyzed by \citet{battaglia2024inference}, for which they find the bias to be negligible. We therefore conclude that this issue is unlikely to be a significant concern for our estimates. We thank our anonymous referee for pointing this out.} To isolate its impact, we explicitly control for sentiment in our analysis, estimate the model following \citet{swanson2021measuring} and \citet{gorodnichenko2023voice}. 
\begin{equation}
	Outcome_{t, t+h} =  b_0^{(h)} + b_1^{(h)}HD^{min}_t + b_2^{(h)}S^{min}_t + error^{(h)}_t
\end{equation}
where $HD^{min}_t$ represents the level of hidden dissent measured from the minutes (i.e., $\hat{HD}_i$ in \autoref{fomc_dl_minutes}), and $S^{min}_t$ denotes the sentiment in the minutes, captured by the intensity of dovish or hawkish language based on the minutes’ content, following the methods of \citet{kozlowski2019geometry} and \citet{jha2021natural}. A higher value of $S^{min}_t$ indicates a more dovish-leaning sentiment in the minutes. This specification allows us to separately examine the effects of hidden dissent and sentiment on financial markets.

Following \citet{gorodnichenko2023voice}, we define the outcome variables as the financial indicators (FI) change over different horizon, $h$. We select the horizon $h = [0, 15]$ where $h=0$ represents the minute release date and estimate coefficients for each horizon separately to illustrate the dynamics of the response of these financial indicators. The list of indicators can be found in \autoref{table_financial_indicators}. Additionally, we employ the bias-corrected and accelerated (BCa) bootstrap method to correct potential biases and construct accurate confidence intervals.\footnote{We report results using the residual-based bootstrap method rather than the data-based bootstrap method. \citet{swanson2021measuring} provides a detailed explanation of why the residual-based bootstrap method is preferred in the current setting. We also conduct the data-based BCa bootstrap, and the results are similar.}

\emph{(1) Stock Market Reactions}

We begin by examining the stock market's response to hidden dissent using daily total returns of the S\&P 500 ETF (SPY). Returns are measured as the log of the close price at date $t+h$ minus the log of the open price at date $t$: $Outcome_{t, t+h}^{SPY} = \log(SPY_{t+h}^{close}) - \log(SPY_{t}^{open})$. \autoref{fig_spy} indicates that higher levels of hidden dissent predict a statistically significant decline in the SPY total return. 

Furthermore, we employ the VIX index to evaluate how stock market volatility expectations react to hidden dissent. Plots in \autoref{fig_vix} shows that on the minute release date, higher hidden dissent detected in the minutes significantly increases market volatility in the ensuing few days. This observation suggests that the stock market responds to newly disclosed information about hidden dissent, which is not conveyed through other communication channels on the FOMC meeting day. In contrast, sentiment information embedded in the minutes does not significantly affect stock market. Meanwhile, as shown in \autoref{section4_1}, VIX has no explanatory power over hidden dissent, our current findings clearly suggest a one-way influence, reinforcing the view that markets react to FOMC communications, not vice versa.

\emph{(2) Bond Market Reactions}

Next, we examine the reaction of U.S. Treasury yields to hidden dissent revealed in FOMC minutes. We analyze the cumulative change in yields over the 14 days following the minutes' release, using the 10-year bond yield (DGS10Y). Overall, the results presented in \autoref{fig_dgs10y} indicate that higher hidden dissent  drives up long-term bond yield changes. Meanwhile, we find that the impact of sentiment on Treasury bond is insignificant, suggesting that hidden dissent and sentiment transmit information through distinct channels.

Following \citet{gorodnichenko2023voice}, we further measure interest rate risk using the spread: $\log(\frac{P^{LQD}_{t+h, close}}{P^{LQD}_{t, open}}) - \log(\frac{P^{LQDH}_{t+h, close}}{P^{LQDH}_{t, open}})$, where an increase in this spread indicates higher perceived interest rate risk. \autoref{fig_lqd_lqdh} suggests that higher hidden dissent in FOMC meetings is associated with an increase in perceived interest rate risk. This suggests that greater hidden dissent may increase uncertainty about future monetary policy, economic growth, or market conditions, ultimately driving up interest rate risk. 

The above analysis demonstrates that hidden dissent, as measured through the minutes, provides additional information not captured by other FOMC communication channels and has a measurable impact on various financial markets.

\section{Conclusion}
\label{conclusion}

\noindent In this paper, we propose a deep learning model based on transfer learning and the self-attention mechanism to predict hidden dissent scores for FOMC meeting members using their transcripts. Leveraging these hidden dissent scores, we construct a measure to evaluate meeting-level hidden dissent and compare it with dissent observed in the actual voting records.

This paper has three main findings. First, hidden dissent is strongly correlated with macroeconomic variables and the education background of members, while other personal characteristics have little influence. Second, hidden dissent within FOMC meeting appears to stem from heterogeneity in both members' preferences and their projections. Third, and critically, we establish hidden dissent as a distinct information channel for financial markets, separate from overall policy sentiment. By analyzing market reactions to timely minutes releases while controlling for sentiment, we show that higher hidden dissent significantly impacts both major financial markets: in the stock market, it tends to increase volatility expectations and decrease share prices; in the bond market, it drives up yields and perceived interest rate risk. This confirms that how the committee reaches a decision, specifically the degree of underlying consensus or dissent, is valuable information for market participants beyond the stated policy or general tone.

One notable limitation of this study is that we do not differentiate the direction of hidden dissent  (i.e., whether members advocate for looser or tighter policy). Hence, the underlying reasons for some members' hidden dissent remain ambiguous.  More data, particularly additional NO votes, could help the deep learning model capture these subtle distinctions.

Another limitation comes from the change in monetary policy framework after 2008. The Federal Reserve no longer just focuses on the federal funds rate, and policy discussion touches on other tools like quantitative easing and forward guidance. FOMC members may disagree not only on the direction of monetary policy but also on the approach. With fewer than two decades of post-2008 data, addressing disagreements over not just the direction but also the approach to monetary policy remains challenging for the deep learning model.
\clearpage

\onehalfspacing

\bibliographystyle{chicago}

\bibliography{fomc_bib}

\begin{thebibliography}{}

\bibitem[\protect\citeauthoryear{Chen, Cong, and Lv}{Chen et~al.}{2022}]{chen2022long}
Chen, X., P.~Cong, and S.~Lv (2022).
\newblock A long-text classification method of chinese news based on bert and cnn.
\newblock {\em IEEE Access\/}~{\em 10}, 34046--34057.

\bibitem[\protect\citeauthoryear{Cheng, Dong, and Lapata}{Cheng et~al.}{2016}]{cheng2016long}
Cheng, J., L.~Dong, and M.~Lapata (2016).
\newblock Long short-term memory-networks for machine reading.
\newblock {\em arXiv preprint arXiv:1601.06733\/}.

\bibitem[\protect\citeauthoryear{Devlin, Chang, Lee, and Toutanova}{Devlin et~al.}{2018}]{devlin2018bert}
Devlin, J., M.-W. Chang, K.~Lee, and K.~Toutanova (2018).
\newblock Bert: Pre-training of deep bidirectional transformers for language understanding.
\newblock {\em arXiv preprint arXiv:1810.04805\/}.

\bibitem[\protect\citeauthoryear{Ding, Zhou, Yang, and Tang}{Ding et~al.}{2020}]{ding2020cogltx}
Ding, M., C.~Zhou, H.~Yang, and J.~Tang (2020).
\newblock Cogltx: Applying bert to long texts.
\newblock {\em Advances in Neural Information Processing Systems\/}~{\em 33}, 12792--12804.

\bibitem[\protect\citeauthoryear{Gentzkow, Kelly, and Taddy}{Gentzkow et~al.}{2019}]{gentzkow2019text}
Gentzkow, M., B.~Kelly, and M.~Taddy (2019).
\newblock Text as data.
\newblock {\em Journal of Economic Literature\/}~{\em 57\/}(3), 535--74.

\bibitem[\protect\citeauthoryear{Kitaev, Kaiser, and Levskaya}{Kitaev et~al.}{2020}]{kitaev2020reformer}
Kitaev, N., {\L}.~Kaiser, and A.~Levskaya (2020).
\newblock Reformer: The efficient transformer.
\newblock {\em arXiv preprint arXiv:2001.04451\/}.

\bibitem[\protect\citeauthoryear{Minaee, Kalchbrenner, Cambria, Nikzad, Chenaghlu, and Gao}{Minaee et~al.}{2021}]{minaee2021deep}
Minaee, S., N.~Kalchbrenner, E.~Cambria, N.~Nikzad, M.~Chenaghlu, and J.~Gao (2021).
\newblock Deep learning--based text classification: a comprehensive review.
\newblock {\em ACM computing surveys (CSUR)\/}~{\em 54\/}(3), 1--40.

\bibitem[\protect\citeauthoryear{Paulus, Xiong, and Socher}{Paulus et~al.}{2017}]{paulus2017deep}
Paulus, R., C.~Xiong, and R.~Socher (2017).
\newblock A deep reinforced model for abstractive summarization.
\newblock {\em arXiv preprint arXiv:1705.04304\/}.

\bibitem[\protect\citeauthoryear{Rae, Potapenko, Jayakumar, and Lillicrap}{Rae et~al.}{2019}]{rae2019compressive}
Rae, J.~W., A.~Potapenko, S.~M. Jayakumar, and T.~P. Lillicrap (2019).
\newblock Compressive transformers for long-range sequence modelling.
\newblock {\em arXiv preprint arXiv:1911.05507\/}.

\bibitem[\protect\citeauthoryear{Vaswani, Shazeer, Parmar, Uszkoreit, Jones, Gomez, Kaiser, and Polosukhin}{Vaswani et~al.}{2017}]{vaswani2017attention}
Vaswani, A., N.~Shazeer, N.~Parmar, J.~Uszkoreit, L.~Jones, A.~N. Gomez, {\L}.~Kaiser, and I.~Polosukhin (2017).
\newblock Attention is all you need.
\newblock {\em Advances in neural information processing systems\/}~{\em 30}.

\bibitem[\protect\citeauthoryear{Wang, Ng, Ma, Nallapati, and Xiang}{Wang et~al.}{2019}]{wang2019multi}
Wang, Z., P.~Ng, X.~Ma, R.~Nallapati, and B.~Xiang (2019).
\newblock Multi-passage bert: A globally normalized bert model for open-domain question answering.
\newblock {\em arXiv preprint arXiv:1908.08167\/}.

\end{thebibliography}


\begin{thebibliography}{}

\bibitem[\protect\citeauthoryear{Akiba, Sano, Yanase, Ohta, and Koyama}{Akiba et~al.}{2019}]{akiba2019optuna}
Akiba, T., S.~Sano, T.~Yanase, T.~Ohta, and M.~Koyama (2019).
\newblock Optuna: A next-generation hyperparameter optimization framework.
\newblock In {\em Proceedings of the 25th ACM SIGKDD international conference on knowledge discovery \& data mining}, pp.\  2623--2631.

\bibitem[\protect\citeauthoryear{Alexopoulos, Han, Kryvtsov, and Zhang}{Alexopoulos et~al.}{2024}]{alexopoulos2024more}
Alexopoulos, M., X.~Han, O.~Kryvtsov, and X.~Zhang (2024).
\newblock More than words: Fed chairs’ communication during congressional testimonies.
\newblock {\em Journal of Monetary Economics\/}~{\em 142}, 103515.

\bibitem[\protect\citeauthoryear{Ang, Piazzesi, and Wei}{Ang et~al.}{2006}]{ang2006does}
Ang, A., M.~Piazzesi, and M.~Wei (2006).
\newblock What does the yield curve tell us about gdp growth?
\newblock {\em Journal of Econometrics\/}~{\em 131\/}(1-2), 359--403.

\bibitem[\protect\citeauthoryear{Apel, Blix~Grimaldi, and Hull}{Apel et~al.}{2022}]{apel2022much}
Apel, M., M.~Blix~Grimaldi, and I.~Hull (2022).
\newblock How much information do monetary policy committees disclose? evidence from the fomc's minutes and transcripts.
\newblock {\em Journal of Money, Credit and Banking\/}~{\em 54\/}(5), 1459--1490.

\bibitem[\protect\citeauthoryear{Arismendi-Zambrano, Kypraios, and Paccagnini}{Arismendi-Zambrano et~al.}{2021}]{arismendi2021federal}
Arismendi-Zambrano, J., E.~Kypraios, and A.~Paccagnini (2021).
\newblock Federal reserve chair communication sentiments’ heterogeneity, personal characteristics, and their impact on uncertainty and target rate discovery.
\newblock Technical report, Technical Report ICM-2021-01, Henley Business School, ICMA Centre.

\bibitem[\protect\citeauthoryear{Ball}{Ball}{1992}]{ball1992does}
Ball, L. (1992).
\newblock Why does high inflation raise inflation uncertainty?
\newblock {\em Journal of Monetary Economics\/}~{\em 29\/}(3), 371--388.

\bibitem[\protect\citeauthoryear{Barnichon and Mesters}{Barnichon and Mesters}{2023}]{barnichon2023sufficient}
Barnichon, R. and G.~Mesters (2023).
\newblock A sufficient statistics approach for macro policy.
\newblock {\em American Economic Review\/}~{\em 113\/}(11), 2809--2845.

\bibitem[\protect\citeauthoryear{Battaglia, Christensen, Hansen, and Sacher}{Battaglia et~al.}{2024}]{battaglia2024inference}
Battaglia, L., T.~Christensen, S.~Hansen, and S.~Sacher (2024).
\newblock Inference for regression with variables generated from unstructured data.

\bibitem[\protect\citeauthoryear{Bekaert, Hoerova, and Duca}{Bekaert et~al.}{2013}]{bekaert2013risk}
Bekaert, G., M.~Hoerova, and M.~L. Duca (2013).
\newblock Risk, uncertainty and monetary policy.
\newblock {\em Journal of Monetary Economics\/}~{\em 60\/}(7), 771--788.

\bibitem[\protect\citeauthoryear{Bertsch, Hull, Lumsdaine, and Zhang}{Bertsch et~al.}{2022}]{bertsch2022central}
Bertsch, C., I.~Hull, R.~L. Lumsdaine, and X.~Zhang (2022).
\newblock Central bank mandates and monetary policy stances: through the lens of federal reserve speeches.
\newblock {\em Available at SSRN 4255978\/}.

\bibitem[\protect\citeauthoryear{Bordo and Istrefi}{Bordo and Istrefi}{2023}]{bordo2023perceived}
Bordo, M. and K.~Istrefi (2023).
\newblock Perceived fomc: The making of hawks, doves and swingers.
\newblock {\em Journal of Monetary Economics\/}~{\em 136}, 125--143.

\bibitem[\protect\citeauthoryear{Born, Ehrmann, and Fratzscher}{Born et~al.}{2014}]{born2014central}
Born, B., M.~Ehrmann, and M.~Fratzscher (2014).
\newblock Central bank communication on financial stability.
\newblock {\em The Economic Journal\/}~{\em 124\/}(577), 701--734.

\bibitem[\protect\citeauthoryear{Chernozhukov, Chetverikov, Demirer, Duflo, Hansen, and Newey}{Chernozhukov et~al.}{2017}]{chernozhukov2017double}
Chernozhukov, V., D.~Chetverikov, M.~Demirer, E.~Duflo, C.~Hansen, and W.~Newey (2017).
\newblock Double/debiased/neyman machine learning of treatment effects.
\newblock {\em American Economic Review\/}~{\em 107\/}(5), 261--265.

\bibitem[\protect\citeauthoryear{Cieslak, Hansen, McMahon, and Xiao}{Cieslak et~al.}{2021}]{cieslak2021policymakers}
Cieslak, A., S.~Hansen, M.~McMahon, and S.~Xiao (2021).
\newblock Policymakers' uncertainty.
\newblock {\em Available at SSRN 3936999\/}.

\bibitem[\protect\citeauthoryear{Curti and Kazinnik}{Curti and Kazinnik}{2023}]{curti2023let}
Curti, F. and S.~Kazinnik (2023).
\newblock Let's face it: Quantifying the impact of nonverbal communication in fomc press conferences.
\newblock {\em Journal of Monetary Economics\/}.

\bibitem[\protect\citeauthoryear{Dell}{Dell}{2024}]{dell2024deep}
Dell, M. (2024).
\newblock Deep learning for economists.
\newblock {\em Journal of Economic Literature\/}.

\bibitem[\protect\citeauthoryear{Foerster and Martinez}{Foerster and Martinez}{2023}]{foerster2023evolution}
Foerster, A. and Z.~Martinez (2023).
\newblock The evolution of disagreement in the dot plot.
\newblock {\em Evolution\/}~{\em 2023}, 21.

\bibitem[\protect\citeauthoryear{Gordon and Lunsford}{Gordon and Lunsford}{2024}]{gordon2024effects}
Gordon, M.~V. and K.~G. Lunsford (2024).
\newblock The effects of the federal reserve chair’s testimony on interest rates and stock prices.
\newblock {\em Economics Letters\/}~{\em 235}, 111537.

\bibitem[\protect\citeauthoryear{Gorodnichenko, Pham, and Talavera}{Gorodnichenko et~al.}{2023}]{gorodnichenko2023voice}
Gorodnichenko, Y., T.~Pham, and O.~Talavera (2023).
\newblock The voice of monetary policy.
\newblock {\em American Economic Review\/}~{\em 113\/}(2), 548--584.

\bibitem[\protect\citeauthoryear{G{\"u}rkaynak, Sack, and Swansonc}{G{\"u}rkaynak et~al.}{2005}]{gurkaynak2005actions}
G{\"u}rkaynak, R.~S., B.~Sack, and E.~T. Swansonc (2005).
\newblock Do actions speak louder than words? the response of asset prices to monetary policy actions and statements.
\newblock {\em International Journal of Central Banking\/}.

\bibitem[\protect\citeauthoryear{Handlan}{Handlan}{2020}]{handlan2020text}
Handlan, A. (2020).
\newblock Text shocks and monetary surprises: Text analysis of fomc statements with machine learning.
\newblock {\em Published Manuscript\/}.

\bibitem[\protect\citeauthoryear{Hansen and McMahon}{Hansen and McMahon}{2016}]{hansen2016shocking}
Hansen, S. and M.~McMahon (2016).
\newblock Shocking language: Understanding the macroeconomic effects of central bank communication.
\newblock {\em Journal of International Economics\/}~{\em 99}, S114--S133.

\bibitem[\protect\citeauthoryear{Hansen, McMahon, and Prat}{Hansen et~al.}{2018}]{hansen2018transparency}
Hansen, S., M.~McMahon, and A.~Prat (2018).
\newblock Transparency and deliberation within the fomc: A computational linguistics approach.
\newblock {\em The Quarterly Journal of Economics\/}~{\em 133\/}(2), 801--870.

\bibitem[\protect\citeauthoryear{Hansen, McMahon, and Tong}{Hansen et~al.}{2019}]{hansen2019long}
Hansen, S., M.~McMahon, and M.~Tong (2019).
\newblock The long-run information effect of central bank communication.
\newblock {\em Journal of Monetary Economics\/}~{\em 108}, 185--202.

\bibitem[\protect\citeauthoryear{Husted, Rogers, and Sun}{Husted et~al.}{2020}]{husted2020monetary}
Husted, L., J.~Rogers, and B.~Sun (2020).
\newblock Monetary policy uncertainty.
\newblock {\em Journal of Monetary Economics\/}~{\em 115}, 20--36.

\bibitem[\protect\citeauthoryear{Istrefi}{Istrefi}{2019}]{istrefi2019fed}
Istrefi, K. (2019).
\newblock In fed watchers’ eyes: Hawks, doves and monetary policy.

\bibitem[\protect\citeauthoryear{Jha, Liu, and Manela}{Jha et~al.}{2021}]{jha2021natural}
Jha, M., H.~Liu, and A.~Manela (2021).
\newblock Natural disaster effects on popular sentiment toward finance.
\newblock {\em Journal of Financial and Quantitative Analysis\/}~{\em 56\/}(7), 2584--2604.

\bibitem[\protect\citeauthoryear{Jurafsky and Martin}{Jurafsky and Martin}{2023}]{jurafsky2023speech}
Jurafsky, D. and J.~H. Martin (2023).
\newblock Speech and language processing (3rd ed. draft).

\bibitem[\protect\citeauthoryear{Kahn and Oksol}{Kahn and Oksol}{2018}]{kahn2018understanding}
Kahn, G.~A. and A.~Oksol (2018).
\newblock Understanding hawks and doves.
\newblock {\em Macro Bulletin\/}~(June 27, 2018), 1--4.

\bibitem[\protect\citeauthoryear{Kamath, Liu, and Whitaker}{Kamath et~al.}{2019}]{kamath2019deep}
Kamath, U., J.~Liu, and J.~Whitaker (2019).
\newblock {\em Deep learning for NLP and speech recognition}, Volume~84.
\newblock Springer.

\bibitem[\protect\citeauthoryear{Kozlowski, Taddy, and Evans}{Kozlowski et~al.}{2019}]{kozlowski2019geometry}
Kozlowski, A.~C., M.~Taddy, and J.~A. Evans (2019).
\newblock The geometry of culture: Analyzing the meanings of class through word embeddings.
\newblock {\em American Sociological Review\/}~{\em 84\/}(5), 905--949.

\bibitem[\protect\citeauthoryear{Lauriola, Lavelli, and Aiolli}{Lauriola et~al.}{2022}]{lauriola2022introduction}
Lauriola, I., A.~Lavelli, and F.~Aiolli (2022).
\newblock An introduction to deep learning in natural language processing: Models, techniques, and tools.
\newblock {\em Neurocomputing\/}~{\em 470}, 443--456.

\bibitem[\protect\citeauthoryear{Leombroni, Vedolin, Venter, and Whelan}{Leombroni et~al.}{2021}]{leombroni2021central}
Leombroni, M., A.~Vedolin, G.~Venter, and P.~Whelan (2021).
\newblock Central bank communication and the yield curve.
\newblock {\em Journal of Financial Economics\/}~{\em 141\/}(3), 860--880.

\bibitem[\protect\citeauthoryear{Lunsford}{Lunsford}{2020}]{lunsford2020policy}
Lunsford, K.~G. (2020).
\newblock Policy language and information effects in the early days of federal reserve forward guidance.
\newblock {\em American Economic Review\/}~{\em 110\/}(9), 2899--2934.

\bibitem[\protect\citeauthoryear{Madeira, Madeira, and Monteiro}{Madeira et~al.}{2023}]{madeira2023origins}
Madeira, C., J.~Madeira, and P.~S. Monteiro (2023).
\newblock The origins of monetary policy disagreement: the role of supply and demand shocks.
\newblock {\em Review of economics and statistics\/}, 1--45.

\bibitem[\protect\citeauthoryear{Malmendier, Nagel, and Yan}{Malmendier et~al.}{2021}]{malmendier2021making}
Malmendier, U., S.~Nagel, and Z.~Yan (2021).
\newblock The making of hawks and doves.
\newblock {\em Journal of Monetary Economics\/}~{\em 117}, 19--42.

\bibitem[\protect\citeauthoryear{Reimers and Gurevych}{Reimers and Gurevych}{2019}]{reimers2019sentence}
Reimers, N. and I.~Gurevych (2019).
\newblock Sentence-bert: Sentence embeddings using siamese bert-networks.
\newblock {\em arXiv preprint arXiv:1908.10084\/}.

\bibitem[\protect\citeauthoryear{Riboni and Ruge-Murcia}{Riboni and Ruge-Murcia}{2014}]{riboni2014dissent}
Riboni, A. and F.~Ruge-Murcia (2014).
\newblock Dissent in monetary policy decisions.
\newblock {\em Journal of Monetary Economics\/}~{\em 66}, 137--154.

\bibitem[\protect\citeauthoryear{Romer and Romer}{Romer and Romer}{2008}]{romer2008fomc}
Romer, C.~D. and D.~H. Romer (2008).
\newblock The fomc versus the staff: where can monetary policymakers add value?
\newblock {\em American Economic Review\/}~{\em 98\/}(2), 230--235.

\bibitem[\protect\citeauthoryear{Rosa}{Rosa}{2011}]{rosa2011words}
Rosa, C. (2011).
\newblock Words that shake traders: The stock market's reaction to central bank communication in real time.
\newblock {\em Journal of Empirical Finance\/}~{\em 18\/}(5), 915--934.

\bibitem[\protect\citeauthoryear{Rosa}{Rosa}{2016}]{rosa2016fedspeak}
Rosa, C. (2016).
\newblock Fedspeak: Who moves us asset prices?
\newblock {\em International Journal of Central Banking\/}~{\em 12\/}(4), 223--261.

\bibitem[\protect\citeauthoryear{Schmeling and Wagner}{Schmeling and Wagner}{2016}]{schmeling2016does}
Schmeling, M. and C.~Wagner (2016).
\newblock Does central bank tone move asset prices?
\newblock {\em Journal of Financial and Quantitative Analysis\/}, 1--48.

\bibitem[\protect\citeauthoryear{Shah, Paturi, and Chava}{Shah et~al.}{2023}]{shah2023trillion}
Shah, A., S.~Paturi, and S.~Chava (2023).
\newblock Trillion dollar words: A new financial dataset, task \& market analysis.
\newblock {\em arXiv preprint arXiv:2305.07972\/}.

\bibitem[\protect\citeauthoryear{Shapiro, Sudhof, and Wilson}{Shapiro et~al.}{2022}]{shapiro2022measuring}
Shapiro, A.~H., M.~Sudhof, and D.~J. Wilson (2022).
\newblock Measuring news sentiment.
\newblock {\em Journal of Econometrics\/}~{\em 228\/}(2), 221--243.

\bibitem[\protect\citeauthoryear{Shapiro and Wilson}{Shapiro and Wilson}{2022}]{shapiro2022taking}
Shapiro, A.~H. and D.~J. Wilson (2022).
\newblock Taking the fed at its word: A new approach to estimating central bank objectives using text analysis.
\newblock {\em The Review of Economic Studies\/}~{\em 89\/}(5), 2768--2805.

\bibitem[\protect\citeauthoryear{Swanson}{Swanson}{2021}]{swanson2021measuring}
Swanson, E.~T. (2021).
\newblock Measuring the effects of federal reserve forward guidance and asset purchases on financial markets.
\newblock {\em Journal of Monetary Economics\/}~{\em 118}, 32--53.

\bibitem[\protect\citeauthoryear{Thornton and Wheelock}{Thornton and Wheelock}{2014}]{thornton2014making}
Thornton, D.~L. and D.~C. Wheelock (2014).
\newblock Making sense of dissents: a history of fomc dissents.
\newblock {\em Federal Reserve Bank of St. Louis Review\/}~{\em 96\/}(3), 213--227.

\bibitem[\protect\citeauthoryear{Tobback, Nardelli, and Martens}{Tobback et~al.}{2017}]{tobback2017between}
Tobback, E., S.~Nardelli, and D.~Martens (2017).
\newblock Between hawks and doves: measuring central bank communication.

\bibitem[\protect\citeauthoryear{Todd}{Todd}{2016}]{timtodd2016}
Todd, T. (2016).
\newblock {\em A Corollary of Accountability: A History of FOMC Policy Communication}.
\newblock Federal Reserve Bank of Kansas City.

\bibitem[\protect\citeauthoryear{Vaswani, Shazeer, Parmar, Uszkoreit, Jones, Gomez, Kaiser, and Polosukhin}{Vaswani et~al.}{2017}]{vaswani2017attention}
Vaswani, A., N.~Shazeer, N.~Parmar, J.~Uszkoreit, L.~Jones, A.~N. Gomez, {\L}.~Kaiser, and I.~Polosukhin (2017).
\newblock Attention is all you need.
\newblock {\em Advances in neural information processing systems\/}~{\em 30}.

\end{thebibliography}

\clearpage

\begin{figure}[hbtp!]
    \begin{center}
	\includegraphics[width=15cm]{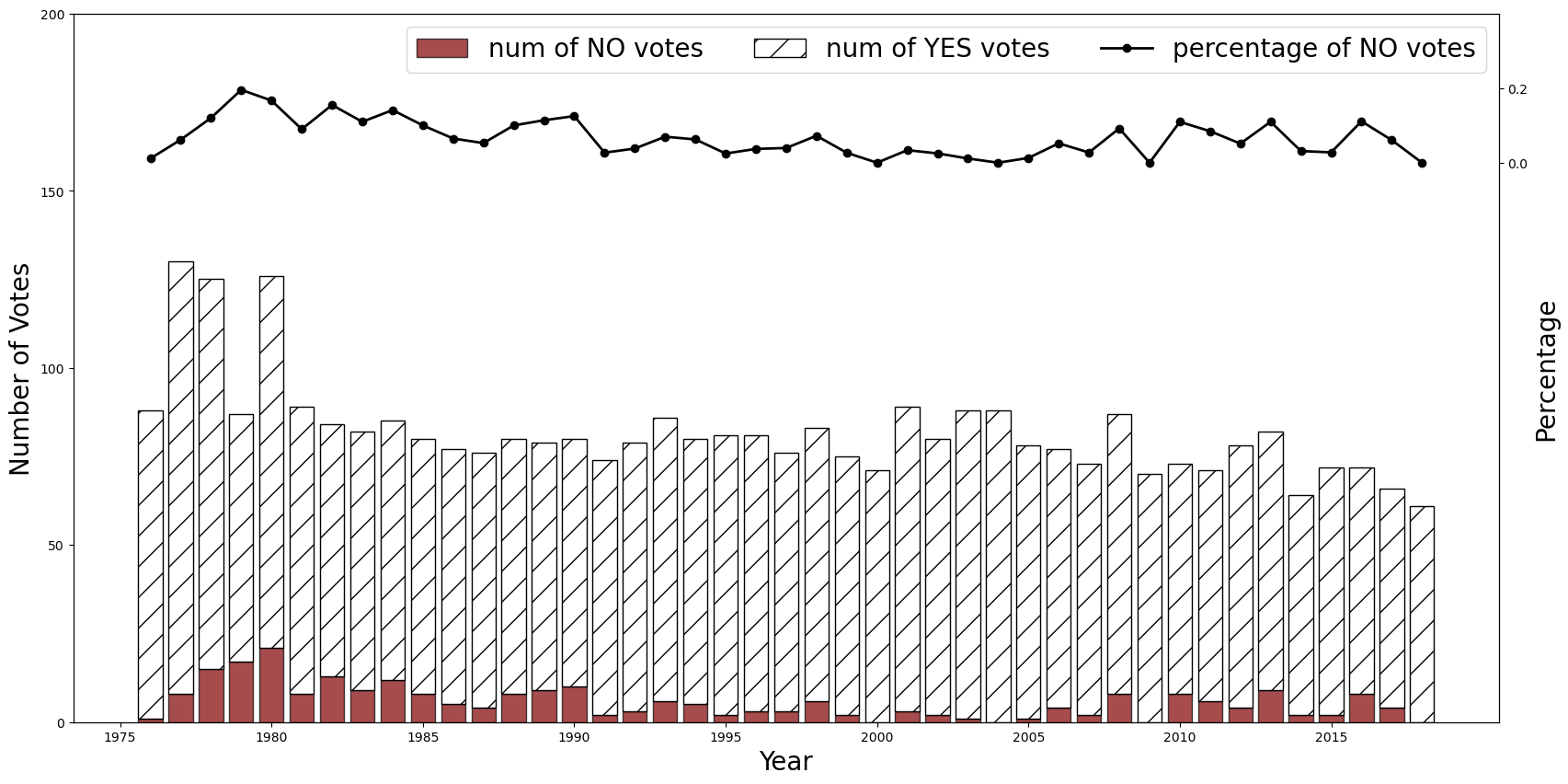}
	\caption{Number of YES and NO Votes in Each Year}
    \label{vote_stat}
    \end{center}
	\emph{Note:} The bar chart illustrates the annual distribution of YES and NO votes by FOMC members. Red bars indicate NO votes, while transparent, striped bars represent YES votes. The dotted line shows the annual percentage of NO votes.
\end{figure}

\begin{figure}[hbtp!]
    \begin{center}
	\includegraphics[width=15cm]{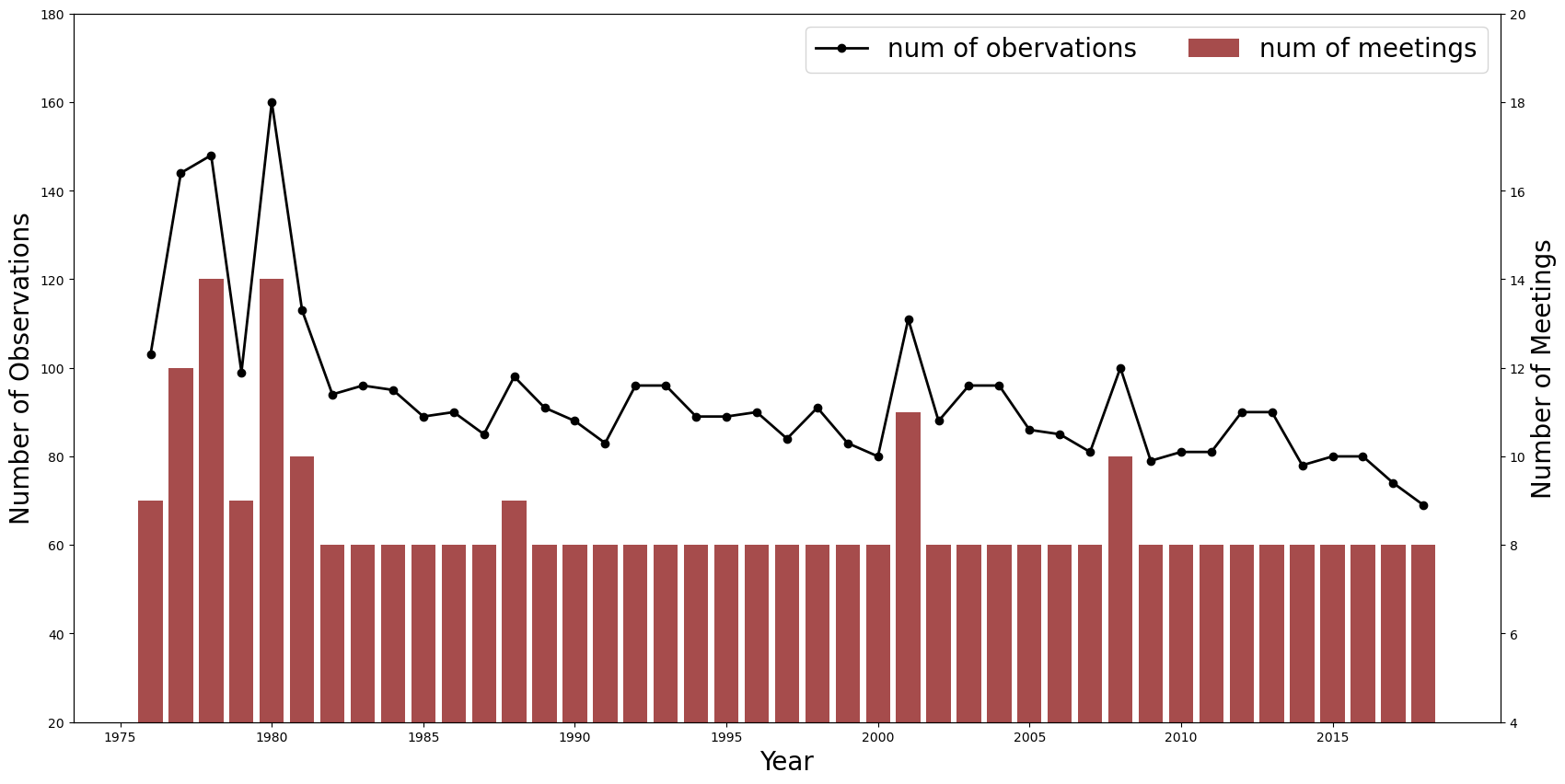}
	\caption{Numbers of Observations and Meetings in Each Year} 
    \label{num_obs_meetings}
    \end{center}
	\emph{Note:} The bar chart displays the annual count of FOMC meetings, with the black dotted line showing the number of voting members (observations) per year. While the number of meetings remains relatively stable in recent years, slight fluctuations in observations reflect varying attendance across meetings.
\end{figure}

\clearpage
\begin{landscape}
\vspace* {\fill}
	\begin{figure}[hbtp!]
        \begin{center}
		\includegraphics[width=1.4\textwidth]{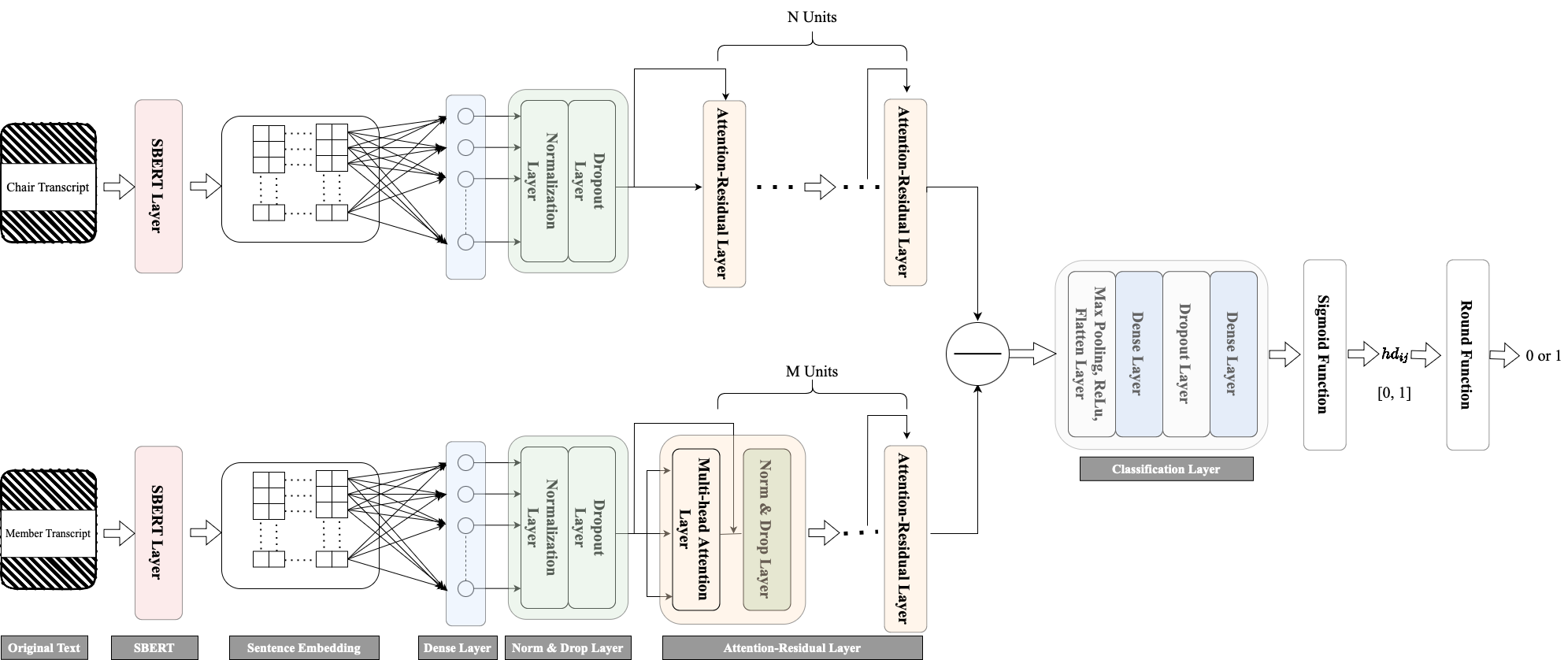}
		\caption{FOMC Deep Learning Model} 
        \label{fomc_dl}
        \end{center}
		\emph{Note:} This figure illustrates the Deep Learning model used in this paper. The SBERT layer transforms the original texts into sentence embeddings, high-dimensional arrays that can be processed by subsequent modules. Multiple multi-head self-attention modules, dense layers, and dropout layers are then applied to extract relevant features. For a detailed discussion, see \autoref{model structure}.
	\end{figure}
	\vfill
\end{landscape}	
\clearpage

\begin{figure}[hbtp!]
    \begin{center}
	\includegraphics[width=16cm]{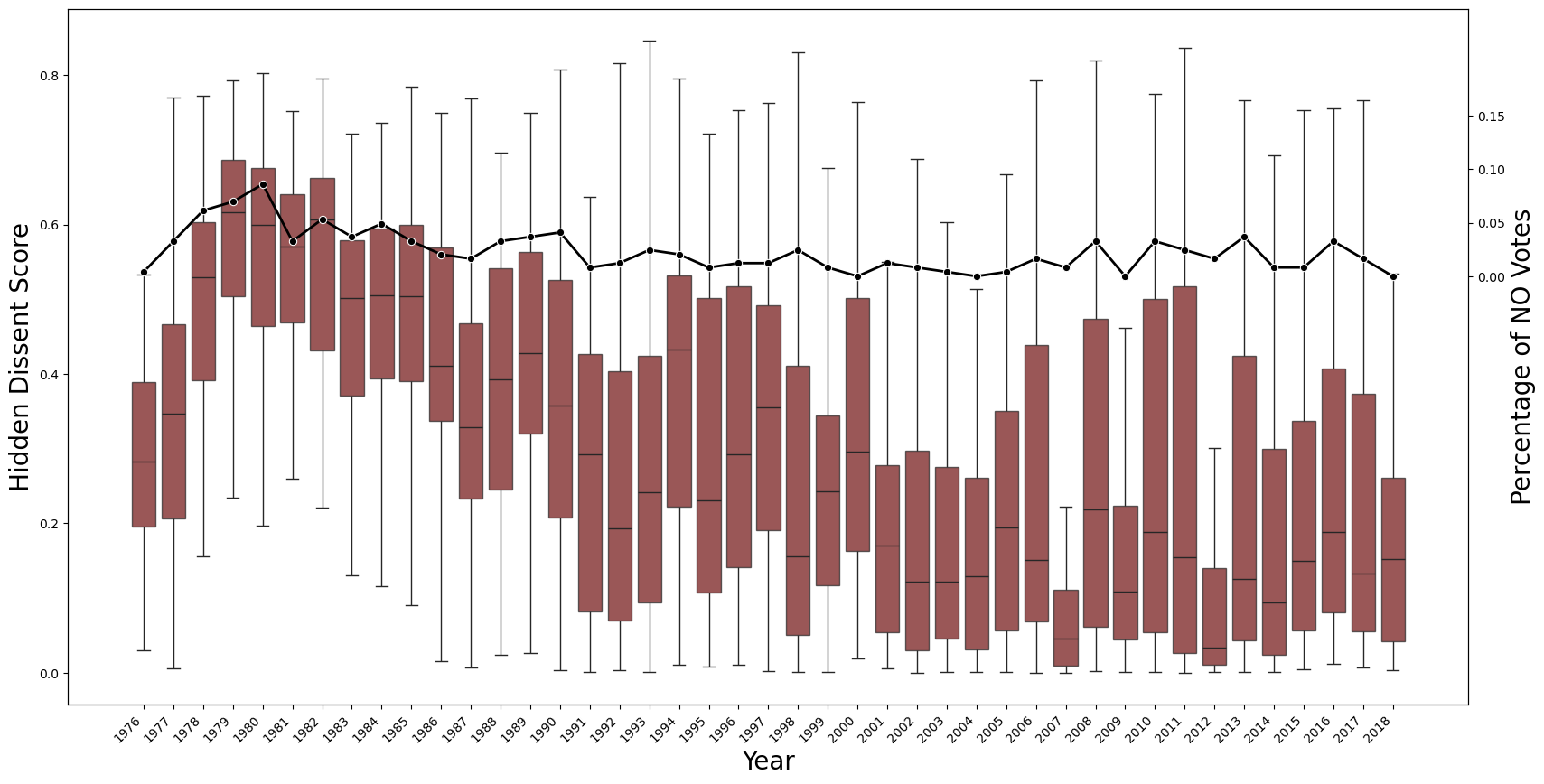}
	\caption{Hidden Dissent Score Distribution} 
    \label{pred_distr}
    \end{center}
	\emph{Note:} This figure presents the boxplot of hidden dissent scores by year. The red box shows the interquartile range (Q1 to Q3), with the black line indicating the median (Q2), and whiskers extending up to 1.5 times the interquartile range (Q3-Q1). The black dotted line reflects the percentage of total NO votes cast in each year.
\end{figure}

\begin{figure}[hbtp!]
	\begin{subfigure}{0.48\linewidth}
		\centering
		\includegraphics[width=1\textwidth]{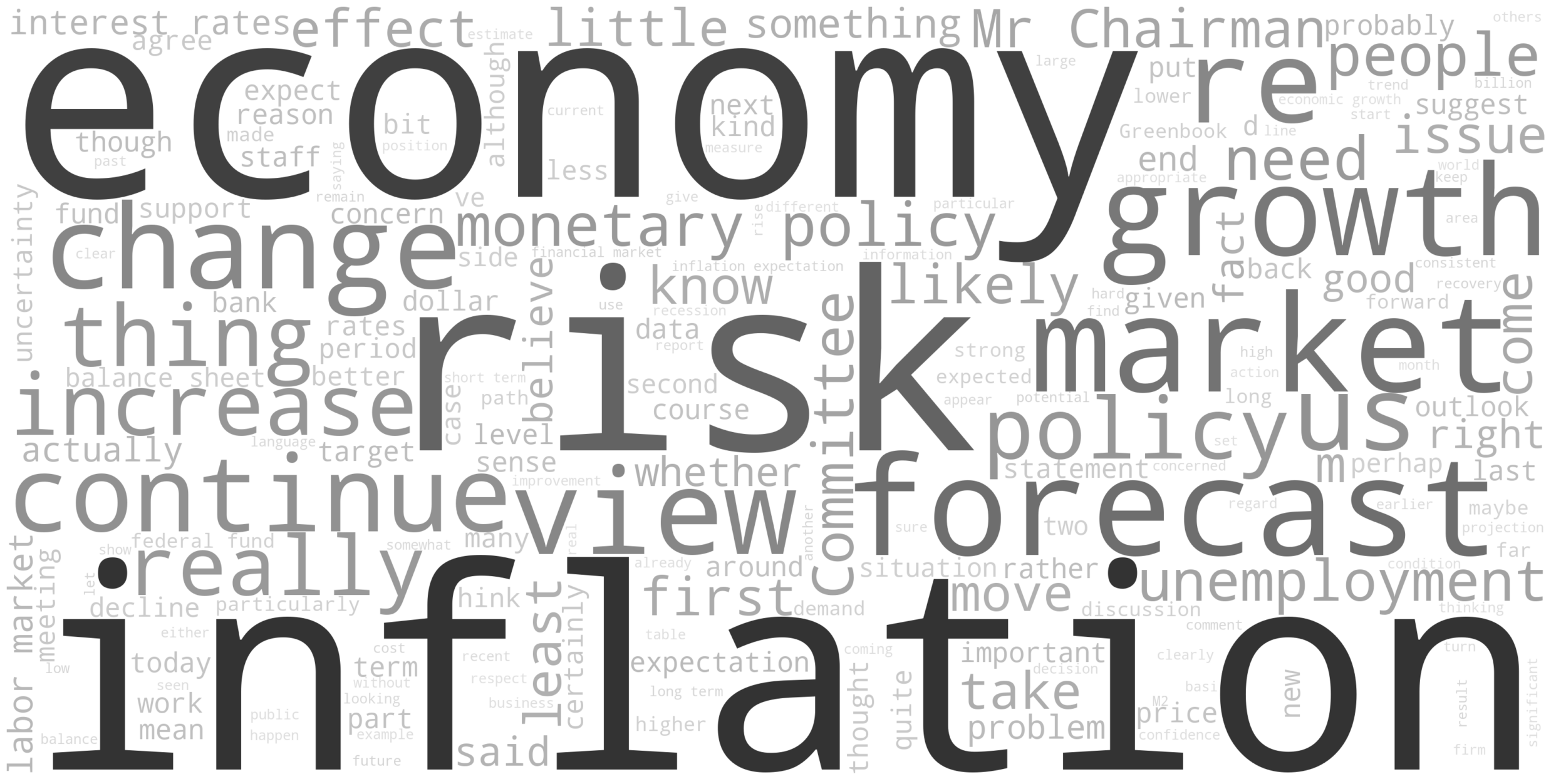}
		\caption{Hidden Dissent Scores ($<0.5$), Full Sample}
	\end{subfigure}
	\begin{subfigure}{0.48\linewidth}
		\centering
		\includegraphics[width=1\textwidth]{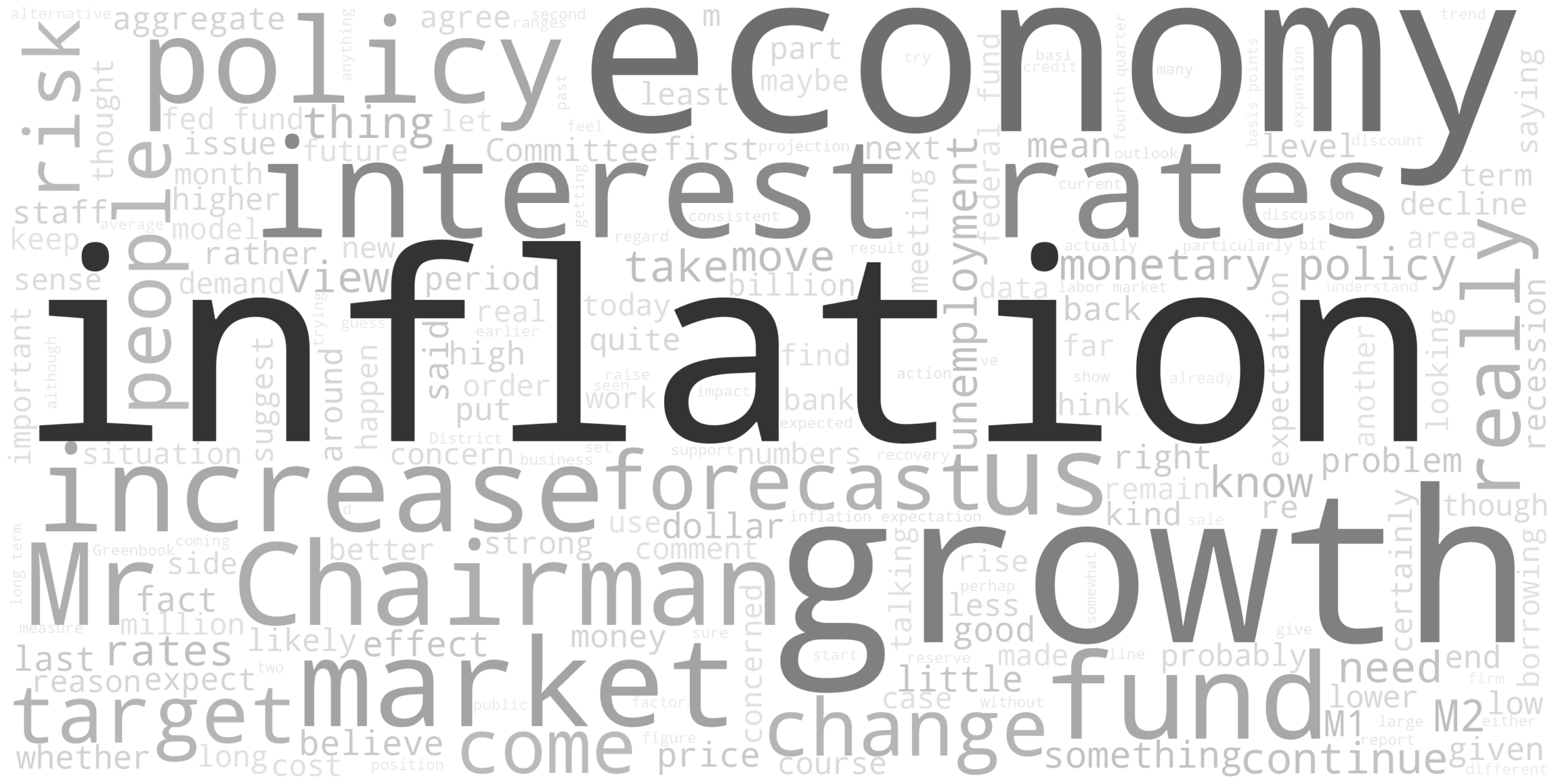}
		\caption{Hidden Dissent Scores ($\geq 0.5$), Full Sample}	
	\end{subfigure}
	\begin{subfigure}{0.48\linewidth}
		\centering
		\includegraphics[width=1\textwidth]{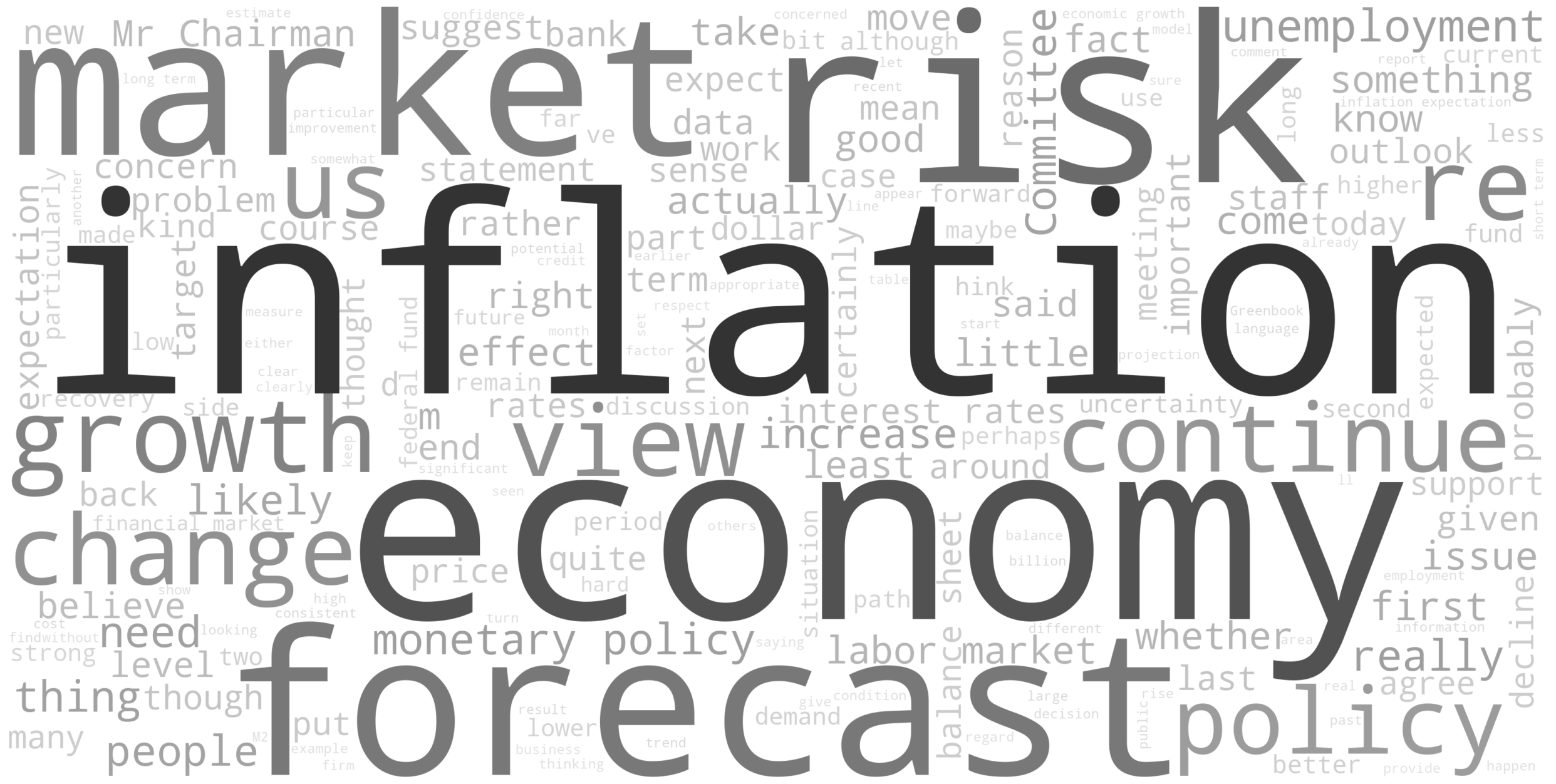}
		\caption{Hidden Dissent Scores ($<0.5$), Rate Unchanged}
	\end{subfigure}
	\begin{subfigure}{0.48\linewidth}
		\centering
		\includegraphics[width=1\textwidth]{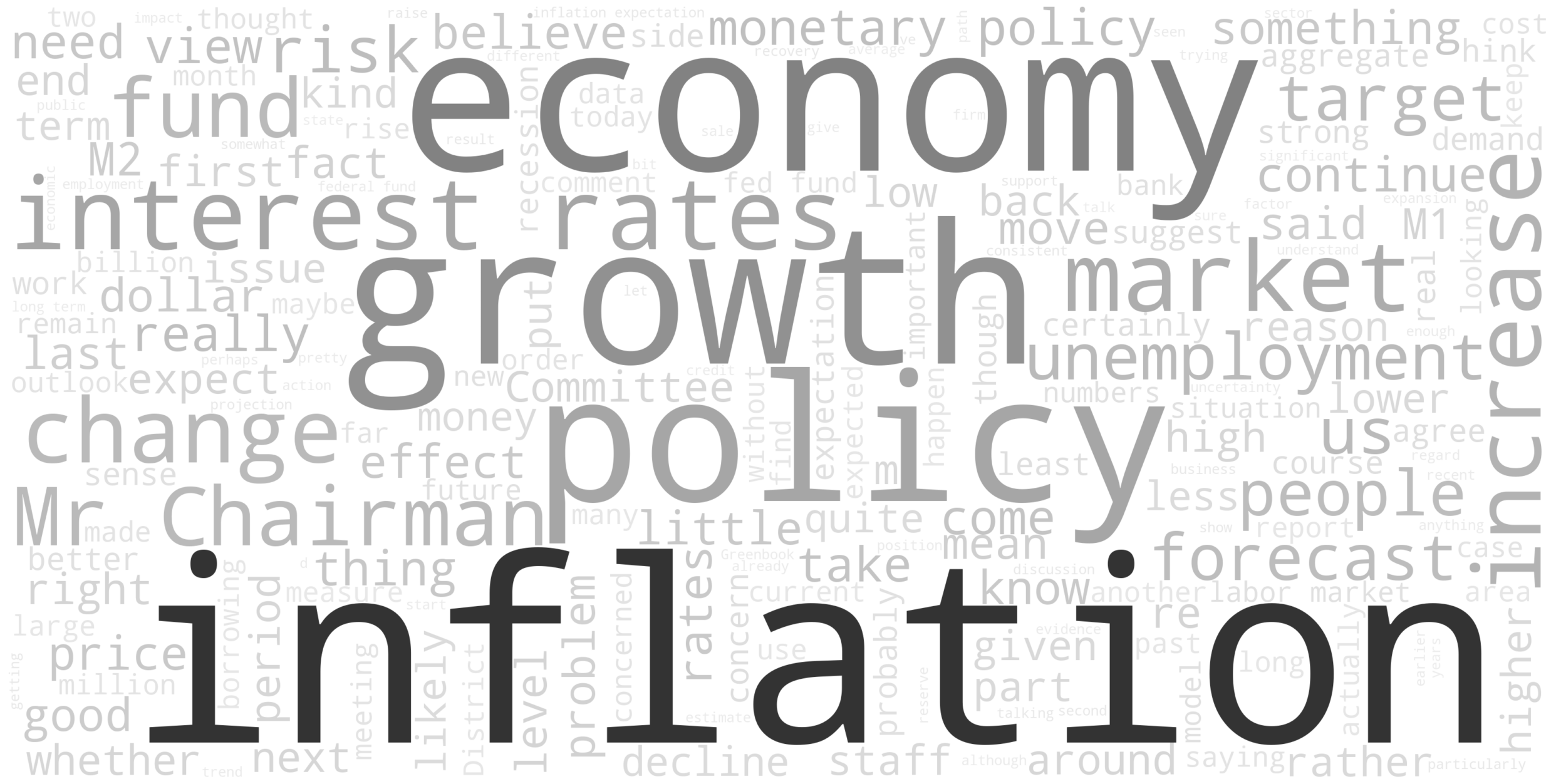}
		\caption{Hidden Dissent Scores ($\geq 0.5$), Rate Unchanged}
	\end{subfigure}
	\begin{subfigure}{0.48\linewidth}
		\centering
		\includegraphics[width=1\textwidth]{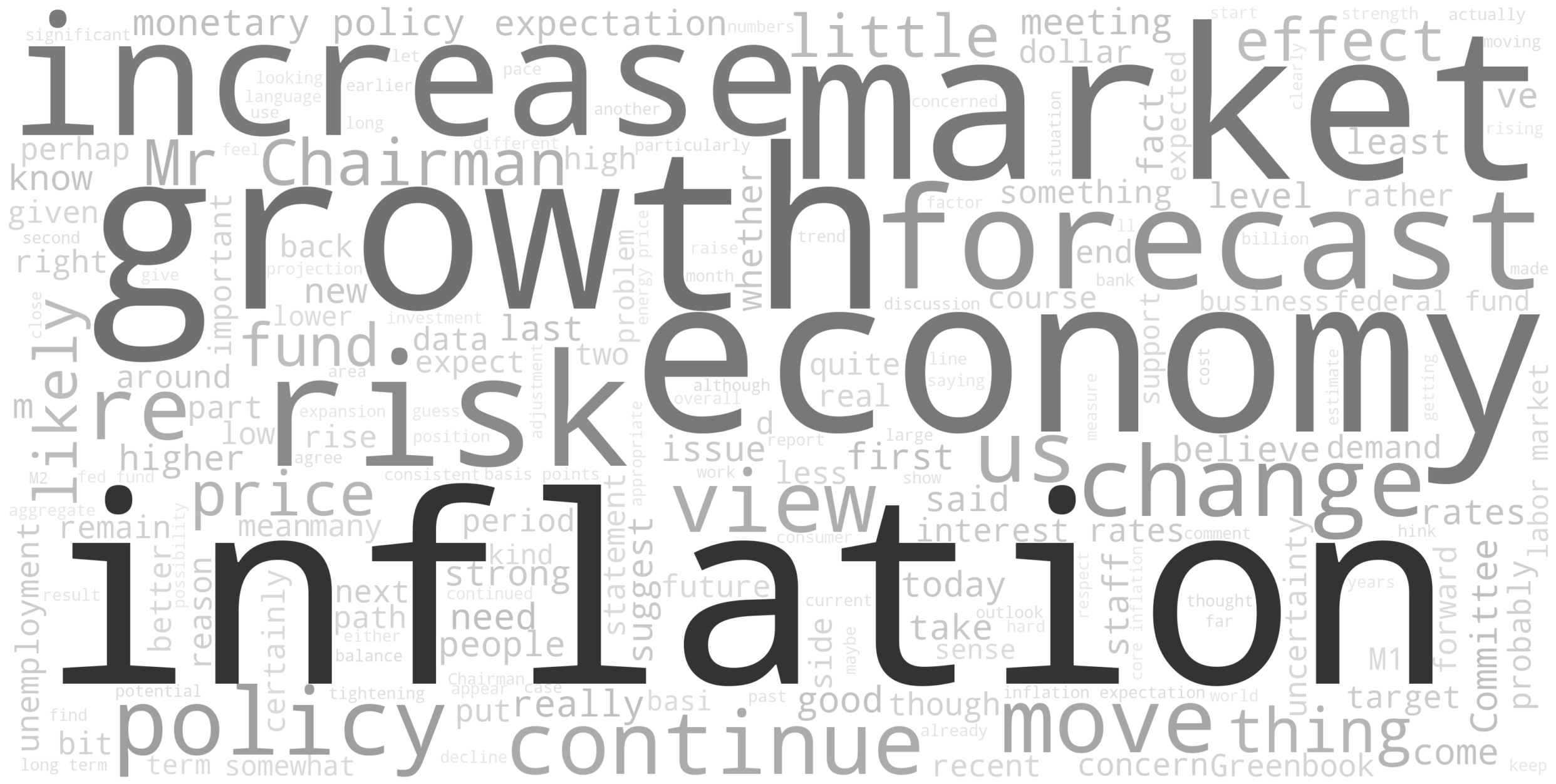}
		\caption{Hidden Dissent Scores ($<0.5$), Rate Increases}
	\end{subfigure}
	\begin{subfigure}{0.48\linewidth}
		\centering
		\includegraphics[width=1\textwidth]{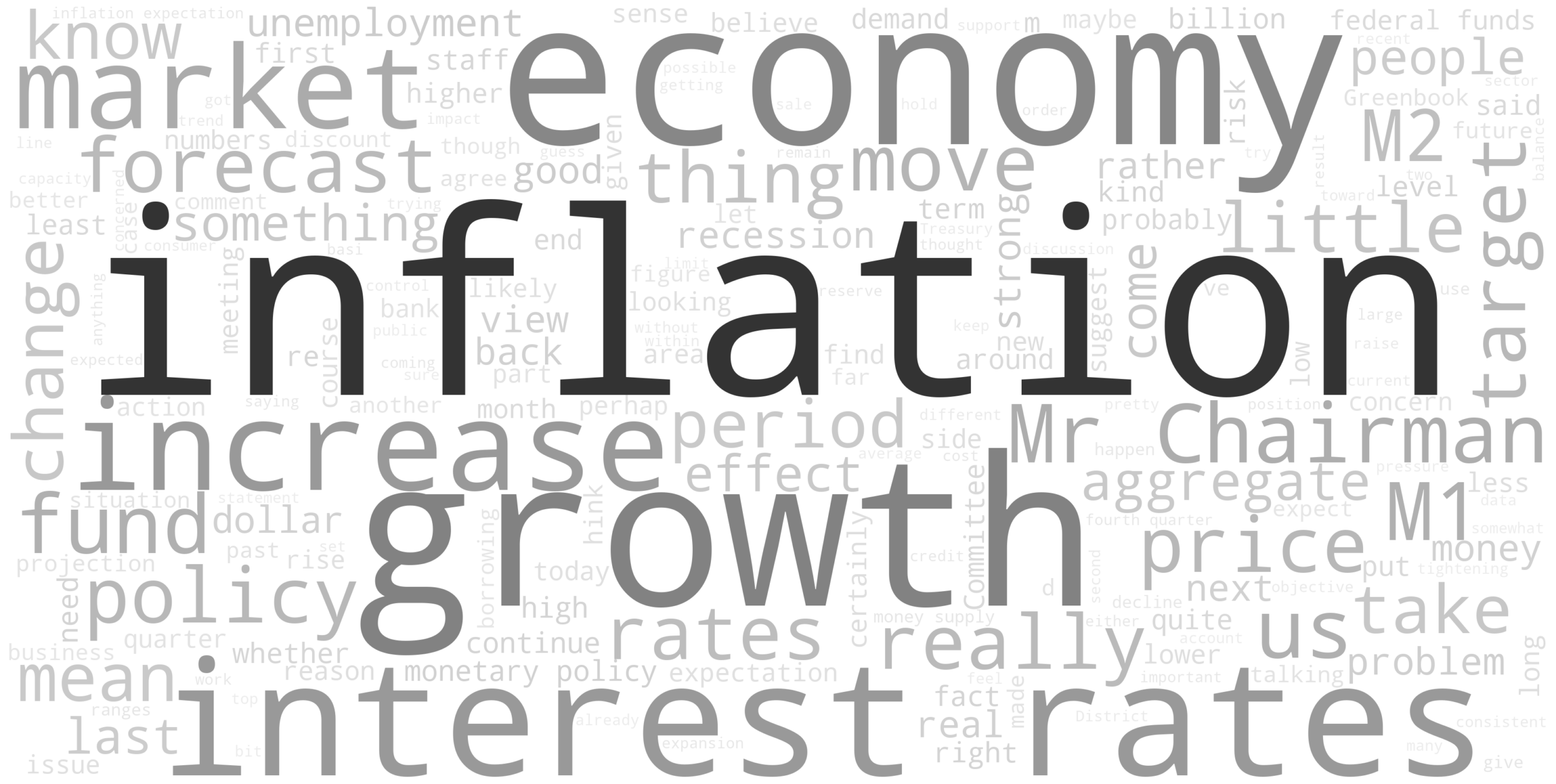}
		\caption{Hidden Dissent Scores ($\geq 0.5$), Rate Increases}
	\end{subfigure}
	\begin{subfigure}{0.48\linewidth}
		\centering
		\includegraphics[width=1\textwidth]{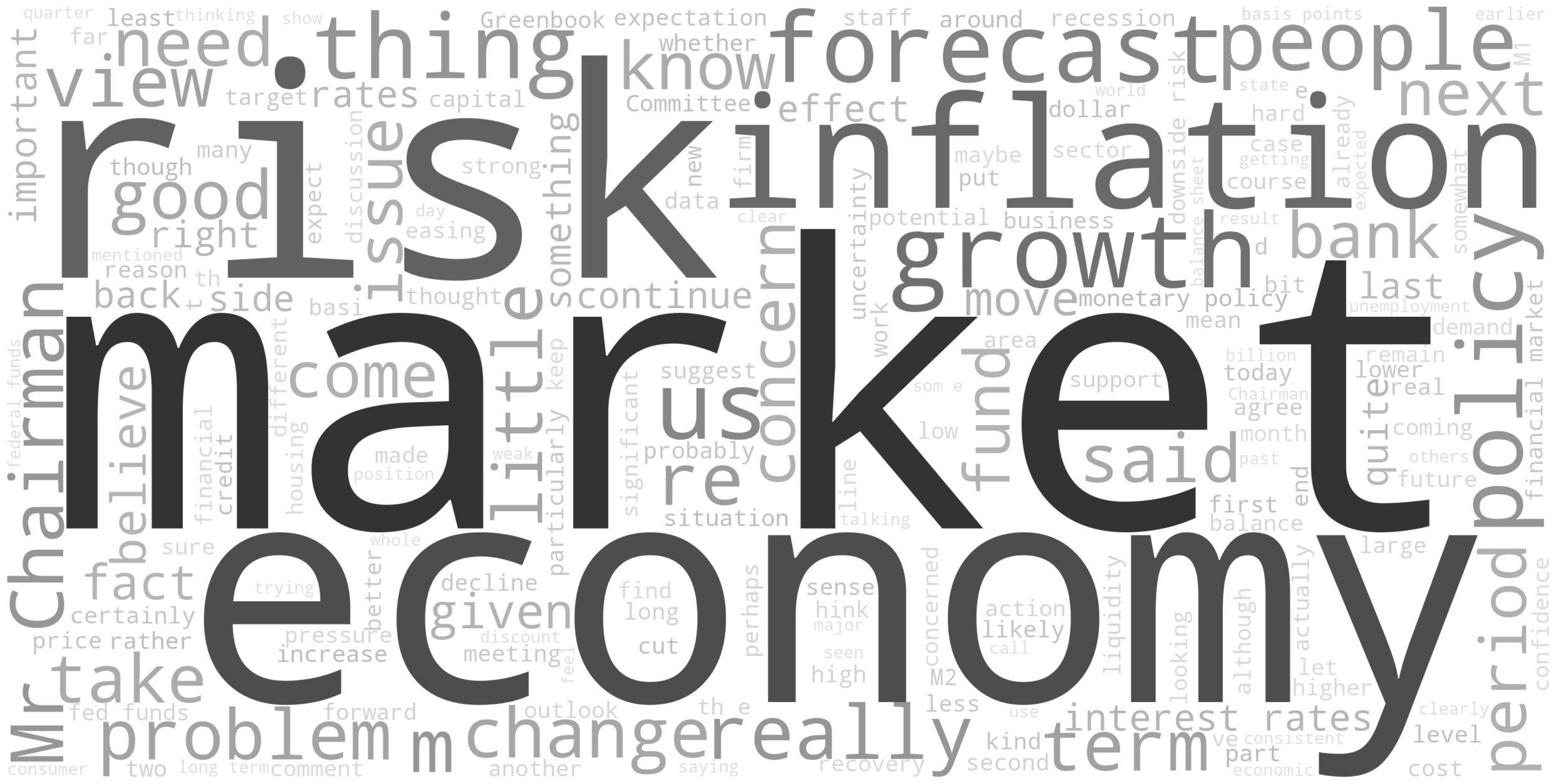}
		\caption{Hidden Dissent Scores ($<0.5$), Rate Decreases}
	\end{subfigure}
	\begin{subfigure}{0.48\linewidth}
		\centering
		\includegraphics[width=1\textwidth]{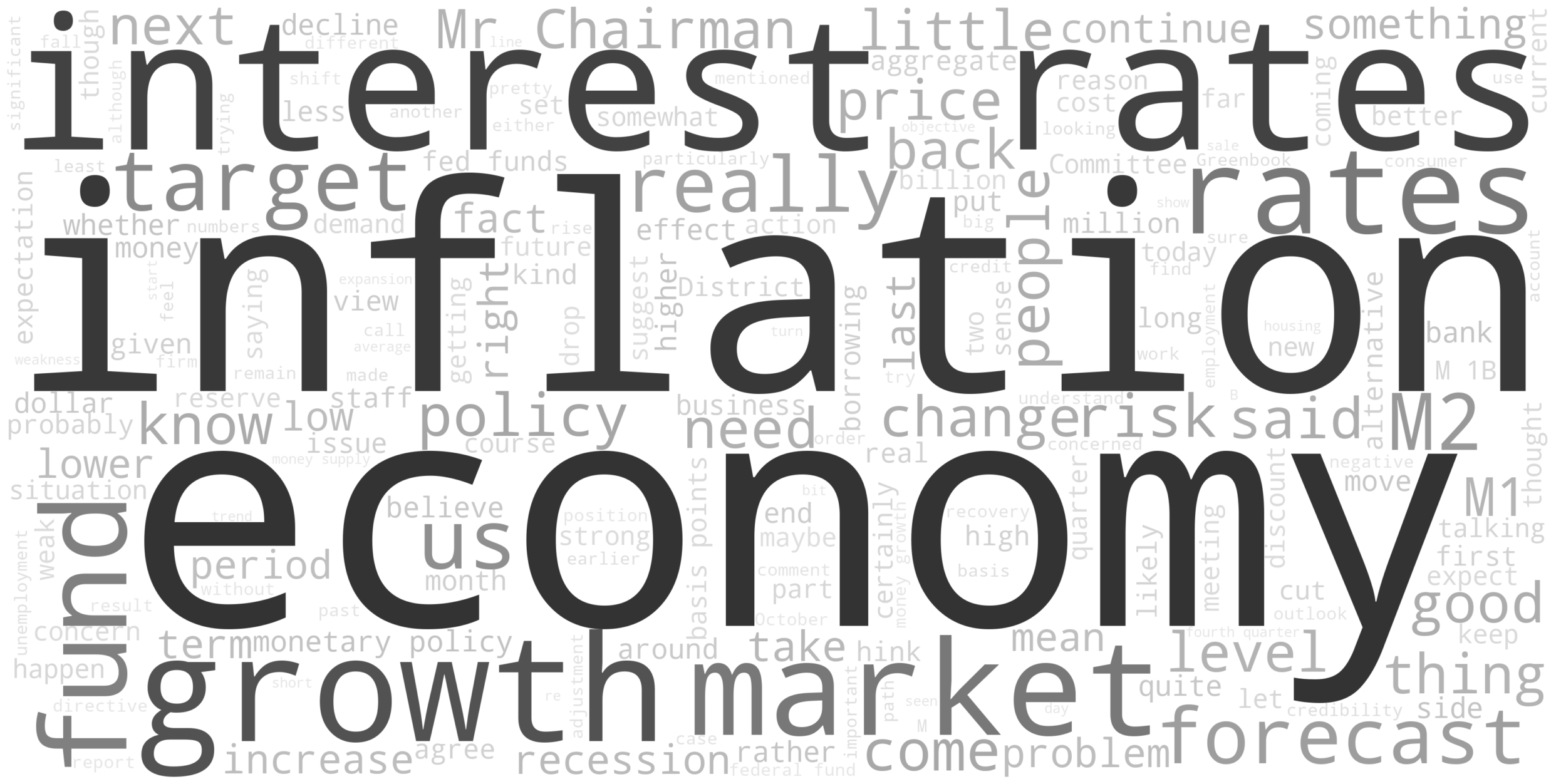}
		\caption{Hidden Dissent Scores ($\geq 0.5$), Rate Decreases}
	\end{subfigure}
	\caption{The Word Cloud for Groups of Different Hidden Dissent Scores}
	\emph{Note:} This figure displays word clouds for different groups categorized by hidden dissent scores and FOMC policy actions—rate increase, decrease, or no change. In each word cloud, both the gray shading and word size reflect their frequency in meeting transcripts.
	\label{wordcloud_voting_prediciton}
\end{figure}

\begin{figure}[hbtp!]
	\begin{center}
		\includegraphics[width=15cm]{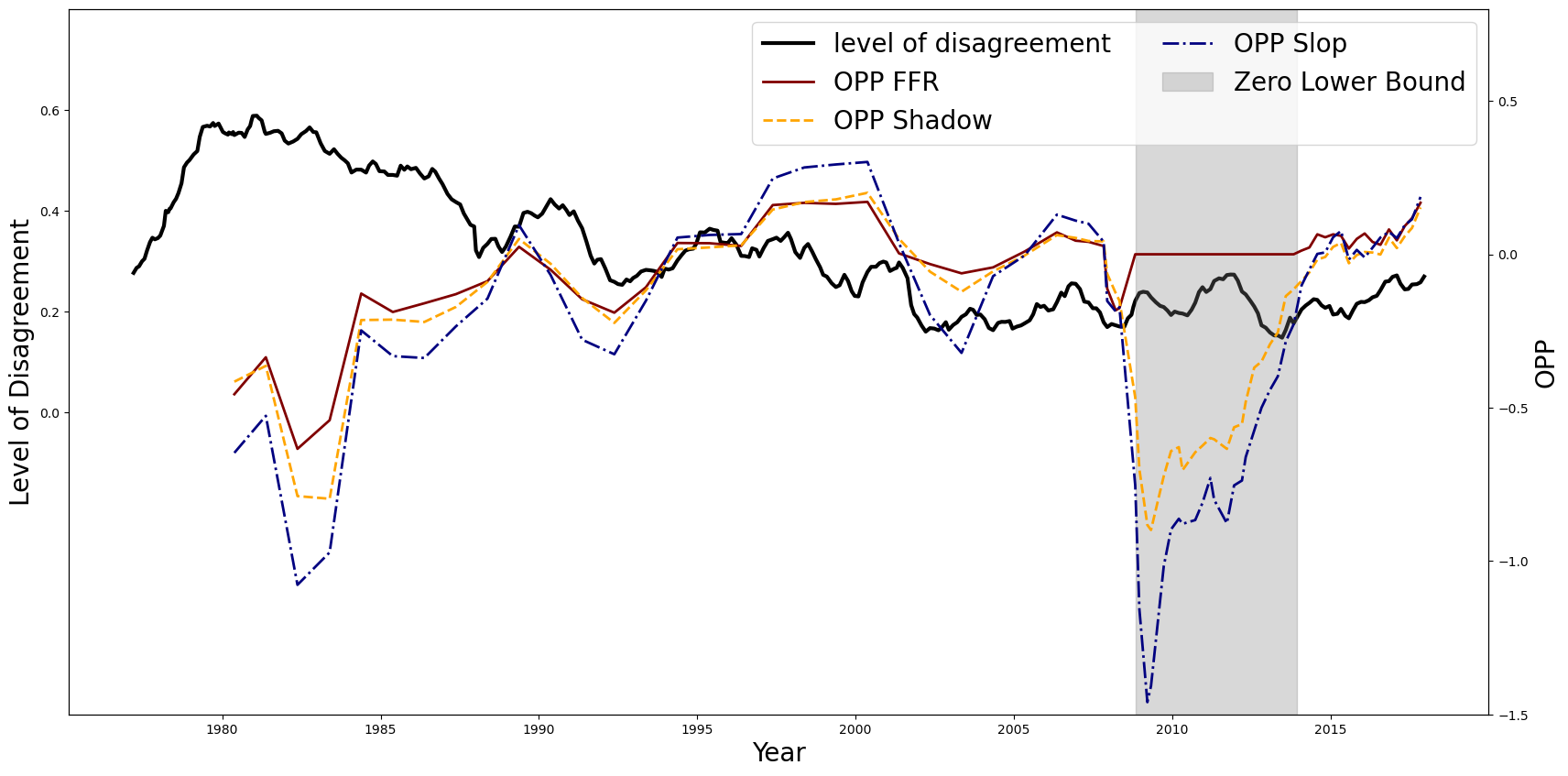}
		\caption{Level of Hidden Dissent and Optimal Policy Perturbation} 
		\label{disagree_opp}
	\end{center}
	\emph{Note:} This figure shows the trends of hidden dissent levels alongside various Optimal Policy Perturbation (OPP) measures introduced by  \citet{barnichon2023sufficient}. The shaded gray area highlights the Financial Crisis period, during which the nominal interest rate reached 0\%.
\end{figure}

\clearpage
\begin{landscape}
	\vspace* {\fill}
	\begin{figure}[hbtp!]
		\begin{center}
			\includegraphics[width=1.4\textwidth]{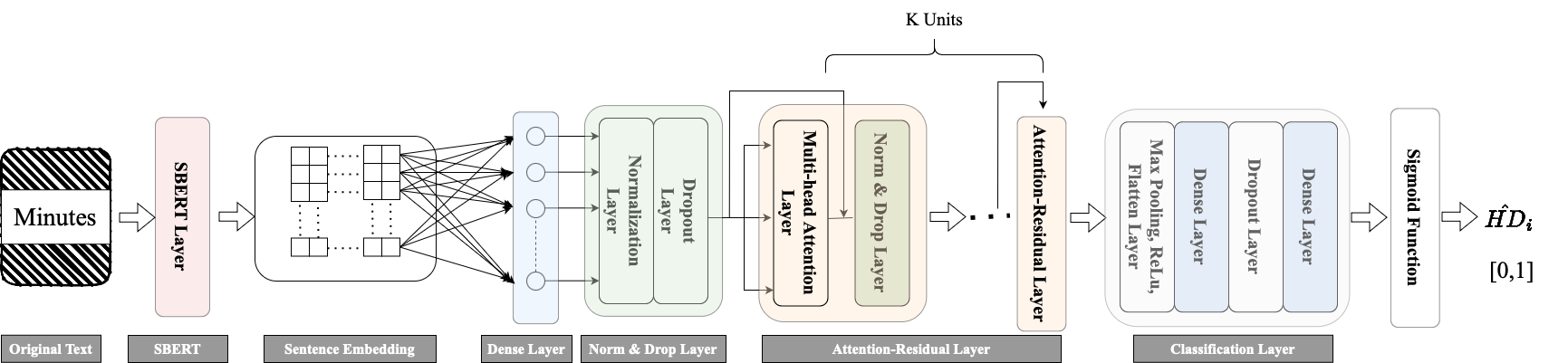}
			\caption{Deep Learning Model For Minutes} 
			\label{fomc_dl_minutes}
		\end{center}
		\emph{Note:} This figure illustrates the Deep Learning model used to predict hidden dissent levels in meeting minutes. This model is a simplified version of the Deep Learning model in \autoref{fomc_dl}. 
	\end{figure}
	\vfill
\end{landscape}	


\begin{figure}[hbtp!]
	\begin{center}
		\includegraphics[width=15cm]{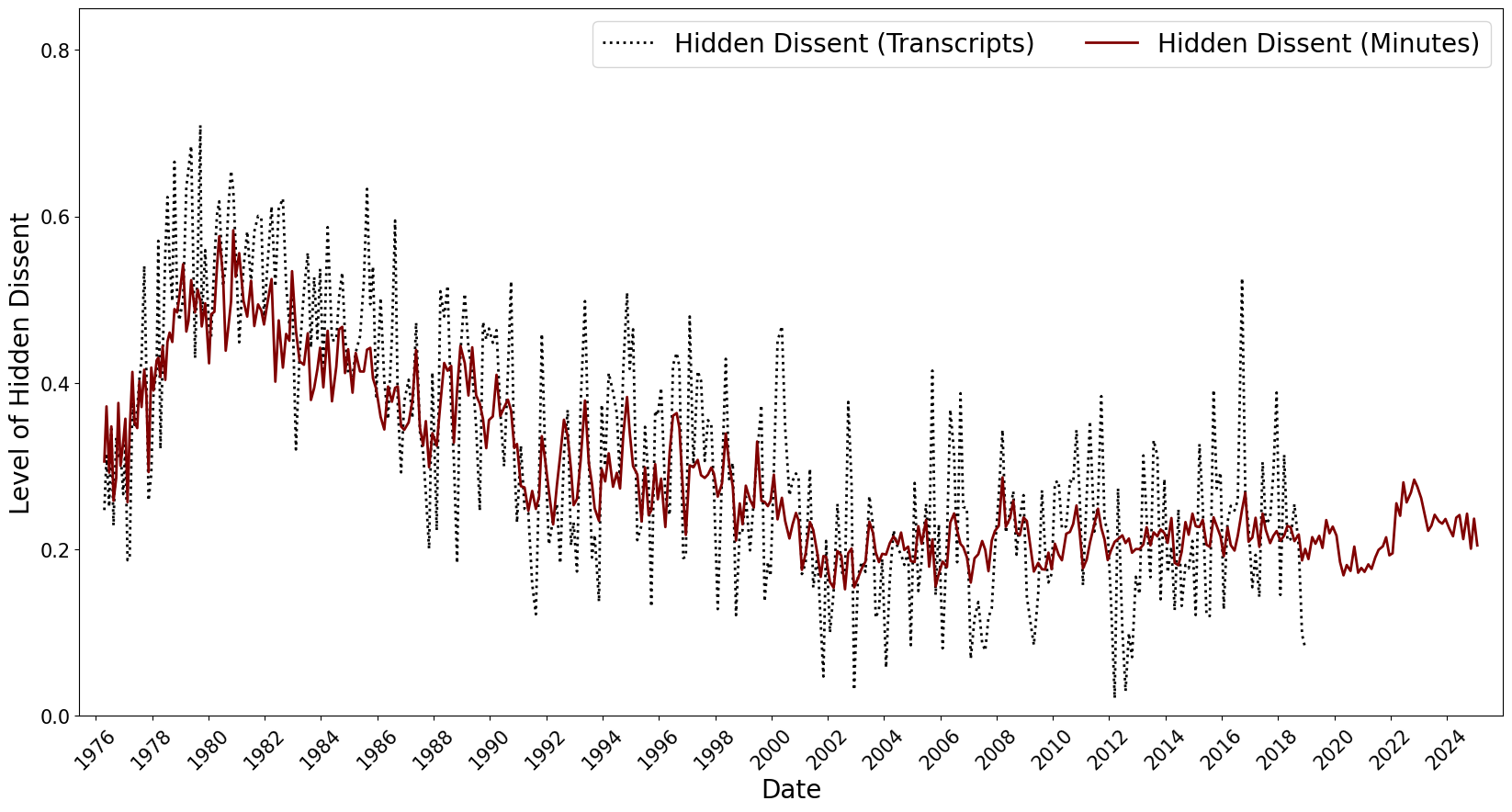}
		\caption{Level of Hidden Dissent in transcripts and minutes} 
		\label{disagree_trans_min}
	\end{center}
	\emph{Note:} This figure displays the trend of hidden dissent levels in transcripts and minutes, with transcript data ending in December 2018 and FOMC minutes data extending to December 2024.
\end{figure}

\begin{figure}[hbtp!]
	\centering
	\includegraphics[width=15cm]{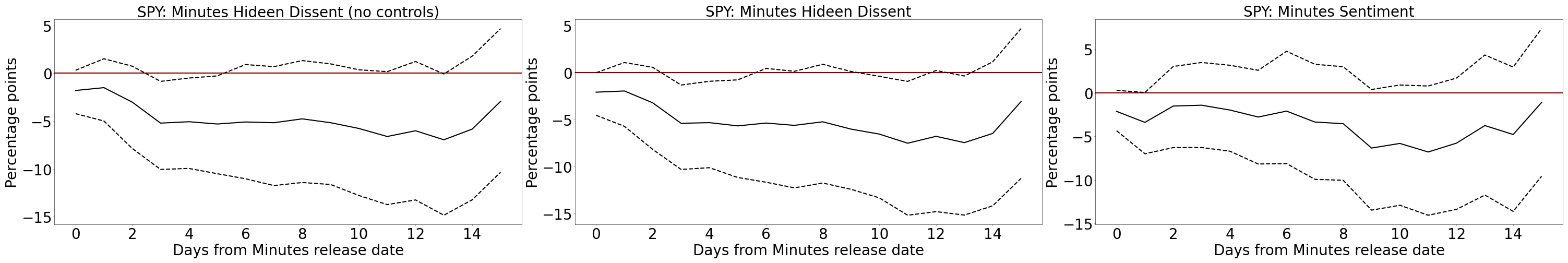}
	\caption{Response of S\&P 500 ETF to Hidden Dissent}
	\begin{minipage}{\textwidth}
		\small \emph{Note:} This figure shows the estimated 15-day dynamic impact of hidden dissent levels on SPDR S\&P 500 ETF Trust (SPY). The dashed lines represent 90\% bias-corrected and accelerated bootstrap confidence intervals. The sample covers 257 FOMC meetings from December 1992 to December 2024.
	\end{minipage}
	\label{fig_spy}
\end{figure}

\begin{figure}[hbtp!]
	\centering
	\includegraphics[width=15cm]{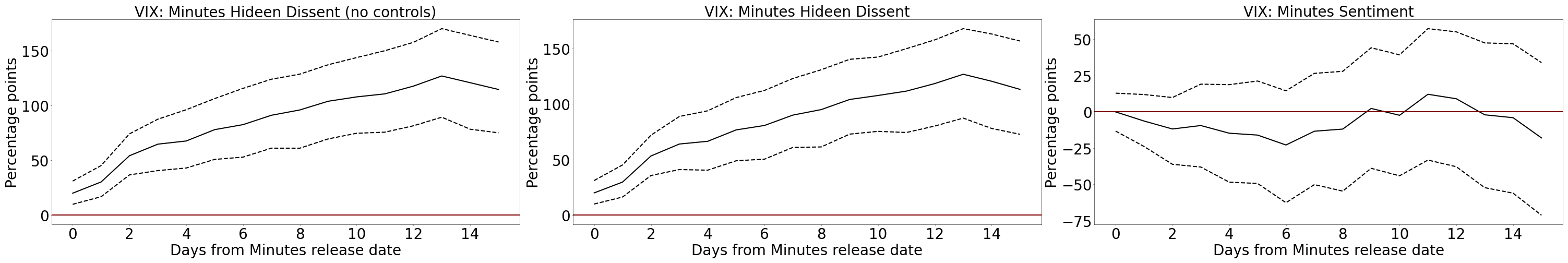}
	\caption{Response of CBOE Volatility Index to Hidden Dissent}
	\begin{minipage}{\textwidth}
		\small \emph{Note:} This figure shows the estimated 15-day dynamic impact of hidden dissent levels on CBOE Volatility Index (VIX). The dashed lines represent 90\% bias-corrected and accelerated bootstrap confidence intervals.  The sample covers 280 FOMC meetings from December 1989 to December 2024
	\end{minipage}
	\label{fig_vix}
\end{figure}

\begin{figure}[hbtp!]
	\centering
	\includegraphics[width=15cm]{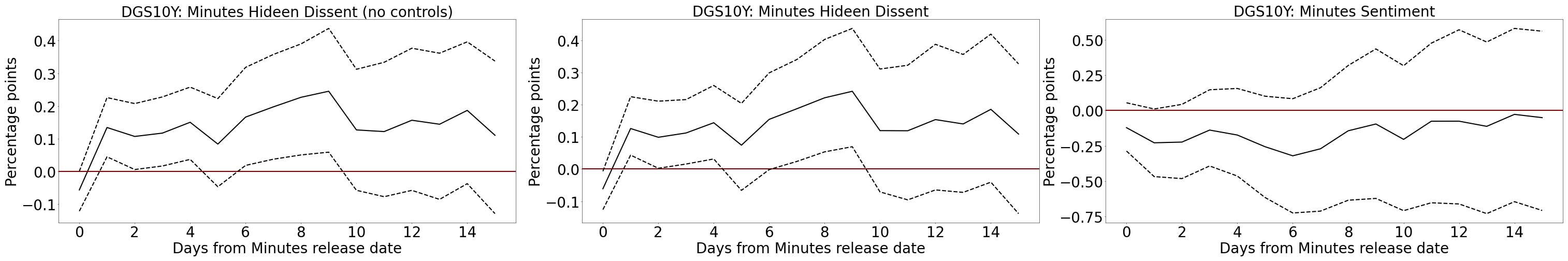}
	\caption{Response of 10-Year Treasury Yield to Hidden Dissent}
	\begin{minipage}{\textwidth}
		\small \emph{Note:} This figure shows the estimated 15-day dynamic impact of hidden dissent levels on 10-Year Treasury Yield (DGS10Y). The dashed lines represent 90\% bias-corrected and accelerated bootstrap confidence intervals. The sample covers 408 FOMC meetings from January 1976 to December 2024.
	\end{minipage}
	\label{fig_dgs10y}
\end{figure}

\begin{figure}[hbtp!]
	\centering
	\includegraphics[width=15cm]{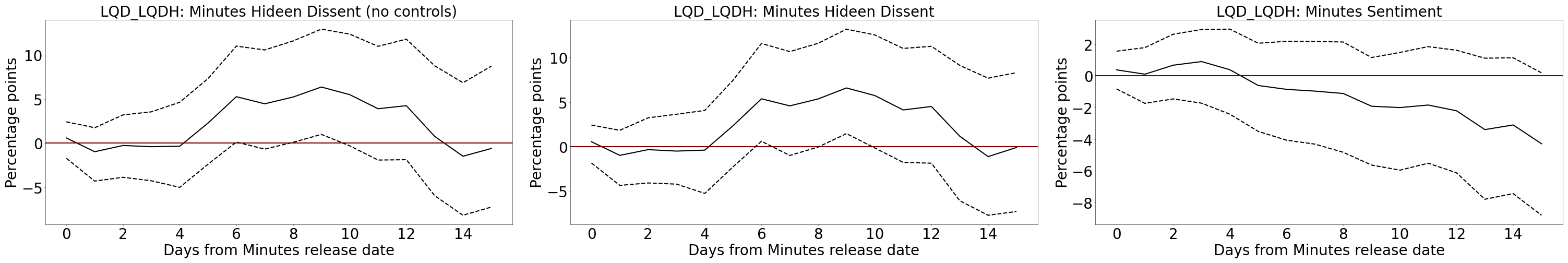}
	\caption{Response of Interest Rate Risk to Hidden Dissent}
	\begin{minipage}{\textwidth}
		\small \emph{Note:} This figure shows the estimated 15-day dynamic impact of hidden dissent levels on Interest Rate Risk (LQD-LQDH), measured as the return spread between corporate bonds and hedged corporate bonds. The dashed lines represent 90\% bias-corrected and accelerated bootstrap confidence intervals. The sample covers 85 FOMC meetings from June 2014 to December 2024.
	\end{minipage}
	\label{fig_lqd_lqdh}
\end{figure}

\clearpage

\begin{table}[!htbp] 
	\begin{center}
		\caption{FOMC Member Personal Information} 
		\label{personal_info}
		\begin{tabular}{l|l}
			\hline
			\hline
			Collected Information & Description \\
			\hline
			Birth Date      & Date of birth of the FOMC member \\
			Gender          & Gender of the FOMC member \\
			Hometown Region & Geographic region of the member's hometown \\
			Term Starts     & Year the member began serving as governor or president \\
			Term Ends       & Year the member left their current position \\
			School Region   & Region of the institution granting the member’s highest degree \\
			School Wealth   & Endowment per student at the institution granting the highest degree \\
			Degree Major    & Major field of the member's highest degree \\
			Great Depression & Equals 1 if the member experienced the Great Depression \\
			Great Inflation & Equals 1 if the member experienced the Great Inflation \\
			WWII            & Equals 1 if the member experienced World War II \\
			Party           & (1) Party of the sitting U.S. President (for governors) or \\
			& (2) Party of the state-level presidential election winner (for presidents) \\
			
			\hline
			\hline
		\end{tabular}
	\end{center}
	\emph{Note:} This table describes the personal information collected for each FOMC member, sourced from the Federal Reserve History website, Wikipedia, and various news articles.
\end{table}

\begin{table}[!htbp] 
	\begin{center}
		\caption{Summary Statistics of Selected FOMC Member Characteristics} 
		\label{personal_info_summary}
		\begin{tabular}{lcc}
			\hline
			\hline
			Personal Char. & Count & Proportion (\%) \\
			\hline
			Hometown & & \\
			\cline{1-1}
			Northeast & 32 & 32.99 \\
			Midwest  & 23 & 23.71 \\
			South & 22  &  22.68\\
			West & 13 & 13.40\\
			Other & 7 & 7.22 \\
			[1.2ex] 
			School & & \\
			\cline{1-1}
			Northeast & 43 & 44.33\\
			Midwest & 26 & 26.80\\
			South & 18 & 18.56\\
			West & 10 & 10.31\\
			[1.2ex] 
			Experience & & \\
			\cline{1-1}
			WWII & 52 & 53.61\\
			Great Depression & 41 & 42.26 \\
			Great Inflation & 43 & 44.33\\
			[1.2ex] 
			Other Char. & & \\
			\cline{1-1}
			Female   & 14 & 14.43\\
			Economics Major & 69 & 71.13\\
			Appt. Democrat & 44 & 45.36\\
			\hline
			\hline
		\end{tabular}\\
	\end{center}
	\emph{Note:} This table displays the distribution of selected characteristics of FOMC members. ``Hometown Other'' indicates members who grew up outside the U.S., and ``Appt. Democrat'' counts members appointed by a Democratic president or from states that voted Democratic in presidential elections.
\end{table}

\begin{table}[!htbp] 
	\begin{center}
		\caption{FOMC Meeting Information} 
		\label{meeting_info}
		\begin{tabular}{ll}
			\hline
			\hline
			Variable & Description \\
			\hline
			\multicolumn{2}{l}{Macro-level Variables}\\
			\cline{1-1}
			Unemployment Trend & Trend in unemployment data from the Tealbook\\
			Unemployment S.D. & Standard deviation of unemployment data from the Tealbook\\
			CPI Trend & Trend in core CPI data from the Tealbook \\
			CPI S.D. & Standard deviation of core CPI data from the Tealbook\\
			[1.2ex]
			\multicolumn{2}{l}{Meeting-level Variables}\\
			\cline{1-1}
			Experience S.D. & Standard deviation of members' experience\\
			Age S.D. & Standard deviation of members' ages\\
			School Wealth S.D. & Standard deviation of endowments per student \\
			Hometown Div. & Diversity of members' hometown regions\\
			School Div. & Diversity of regions where members earned their highest degrees\\
			Major Pct. & Percentage of members with an economics-related highest degree\\
			Gender Pct. & Percentage of female members present at a meeting\\
			Great Depression Pct. & Percentage of members who experienced the Great Depression\\
			Great Inflation Pct. & Percentage of members who experienced the Great Inflation\\
			World War II Pct. & Percentage of members who experienced World War II\\
			Appt. Democrat Pct. & Percentage of members appointed by a Democratic president\\
			Incumbent Democrat & Dummy variable for the incumbent president's party; 1 if Democrat\\
			\hline
			\hline
		\end{tabular}
	\end{center}
	\emph{Note:} Meeting-level data is derived from FOMC members’ personal information, while macro-level economic data is sourced from the Tealbook (formerly the Greenbook).
\end{table}

\begin{table}[!htbp] 
\begin{center}
\caption{Hyperparameter Value Search Space} 
\label{hyperparameter_search_space}
\begin{tabular}{l|l}
\hline
\hline
Variable & Value Search Space \\
\hline
Number of MHSA Modules in Chair Section  & [1, 12], step = 1 \\
Number of MHSA Modules in Member Section  & [1, 12], step = 1 \\
Number of MHSA Heads in Chair Section  & [4, 12], step = 4\\
Number of MHSA Heads in Member Section  & [4, 12], step = 4\\
Dropout Rate & [0.4, 0.8], step = 0.005\\
Initial Learning Rate & [$1e^{-3}$, $1e^{-6}$]\\
\hline
\hline
\end{tabular}\\
\end{center}
\emph{Note:} The learning rate is sampled from a specified range in the logarithmic domain, and a learning rate scheduler is applied throughout the model training process.
\end{table}

\begin{table}[!htbp] 
	\begin{center}
		\caption{Summary Statistics for Hidden Dissent Measures} 
		\label{disagree_measure}
		\begin{tabular}{lccccc}
			\hline
			\hline
			Measure            & Mean & S.D. & Min & Max & Count \\
			\hline
			Individual Level & & & & & \\
			\cline{1-1}
		    $hd_{ij}$            & 0.3235 & 0.2282 & 0.0000 & 0.8464 & 3523\\
			$v_{ij}$            & 0.0693 & 0.2539 & 0.0000 & 1.0000 & 3523\\
			[1.2ex] 
			Meeting Level & & & & & \\
			\cline{1-1}
			$HD_{i}$           & 0.3207 & 0.1513 & 0.0209 & 0.7095 & 370\\
			$V_{i}$           & 0.0693 & 0.0993 & 0.0000 & 0.6667 & 370\\
			\hline
			\hline
		\end{tabular}\\
	\end{center}
	\emph{Note:} This table presents summary statistics for the hidden dissent measures used in this study. The variable $x_{ij}$ represents hidden dissent values directly derived from our deep learning model, while $X_{i}$ denotes the aggregated hidden dissent level based on $x_{ij}$.
\end{table}



\begin{table}[!htbp]
	\small 
	\caption{Explaining Hidden Dissent and Revealed Dissent (Individual-level)} 
	\begin{center}
			\begin{tabular}{@{\extracolsep{-10pt}}lD{.}{.}{-3} D{.}{.}{-3} D{.}{.}{-3} D{.}{.}{-3}} 
			\hline 
			\hline
			& \multicolumn{2}{c}{$hd_{ij}$ (\textit{Mixed Effects Beta}) } & \multicolumn{2}{c}{$v_{ij}$ (\textit{Mixed Effects Probit})} \\ 
			\cline{2-5}
			& \multicolumn{1}{c}{(1)} & \multicolumn{1}{c}{(2)} & \multicolumn{1}{c}{(3)}& \multicolumn{1}{c}{(4)} \\
			\multicolumn{2}{l}{Macro Factors}   & & & \\
			\cline{1-1} 
		 $T_{unemp} $    & -0.2286^{*}  & -0.0136       &  0.0039       & 0.0584       \\
		&  (0.1281)      & (0.1289)      & (0.1495)     & (0.1515)      \\
		$D_{unemp} $    & -0.0117      & -0.1056^{**}  & 0.2425^{*}    & 0.1561        \\
		& (0.0443)      & (0.0462)      & (0.1299)      & (0.1345)      \\
		$T_{CPI}$           & 1.0414^{***}  & 1.0096^{***}  & 0.1390        & 0.1433        \\
		& (0.1366)      & (0.1436)      & (0.1161)      & (0.1200)      \\
		$D_{CPI} $         & 0.5246^{***}  & 0.4524^{***}  & 0.1773        & 0.2261        \\
		& (0.0797)      & (0.1029)      & (0.1723)      & (0.2209)      \\
			\multicolumn{2}{l}{Member Char.}   & & & \\
			\cline{1-1}
		$\text{Econ Major}$       &                 & -0.0755       &                 & -0.0955       \\
		&                 & (0.1580)      &                 & (0.6507)      \\
		$\text{School Northeast}$        &                 & -0.3826       &                 & -1.9152^{**}  \\
		&                 & (0.2335)      &                 & (0.9074)      \\
		$\text{School South}$     &                 & -0.7624^{***} &                 & -3.8795^{***} \\
		&                 & (0.2374)      &                 & (1.0422)      \\
		$\text{School West}$      &                 & 0.0669        &                 & -1.4407       \\
		&                 & (0.3065)      &                 & (1.2051)      \\
			\hline 
			Control Var. & \multicolumn{1}{c}{No} & \multicolumn{1}{c}{Yes} & \multicolumn{1}{c}{No} & \multicolumn{1}{c}{Yes} \\
			Log Likelihood & \multicolumn{1}{c}{1622.9400} &\multicolumn{1}{c}{1642.1933} & \multicolumn{1}{c}{-413.5240} & \multicolumn{1}{c}{-393.9099} \\ 
			$N$ &\multicolumn{1}{c}{2,547} &\multicolumn{1}{c}{2,528} & \multicolumn{1}{c}{2,547} & \multicolumn{1}{c}{2,528} \\ 
			\hline 
			\hline 
		\end{tabular} 
	\end{center}
	\begin{tablenotes}
		\small
		\item \emph{Note:} This table examines variables potentially correlated with hidden dissent and dissent scores. Mixed-effects regressions are clustered at the FOMC member level, with member characteristics detailed in \autoref{personal_info}. The sample period spans from the February 1986 FOMC meeting, when Tealbook core CPI data first became available, to the December 2018 meeting.  $^{***}p<0.01$; $^{**}p<0.05$; $^{*}p<0.1$
	\end{tablenotes}
	\label{personal_level_reg_1}
\end{table}

\clearpage
\newgeometry{left=2cm, right=2cm, top=1.5cm, bottom=1.5cm}  
\begin{table}[!htbp]
	\small 
	\caption{Hidden Dissent and Personal Experience}
	\begin{center}
		\begin{tabular}{@{\extracolsep{-10pt}}lD{.}{.}{-3} D{.}{.}{-3} D{.}{.}{-3}} 
			\hline 
			\hline
			& \multicolumn{3}{c}{$hd_{ij}$ (\textit{Beta Model}) } \\ 
			\cline{2-4}
			& \multicolumn{1}{c}{(1)} & \multicolumn{1}{c}{(2)} & \multicolumn{1}{c}{(3)}\\
			\multicolumn{2}{l}{Macro Factors}   & &  \\
			\cline{1-1} 
			$T_{unemp}$                                & -0.0086       & -0.9323^{***} & 0.1602        \\
			& (0.1404)      & (0.2336)      & (0.1573)      \\
			$D_{unemp}$                                & 0.0081        & -0.1530^{*}   & 0.0037        \\
			& (0.0498)      & (0.0882)      & (0.0621)      \\
			$T_{CPI}$                                  & 0.7468^{***}  & 1.1024^{***}  & 0.5810^{**}   \\
			& (0.1758)      & (0.1863)      & (0.2396)      \\
			$D_{CPI}$                                  & 0.5970^{***}  & 0.4810^{***}  & 0.5346^{***}  \\
			& (0.1189)      & (0.1034)      & (0.1986)      \\
			\multicolumn{2}{l}{Interaction with Personal Experience}   & &  \\
			\cline{1-1} 
			$T_{unemp} \times \text{Great Depression}$ & -1.1331^{***} &                 &                 \\
			& (0.3156)      &                 &                 \\
			$D_{unemp} \times \text{Great Depression}$ & -0.3728^{***} &                 &                 \\
			& (0.1258)      &                 &                 \\
			$T_{CPI} \times \text{Great Depression}$   & 0.4688^{*}    &                 &                 \\
			& (0.2843)      &                 &                 \\
			$D_{CPI} \times \text{Great Depression}$   & -0.1709       &                 &                 \\
			& (0.1690)      &                 &                 \\
			$T_{unemp} \times \text{Great Inflation}$  &                 & 1.0191^{***}  &                 \\
			&                 & (0.2775)      &                 \\
			$D_{unemp} \times \text{Great Inflation}$  &                 & 0.1453        &                 \\
			&                 & (0.1040)      &                 \\
			$T_{CPI} \times \text{Great Inflation}$    &                 & -0.3770       &                 \\
			&                 & (0.2787)      &                 \\
			$D_{CPI} \times \text{Great Inflation}$    &                 & 0.0583        &                 \\
			&                 & (0.1803)      &                 \\
			$T_{unemp} \times \text{WWII}$             &                 &                 & -1.0511^{***} \\
			&                 &                 & (0.2637)      \\
			$D_{unemp} \times \text{WWII}$             &                 &                 & -0.1838^{*}   \\
			&                 &                 & (0.1035)      \\
			$T_{CPI} \times \text{WWII}$               &                 &                 & 0.5648^{*}    \\
			&                 &                 & (0.2969)      \\
			$D_{CPI} \times \text{WWII}$               &                 &                 & -0.0377       \\
			&                 &                 & (0.2200)      \\
			\hline
			Fixed Effects  & \multicolumn{1}{c}{Yes} & \multicolumn{1}{c}{Yes} & \multicolumn{1}{c}{Yes} \\
			Log Likelihood & \multicolumn{1}{c}{1786.0231} &\multicolumn{1}{c}{1779.7226} & \multicolumn{1}{c}{1784.1786}\\ 
			$N$ &\multicolumn{1}{c}{2,539} &\multicolumn{1}{c}{2,539} & \multicolumn{1}{c}{2,539} \\ 
			\hline 
			\hline 
		\end{tabular}
	\end{center}
	\begin{tablenotes}
		\small
		\item \emph{Note:} This table examines how members' personal experience may interact with macroecomic situations and affect their hidden dissent. The sample period spans from the February 1986 FOMC meeting, when Tealbook core CPI data first became available, to the December 2018 meeting.  $^{***}p<0.01$; $^{**}p<0.05$; $^{*}p<0.1$
	\end{tablenotes}
	\label{personal_level_reg_3}
\end{table}

\clearpage 
\restoregeometry 

\begin{table}[!htbp]
\small
\caption{Explaining Hidden Dissent and Revealed Dissent (Meeting-level)}
\begin{center}
		   \begin{tabular}{@{\extracolsep{-10pt}}lD{.}{.}{-3} D{.}{.}{-3} D{.}{.}{-3} D{.}{.}{-3}}
			\hline
			\hline
			 & \multicolumn{2}{c}{$HD_i$ (\textit{Beta})}  & \multicolumn{2}{c}{$V_i$ (\textit{Fractional Logistic})} \\
			\cline{2-5}
			 & (1) & (2) & (3) & (4) \\
			\multicolumn{2}{l}{Macro Factors}   & & & \\
			\cline{1-1}
            $T_{unemp}$ & -0.5488^{**} & -0.5938^{***} & -0.0914 & -0.0703 \\ 
            & (0.2213) & (0.2285) & (0.6175) & (0.6772) \\ 
            $D_{unemp}$ & 0.0854 & -0.1334 & 0.4631^{**} & -0.1468 \\ 
            & (0.0642) & (0.0842) & (0.1810) & (0.2812) \\ 
            $T_{CPI}$ & 1.1071^{***} & 0.6994^{***} & 1.3953^{**} & 1.2535 \\ 
            & (0.2333) & (0.2621) & (0.6480) & (0.8195) \\ 
            $D_{CPI}$ & 0.6876^{***} & 0.2525 & 0.5953^{***} & 0.7648 \\ 
            & (0.0827) & (0.1746) & (0.2255) & (0.5876) \\ 
            \multicolumn{2}{l}{Member Char.}  & & & \\
			\cline{1-1}
          $P_{major}$ &  & -1.3761^{***} &  & -1.4270 \\ 
          &  & (0.3570) &  & (1.1393) \\ 
          $E_{school}$ &  & 1.0154^{***} &  & 1.5520 \\ 
          &  & (0.3134) &  & (1.0837) \\ 
            \hline
           Control Var. & \multicolumn{1}{c}{No} & \multicolumn{1}{c}{Yes} & \multicolumn{1}{c}{No} & \multicolumn{1}{c}{Yes} \\
           Pseudo $R^{2}$ & \multicolumn{1}{c}{0.2275} & \multicolumn{1}{c}{0.3893} & \multicolumn{1}{c}{0.4547} & \multicolumn{1}{c}{0.4914} \\ 
           Log Likelihood & \multicolumn{1}{c}{241.0162} & \multicolumn{1}{c}{274.6247} & \multicolumn{1}{c}{-14.4783} & \multicolumn{1}{c}{-13.5026} \\ 
           $N$                & 268                  & 268                   & 268          & 268            \\
            \hline
            \hline
            \end{tabular}
\end{center}
\begin{tablenotes}
\small
\item \emph{Note:} This table analyzes factors that may correlate with hidden dissent levels and dissent. Hidden dissent, $HD_i$, is calculated as the average of hidden dissent scores, $hd_{ij}$, while dissent, $V_i$, is defined as the percentage of NO votes in a meeting. In the macro-variable section, $T_x$ represents the trend of variable $x$, and $D_x$ denotes its standard deviation, derived from past and forecast values in the Tealbooks. $E_{school}$ captures education institutional diversity, and $P_{major}$ represents the percentage of FOMC members with a highest degree in an economics-related field. The sample period spans from the February 1986 FOMC meeting, when Tealbook core CPI data first became available, to the December 2018 meeting. $^{***}p<0.01$; $^{**}p<0.05$; $^{*}p<0.1$
\end{tablenotes}
\label{voting_pred_reg1}
\end{table}


\begin{landscape}
\begin{table}
	\caption{Level of Hidden Dissent and Summary of Economic Projections}
	\begin{center}
		\begin{tabular}{@{\extracolsep{-10pt}}lD{.}{.}{-3} D{.}{.}{-3} D{.}{.}{-3} D{.}{.}{-3} D{.}{.}{-3} D{.}{.}{-3}}
			\hline
			\hline
			& \multicolumn{4}{c}{$HD_i$} &  \multicolumn{2}{c}{$\text{Residualized }HD_i$}\\
			& \multicolumn{4}{c}{(\textit{Beta})} &  \multicolumn{2}{c}{(\textit{DML})} \\
			\cline{2-7} 
			                            & (1)          & (2)           & (3)           &  (4)          &  (5)         & (6)    \\              
			 $\text{Policy}_{PC1}$      & 0.1242^{**}  & 0.1262^{**}   &               &               &              &       \\
			                            & (0.0607)     & (0.0609)      &               &               &              &        \\
			 $\text{Policy}_{PC2}$      &              & -0.0447       &               &               &              &      \\
			                            &              & (0.0805)      &               &               &              &        \\
			 $\text{Economy}_{PC1}$     &              &               & -0.0398       & -0.0578^{*}   &              &       \\
			                            &              &               & (0.0256)      & (0.0331)      &              &       \\
			 $\text{Economy}_{PC2}$     &              &               & 0.0058        & -0.0032       &              &      \\
			                            &              &               & (0.0637)      & (0.0603)      &              &       \\
			 $\text{Economy}_{PC3}$     &              &               & 0.0210        & -0.0436       &              &       \\
			                            &              &               & (0.0731)      & (0.0670)      &              &        \\
			$\text{Residualized Policy}_{PC1}$ &       &               &               &               & 0.0242^{**}  &        \\   
			                                   &       &               &               &               & (0.0122)     &        \\  
		    $\text{Residualized Economy}_{PC1}$&       &               &               &               &              & -0.0128  \\   
			                                   &       &               &               &               &              & (0.0109)  \\  
			                                                 
			 \hline
			Pseudo $R^2$ & \multicolumn{1}{c}{0.1257} & \multicolumn{1}{c}{0.1349} & \multicolumn{1}{c}{0.0598} & \multicolumn{1}{c}{0.1057} & & \\
			Log Likelihood & \multicolumn{1}{c}{29.7784} & \multicolumn{1}{c}{29.9327} & \multicolumn{1}{c}{41.1820} & \multicolumn{1}{c}{29.4507} & & \\
			$N$                   & 29    &  29    & 40       & 29  & 29 & 29 \\
			\hline
			\hline
		\end{tabular}
	\end{center}
	\begin{tablenotes}
		\small
		\item \emph{Note:} 	This table investigates the correlation between the level of hidden dissent observed in meeting transcripts and disagreement within the Summary of Economic Projections. Disagreement levels for each projection category are calculated as the absolute sum of deviations from the median projection, the same method employed by \citet{foerster2023evolution}. These levels are subsequently categorized into two distinct groups: disagreement pertaining to current monetary policy, specifically Federal Funds Rate (FFR) projections, and disagreement related to the current state of the economy, encompassing other projection types. Principal component analysis is then applied to each group.  Due to the availability of SEP data, the sample for the macroeconomic variables, which includes long-term projections, begins with the January 2009 meeting, whereas for the federal funds rate, it starts with the January 2012 meeting. Both samples conclude with the December 2018 meeting. In column (4), the sample of macroeconomic variables is maintained within the same period range as that of the federal funds rate sample. Column (5) and (6) show the second stage results from the Double Machine Learning model. $^{***}p<0.01$; $^{**}p<0.05$; $^{*}p<0.1$
	\end{tablenotes}
	\label{voting_pred_reg1_dot_plot}
\end{table}
\end{landscape}


\begin{table}
	\caption{Level of Hidden Dissent and Optimal Policy Perturbation (\cite{barnichon2023sufficient})}
	\begin{center}
		\begin{tabular}{lccc}
			\hline
			\hline
			& $\text{OPP}_{\text{ffr}}$(Abs.) & $\text{OPP}_{\text{shadow}}$(Abs.)  & $\text{OPP}_{\text{slop}}$(Abs.)   \\
			\hline
			\multicolumn{4}{l}{Full OPP Sample}  \\
			\cline{1-2}
			$HD_i$  & 0.4562***     & 0.0176           & 0.0340          \\
			& (0.1313)      & (0.2412)         & (0.3458)        \\
			$V_i$   & 0.2236        & 0.3404           & 0.2818          \\
			& (0.2332)      & (0.4430)         & (0.5630)        \\
			\hline
			Adjusted $R^2$     & 0.3114        & -0.0139          & -0.0220         \\
			N               & 77            & 77               & 77              \\
			\hline
			\multicolumn{4}{l}{OPP Sample, Excludes Zero-Bound Period}  \\
			\cline{1-2}
			$HD_i$  & 0.4156***    & 0.5214***       & 0.8006***       \\
			& (0.1524)      & (0.1800)         & (0.2392)        \\
			$V_i$   & 0.2718        & 0.3715           & 0.3183          \\
			& (0.2396)      & (0.4186)         & (0.4989)        \\
			\hline
			Adjusted $R^2$     & 0.3060        & 0.3189           & 0.3404          \\
			N              & 55           &  55              &  55              \\
			\hline
			\hline
		\end{tabular}
	\end{center}
		\begin{tablenotes}
		\small
		\item \emph{Note:} This table examines the relationship between the level of hidden dissent inside the FOMC meeting and the deviation from the optimal policy after the meeting. Deviation is quantified by Optimal Policy Perturbation (OPP), as defined by \citet{barnichon2023sufficient}. $\text{OPP}_{\text{ffr}}$ ($\text{OPP}_{\text{shadow}}$) denotes the deviation from the optimal Fed funds rate (shadow rate), and $\text{OPP}_{\text{slop}}$  represents the deviation from the optimal slope of the yield curve.  ``Abs.'' means absolute value. The sample spans from 1980, when OPP data became available, to 2018, in line with the availability of transcript data. $^{***}p<0.01$; $^{**}p<0.05$; $^{*}p<0.1$
	\end{tablenotes}
		\label{voting_pred_reg3_OPP}
\end{table}


\begin{table}[!htbp]
	\caption{Financial Indicators}
	\label{table_financial_indicators}
	\begin{tabular}{lll}
		\hline
		\hline
		Symbol       &  Financial Market Indicator                         & Data Period\\
		\hline
		SPY          &   SPDR S\&P 500 ETF Trust                   & 1993-01-29 to 2025-01-10\\
		DGS10Y     &  10-Year Treasury Yield     & 1976-01-01 to 2025-01-10 \\
		VIX          & CBOE Volatility Index                               & 1990-01-02 to 2025-01-10\\
		LQD          & Investment Grade Corporate Bonds ETF                & 2014-06-17 to 2025-01-10\\
		LQDH         & Hedged Investment Grade Corporate Bonds ETF         & 2012-02-24 to 2025-01-10\\
		\hline
		\hline
	\end{tabular}
	\begin{tablenotes}
		\small
		\item \emph{Note:} This table lists the financial indicators used to gauge the reactions of different financial markets to the hidden dissent in meetings. The corresponding data is sourced from Yahoo Finance. The sample period spans 408 FOMC meetings held between 1976 and 2024, subject to data availability for each specific indicator.
	\end{tablenotes}
\end{table}

\clearpage


\appendix
\counterwithin{figure}{section}
\counterwithin{table}{section}
\setcounter{section}{-1}
\setcounter{page}{1}

\section{Online Appendices}

This online appendix contains the following four items. In \autoref{appendix_ml_nlp_intro}, we give a brief introduction to the applications of Deep Learning methods in the natural language processing (NLP) field. Results in \autoref{appendix_robust_check} serve as the robustness checks for the main results in our paper.  In \autoref{appendix_speech_data}, we apply the deep learning model on FOMC members' public speech data. And \autoref{appendix_personal_detailed} shows the complete results for regressions.


\renewcommand{\thesection}{A}
\section{A Brief Introduction of Deep Learning in NLP}
\label{appendix_ml_nlp_intro}

In earlier natural language processing (NLP) models, researchers often rely on manually crafted features to project text data into low-dimensional feature vectors. One representative example is the Bag of Words method. Researchers need to have sufficient domain knowledge to pick out the most relevant features, which can be time-consuming and difficult to generalize. And these NLP models typically neglect intricate interconnections between different words and sentences within an article.. \citetapp{gentzkow2019text} note that, the emergence of new machine learning models for NLP in computer science offers novel solutions for addressing certain economic challenges. In recent years, Deep Learning models have advanced rapidly and enabled us to surmount the constraints of previous-generation NLP models. Moreover, the introduction of Transfer Learning allows us to pre-train a complex model on extensive datasets, so we can achieve high prediction accuracy even if the target dataset is small.

The root of Deep Learning can be traced back to 1943 when Walter Pitts and Warren McCulloch created the first neural network model based on the neural networks of the human brain. Until recently, the commonly used Deep Learning models in text classification can be broadly categorized into two groups: Convolutional Neural Network (CNN) and Recurrent Neural Network (RNN) models.\footnote{There are other equally important deep learning models like MLP, Autoencoder, GAN, etc. However, to our best knowledge, they are less solely used in text classification tasks.} In text classification tasks, CNN models excel in detecting position-related patterns within text, while RNN models, particularly the Long Short-Term Memory (LSTM) model, exhibit robust performance in training with lengthy text. By introducing the Gated Recurrent Unit (GRU), LSTM can effectively address the vanishing gradients problem in long text classification tasks\citepapp{minaee2021deep}.

Nevertheless, both CNN and RNN models exhibit inherent limitations. RNN's sequential processing impedes parallel computing, leading to prolonged model training durations compared to CNN models. Conversely, while CNN models is well suited for parallel computing, pooling operations within CNN models may inadvertently neglect important spatial relationships among text segments. To mitigate the above problems, a novel model called Transformer is introduced and is attracting increasing  attention in the deep learning community. 

In their seminal paper, \citetapp{vaswani2017attention} introduced the Transformer model, with the Self-Attention mechanism being its pivotal component. The basic idea behind the Self-attention module is illustrated in \autoref{self_attention} and can be distilled as follows: the original input is fed into three different dense layers to generate three different datasets Query, Key and, Value. The Query and Key sets are used to calculate which part of the original input should be paid more attention to, then these attention weights are applied to the Value set to extract the salient features that deserve the most attention. The Self-attention module can be written as:
\begin{equation*}
	\text{Self-attention(Q,K,V)}=\text{softmax}(\frac{QK^T}{\sqrt{d}})V
\end{equation*}
where $d$ is the dimension of the vector used to represent one token.

\begin{figure}[hbtp!]
	\begin{center}
		\includegraphics[width=12cm]{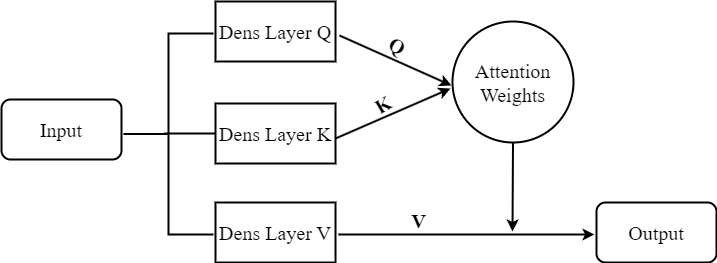}
		\caption{Self-attention module} 
		\label{self_attention}
	\end{center}
	\emph{Note:} This figure shows the basic idea behind the Self-attention module. In the Self-attention module, V, Q, and K are transformed from the same input and usually they have the same shape. 
\end{figure}

In contrast to CNN and RNN models, the Self-attention module can capture long-distance spatial relationships within text component without worrying about the vanishing gradient problem. Meanwhile, the Self-attention module can utilize parallel computing easily and harness the immense computational power of modern GPUs and TPUs. Furthermore, the Self-attention module is versatile and can be integrated seamlessly with other modules to build more complex deep learning models for specific tasks. All these advantages of the Self-attention module make it quickly become one of the well-received innovations in the deep learning research field.   

Prior to the introduction of the Transformer model, earlier NLP models that utilized the Self-attention mechanism still used RNNs for input representations \citepapp{cheng2016long, paulus2017deep}. The Transformer model stands as the first model solely based on the Self-attention modules, forgoing the need for convolutional or recurrent layers. Since its inception, the Transformer model swiftly becomes the premier choice for NLP tasks, and myriad NLP models are built on the Transformer structure. Though the Transformer model was originally proposed to compete with other seq2seq models, it has been proven to be versatile and can be used in a wide range of modern deep learning applications. For example, the word embedding model employed in this paper is built on the encoder component of the Transformer model. 

One of the most influential Transformer-based models is the Bidirectional Encoder Representations from Transformers (BERT) model built by \citetapp{devlin2018bert}.\footnote{Another influential NLP model built on Transformer is Generative Pre-training Transformer (GPT), which is the backbone of Chat-GPT application.} BERT is a pre-trained large language model, and it is trained on Wikipedia and BooksCorpus, collectively encompassing approximately 3.3 billion words. Not surprised, BERT obtains state-of-the-art results on multiple mainstream NLP tasks. 

BERT's exceptional performance hinges on its bidirectional feature. Human languages are complex, and one word usually has multiple different meanings. Traditional NLP methods like Bag of Words, treats a word such as ``run'' identically across different contexts. For example, in the following Sentence 1 and Sentence 2, the word ``runs'' has different meanings, but the Bag of Words algorithm lacks the capacity to distinguish these meanings based solely on the isolated word. One remedy is adopting the n-gram model, which takes the n neighboring words into consideration. However, the vocabulary size in the n-gram method grows exponentially with the value of n, which can lead to high computational costs. Furthermore, this method still cannot capture the long-distance semantic relationships between words. Later on, deep learning methods, like seq2seq, are introduced into the NLP field, and  these language models can capture the underlying context and treat the ``run'' differently. However, due to the autoregressive nature, many such models exclusively process text from left to right, so when the algorithm encounters examples like sentence 3 and sentence 4, the word ``bank'' will be treated as the same in both instances because the context these models can learn from is the same word ``The''. The BERT model, however, operates bidirectionally, and can use the context after the ``bank'' word to distinguish word meanings in different scenarios. So the word ``bank'' can be treated differently in the BERT model as it should be.

\textbf{Sentence 1:} \emph{This is a relatively small number, and it suggests that the snapback when the liquidation finally \textbf{runs} its course potentially will be only modest.} (FOMC Conference Call, April 18, 2001, Chair Greenspan)

\textbf{Sentence 2:} \emph{You would probably have massive bank \textbf{runs}, and the whole system would shut down.} (FOMC Meeting, June 19-20, 2012, Vice Chair Dudley)

\textbf{Sentence 3:} \emph{The \textbf{bank} money is coming in slowly, though not for the full amount, and the IMF money is going ahead.} (FOMC Meeting, December 20-21, 1982, Vice Chair Solomon)

\textbf{Sentence 4:} \emph{The \textbf{bank} of the Hudson River in New York is a popular spot for fishing.}

Another advantage of using large language models like BERT is that we can benefit from transfer learning.  Transfer learning is a machine learning technique that involves using a pre-trained model on a new problem. This technique allows us to leverage the knowledge and features learned by the pre-trained model and apply them to a different but related problem that has less data or computational resources. Usually, there are two ways to use BERT: Feature-based approach and Fine-tuning approach. The Feature-based approach does not change the weights in the pre-trained BERT model, and it uses BERT to translate natural language sentences into vector representations, which serve as input for a subsequent deep learning model tailored to a specific NLP task. Fine-tuning approach changes part of the pre-trained BERT model weights to make it more suitable for a particular NLP task. 

In this study, we take the first approach and use the BERT model directly for word embedding, and the vectorized text content will be the input for a customized deep learning model. Word embedding represent words as vectors of real numbers but still capture these words' meaning and semantic relationships. However, there is one caveat on applying word embedding on a long transcript document: most off-the-shelf pre-trained BERT models can only process up to 512 tokens at a time. This limitation stems from the quadratic complexity ($O(n^2)$) of the self-attention layer in BERT concerning sequence length ($n$). Imposing this token cap prevents excessive memory consumption and computational overheads. There are several strategies that are commonly used to circumvent the limitation: (1) selectively processing the most informative portions of the text to alleviate computational burdens. For example, \citetapp{chen2022long} choose to process the beginning and end of news articles to extract the most informative texts. (2) dividing the original text into segments and processing them individually, as pursued by \citetapp{wang2019multi}.(3) introducing modified transformer models to accommodate longer texts \citepapp{rae2019compressive, kitaev2020reformer, ding2020cogltx}.

Selectively processing certain parts of a long text may work well for news articles that have a well-defined structure to rely on and can easily extract the most informative section. However, this method may not be suitable for meeting transcripts because we do not have a sound foundation to reason that a certain part of the transcript is more informative than the rest. Dividing the original text into chunks also creates another challenge: how to treat the different chunks? First potential solution is to label all the chunks based on the vote results. However, this approach is not ideal given that, during the meeting discussion and some member voted NO, it is hard to argue that these FOMC members disagree on every sentence. A highly likely scenario is that members agree on most of the meeting discussion but disagree on select portions. Simply labeling all the chunks based on the final vote result risks misclassification and reduced predictive accuracy. Second potential solution is to concatenate the vectors from different chunks. Given that each word is converted into a $1\times768$ vector. Yet, this approach can generate a huge matrix for each lengthy text and impose significant computational overhead. Another potential solution is to average vector values of these chunks and generate a $N\times768$ matrix as the word embedding of the whole transcript. The model, Sentence-BERT (SBERT), used in this paper is adopted from this idea. However, benefiting from pre-training on large amounts of text, the specific model used in our paper is fine-tuned to be suitable for deriving a fixed sized embedding not on word level but sentence level.

\bibliographystyleapp{chicago}
\bibliographyapp{fomc_bib}   

\clearpage


\renewcommand{\thesection}{B}

\section{Results for Robustness Checks}
\label{appendix_robust_check}

\begin{figure}[hbtp!]
	\centering
	\begin{subfigure}{0.48\linewidth}
		\centering
		\includegraphics[width=1\textwidth]{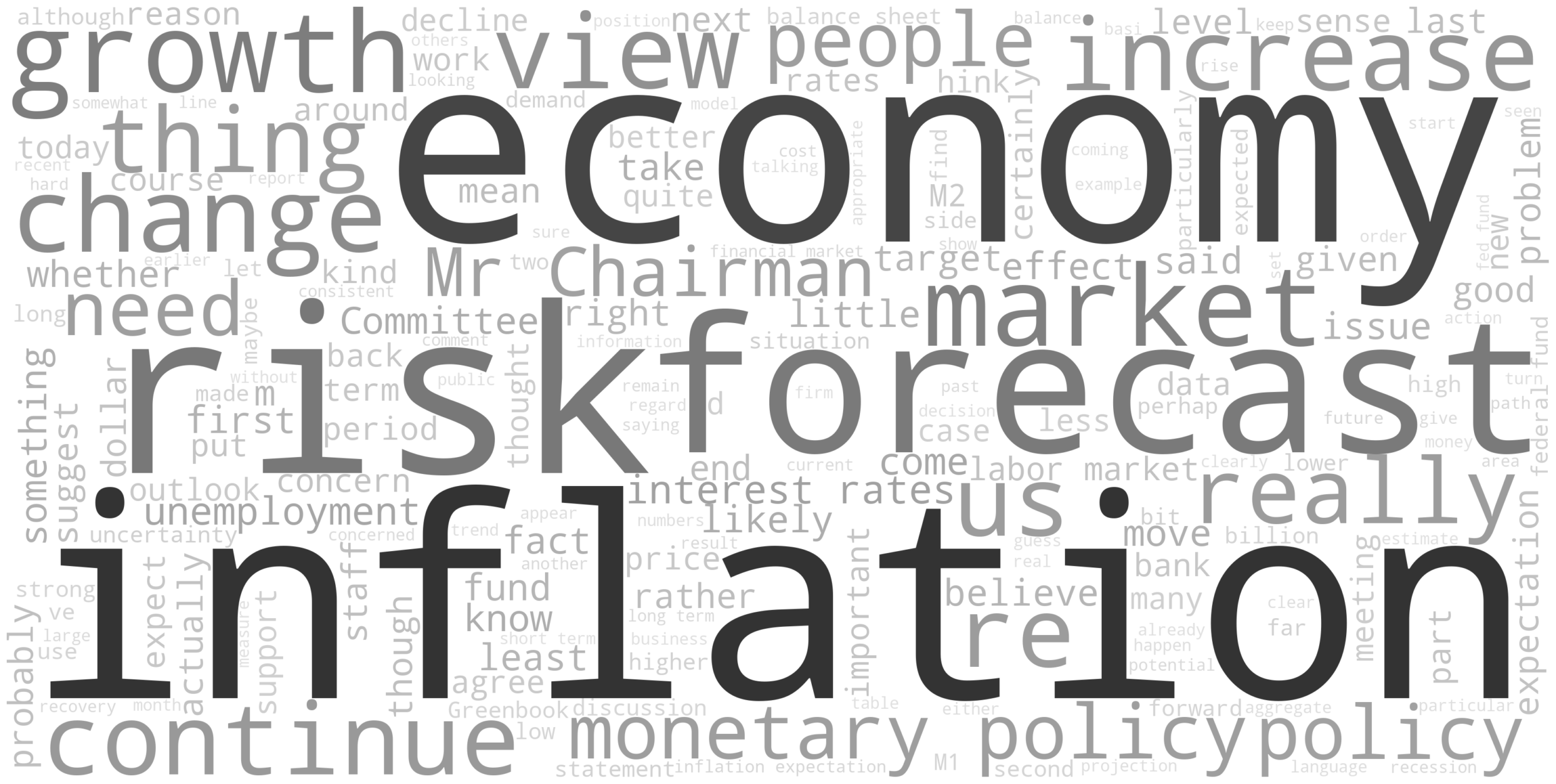}
		\caption{Vote For, Full Sample}
	\end{subfigure}
	\begin{subfigure}{0.48\linewidth}
		\centering
		\includegraphics[width=1\textwidth]{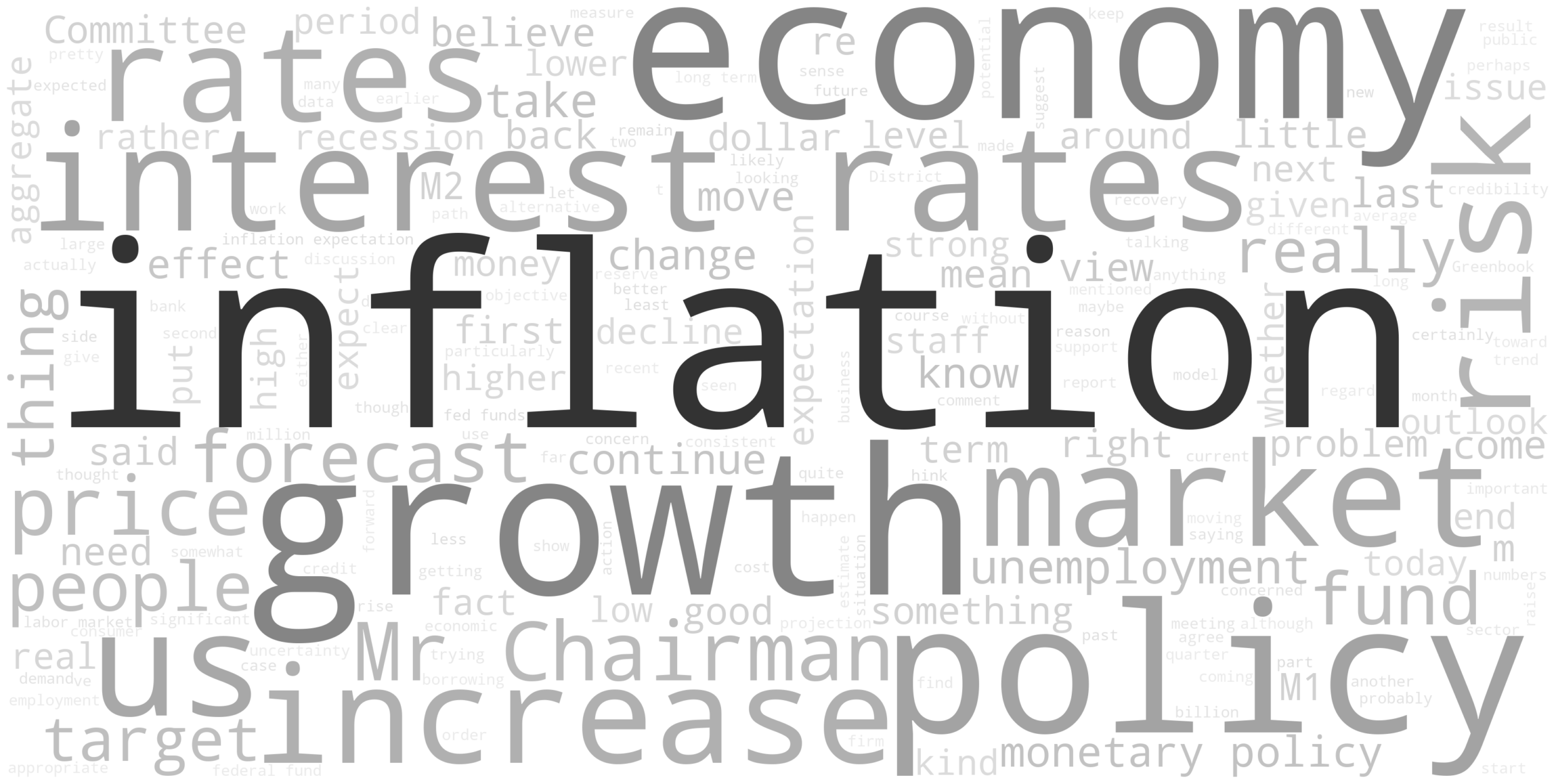}
		\caption{Vote Against, Full Sample}	
	\end{subfigure}
	\begin{subfigure}{0.48\linewidth}
		\centering
		\includegraphics[width=1\textwidth]{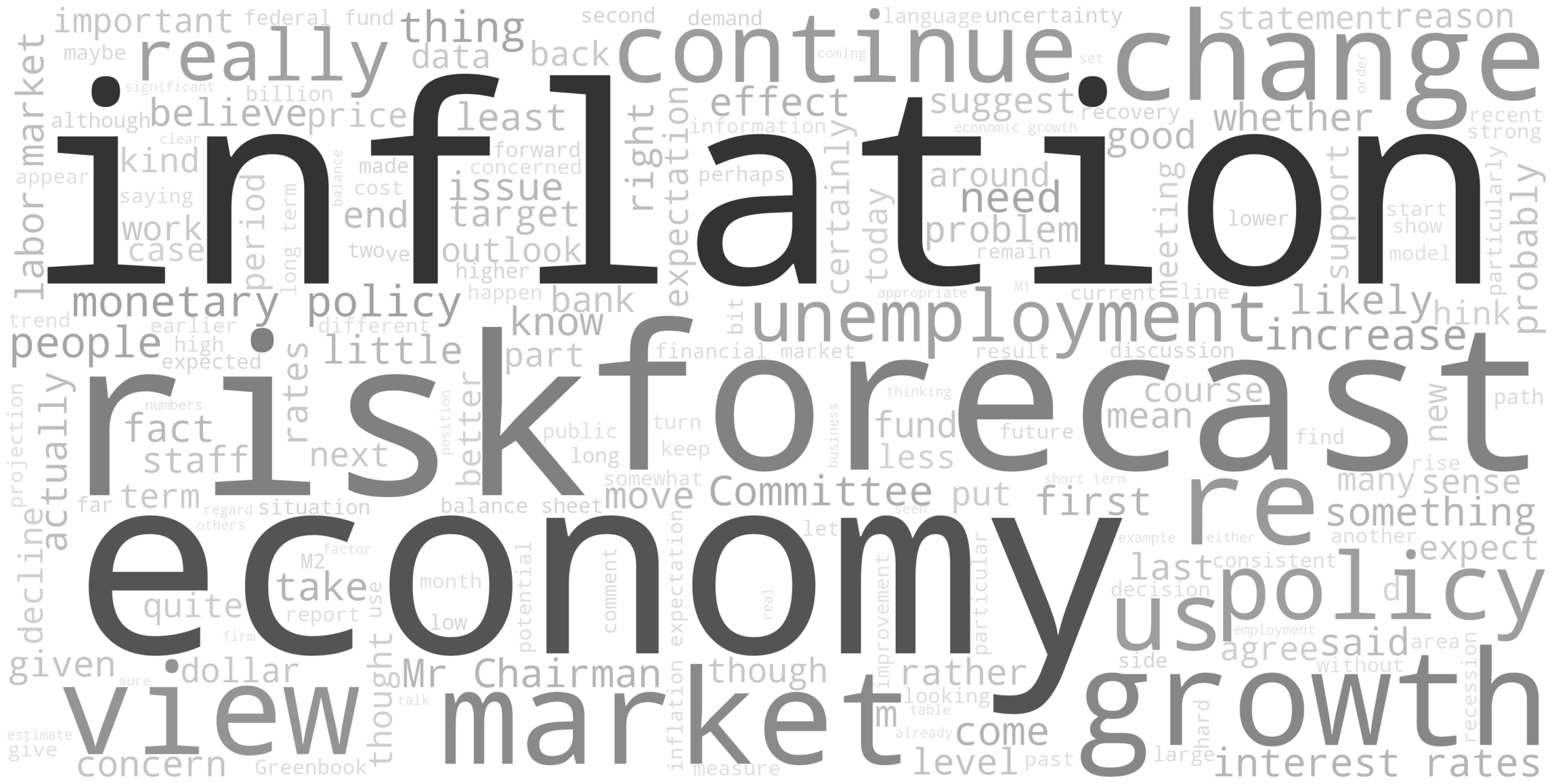}
		\caption{Vote For, Rate Unchanged}
	\end{subfigure}  
	\begin{subfigure}{0.48\linewidth}
		\centering
		\includegraphics[width=1\textwidth]{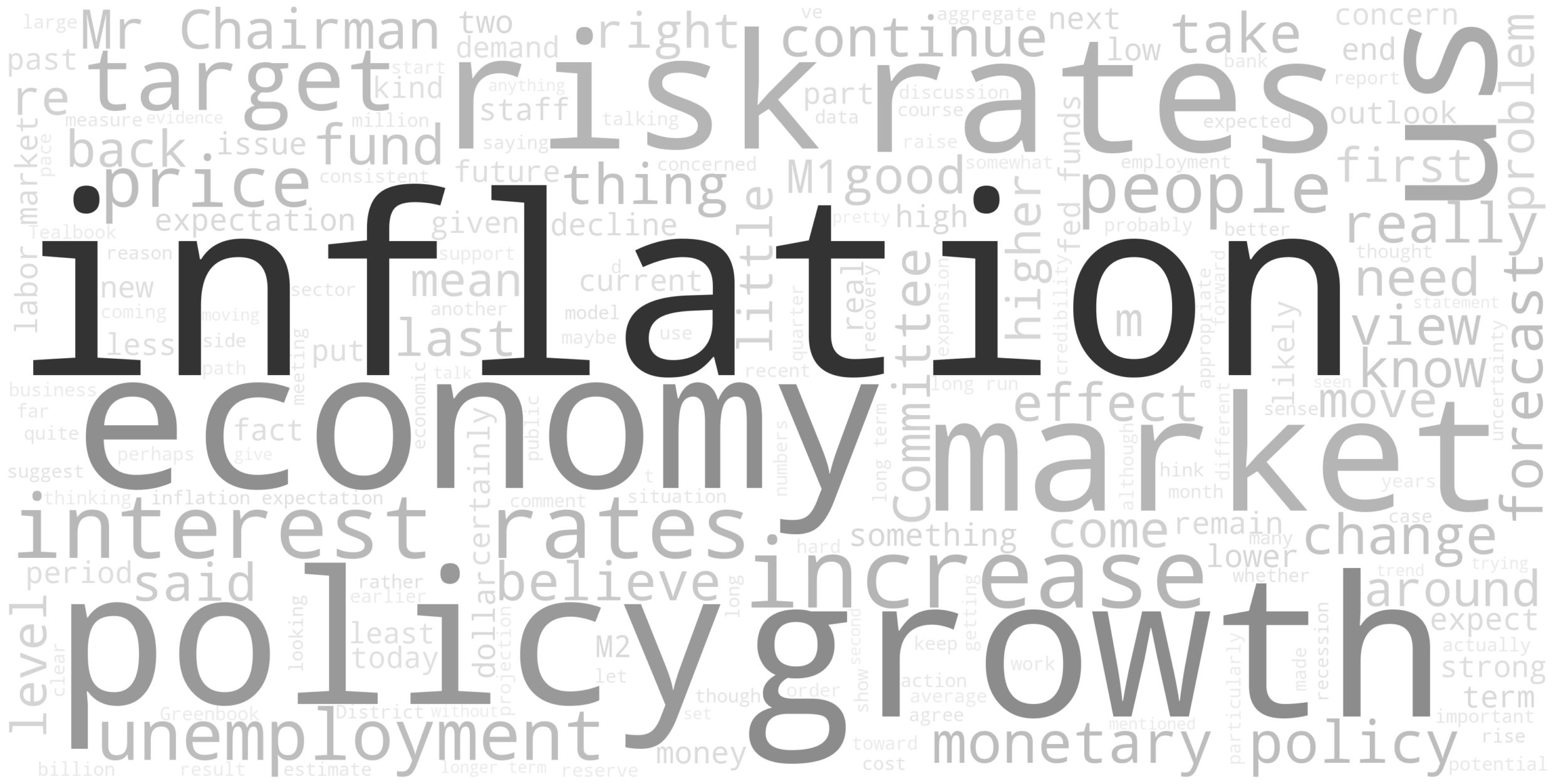}
		\caption{Vote Against, Rate Unchanged}
	\end{subfigure}
	\begin{subfigure}{0.48\linewidth}
		\centering
		\includegraphics[width=1\textwidth]{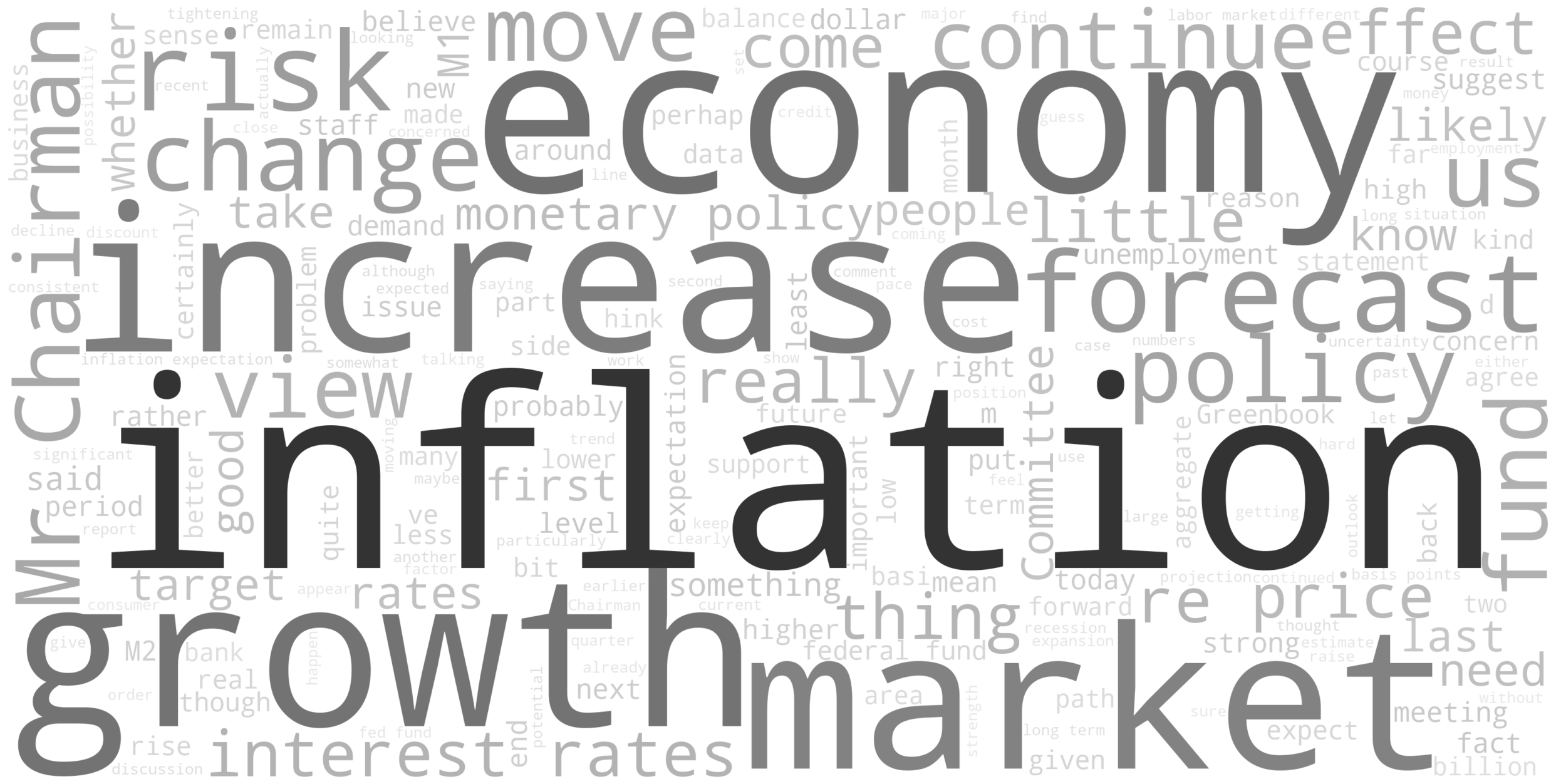}
		\caption{Vote For, Rate Increases}
	\end{subfigure}
	\begin{subfigure}{0.48\linewidth}
		\centering
		\includegraphics[width=1\textwidth]{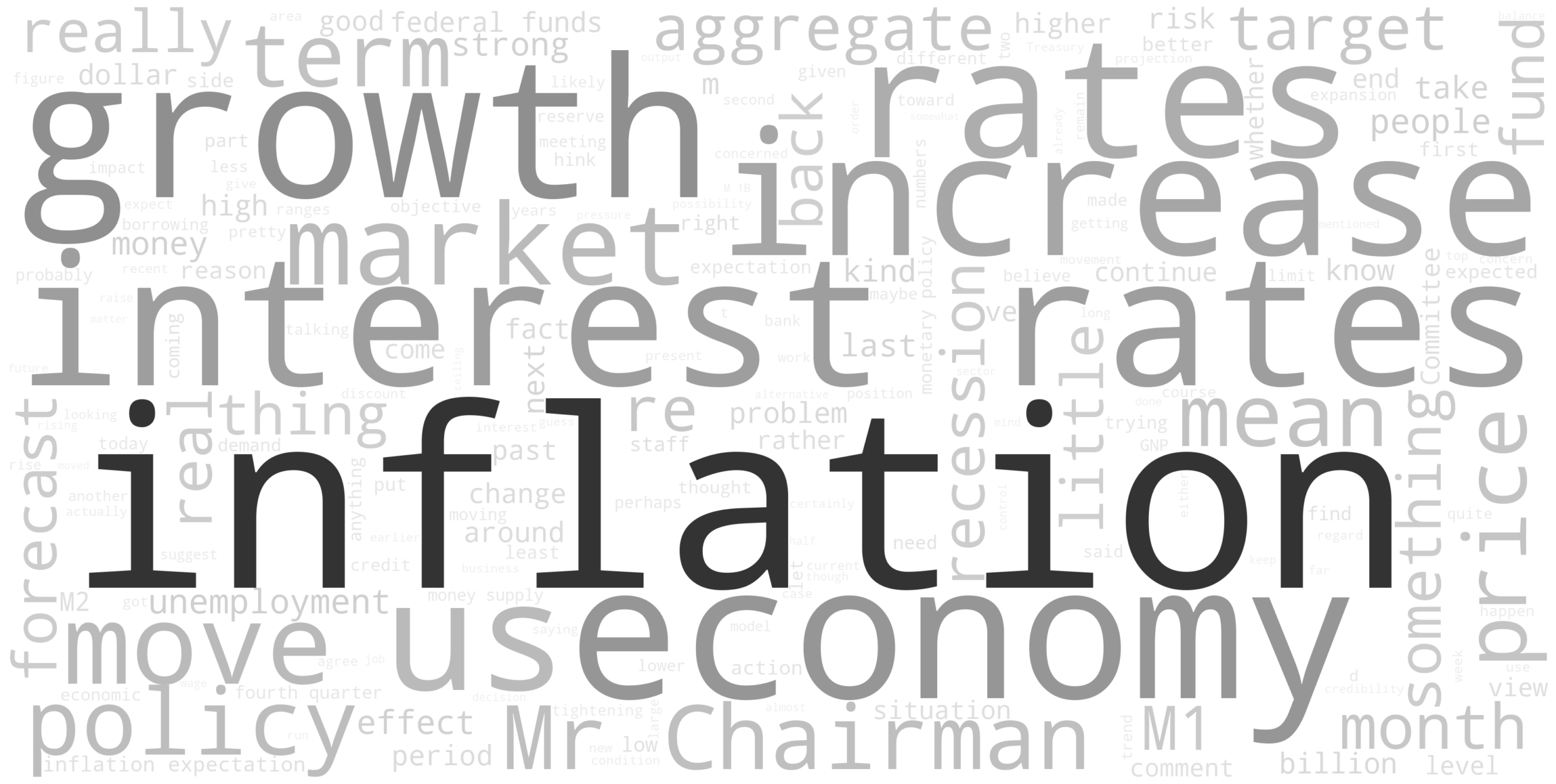}
		\caption{Vote Against, Rate Increases}
	\end{subfigure}
	\begin{subfigure}{0.48\linewidth}
		\centering
		\includegraphics[width=1\textwidth]{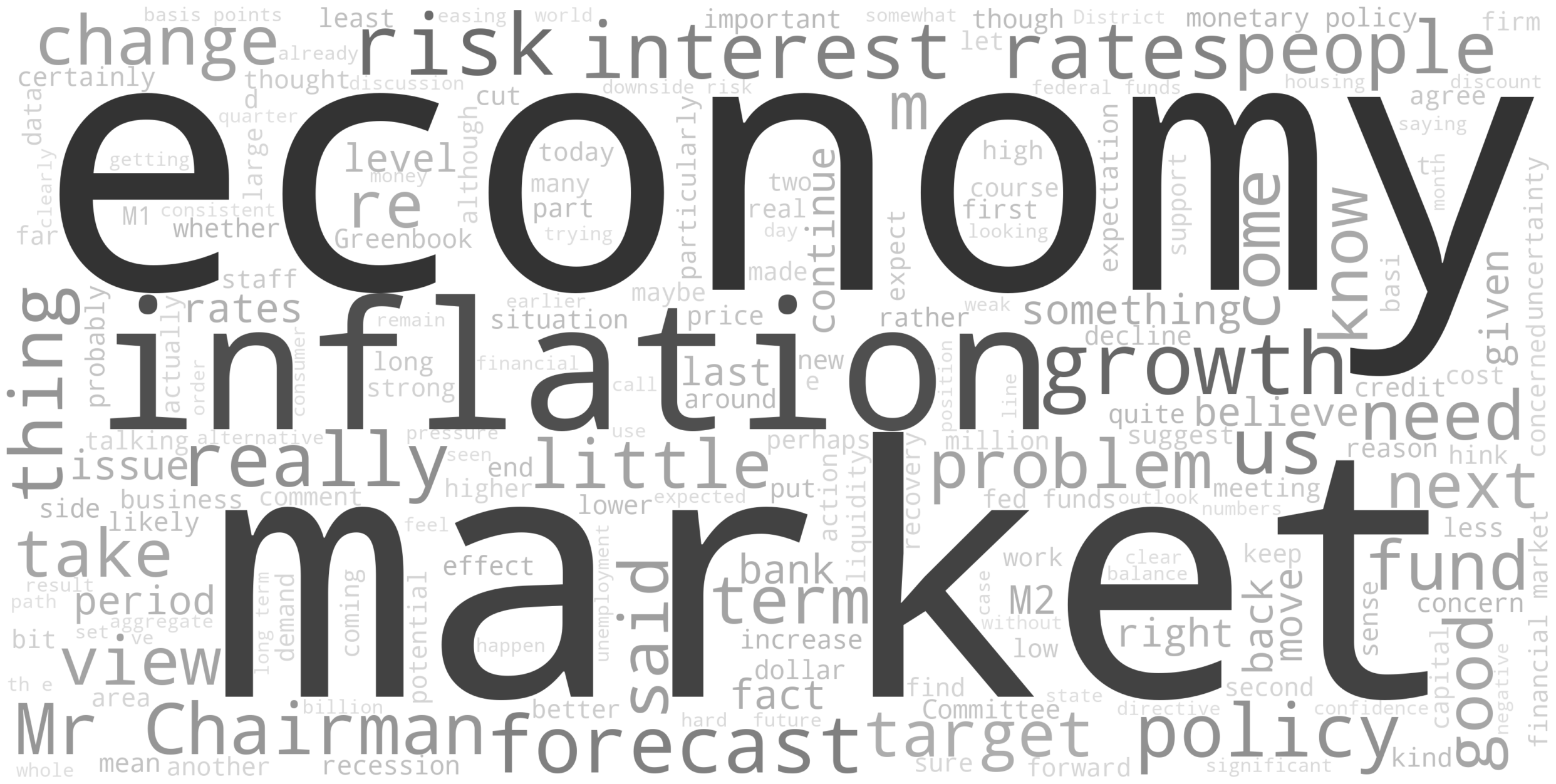}
		\caption{Vote For, Rate Decreases}
	\end{subfigure}
	\begin{subfigure}{0.48\linewidth}
		\centering
		\includegraphics[width=1\textwidth]{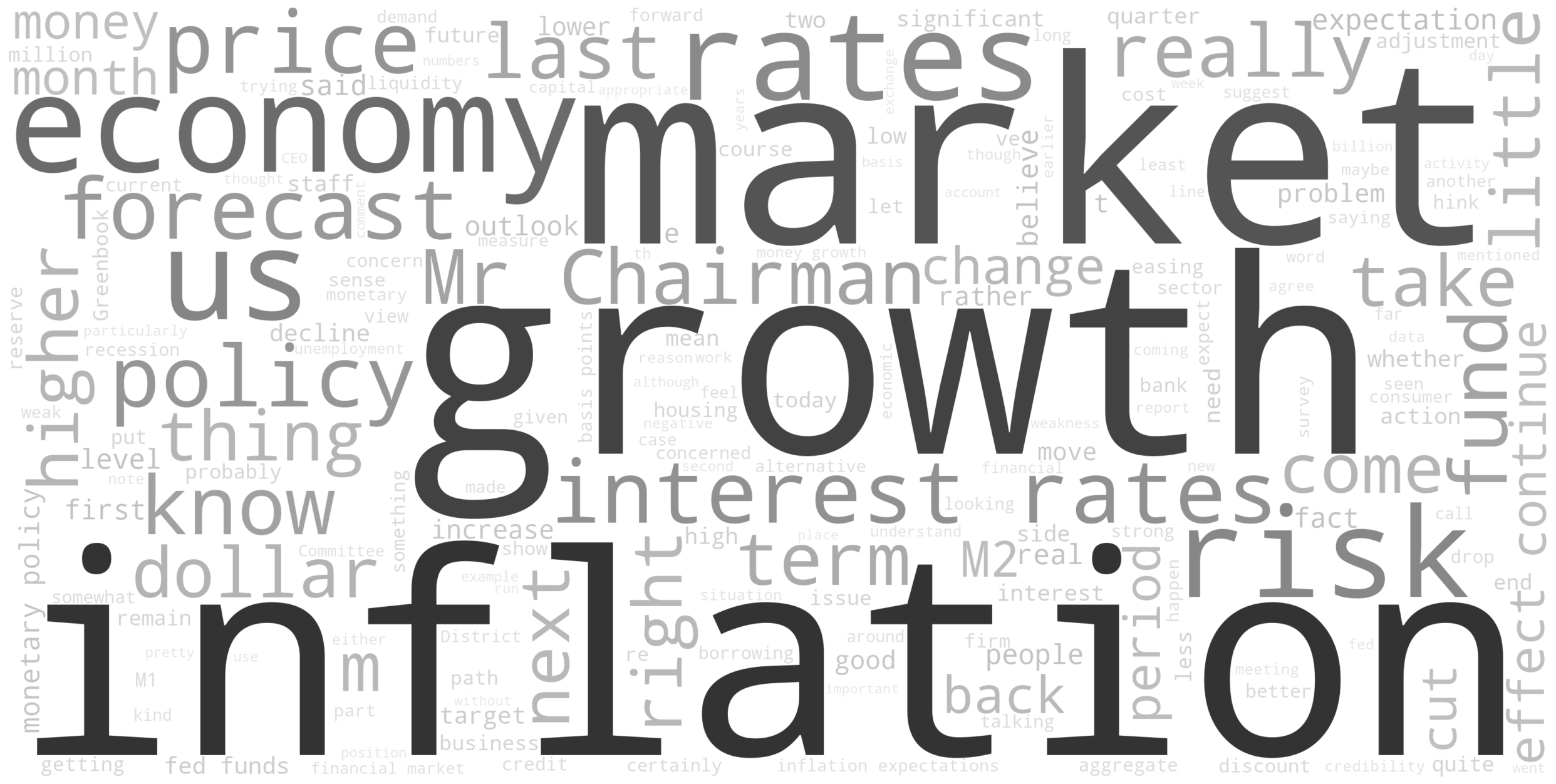}
		\caption{Vote Against, Rate Decreases}
	\end{subfigure}
	\caption{The Word Cloud for Different Vote Groups}
    \raggedright \emph{Note:} This figure shows the word cloud plots for different vote groups and FOMC meeting policy actions. In these word cloud plots, both the gray shades and the sizes of these words represent their frequency in meeting transcripts. 
    \label{wordcloud_vote}
\end{figure}


\clearpage
\newgeometry{left=2cm, right=2cm, top=1.5cm, bottom=1.5cm}  
\begin{table}[!htbp]
	\small 
	\caption{Voting Results and Personal Experience}
\begin{center}
\begin{tabular}{@{\extracolsep{-10pt}}lD{.}{.}{-3} D{.}{.}{-3} D{.}{.}{-3}} 
	\hline 
	\hline
	& \multicolumn{3}{c}{$v_{ij}$ (\textit{Probit Model}) } \\ 
	\cline{2-4}
	& \multicolumn{1}{c}{(1)} & \multicolumn{1}{c}{(2)} & \multicolumn{1}{c}{(3)}\\
	\multicolumn{2}{l}{Macro Factors}   & &  \\
	\cline{1-1} 
	$T_{unemp}$                                & -0.0275       & 0.0049        & 0.0085        \\
											   & (0.1724)      & (0.2984)      & (0.1854)      \\
	$D_{unemp}$                                & 0.1610        & -0.1444       & 0.2228        \\
											   & (0.1446)      & (0.3071)      & (0.1750)      \\
	$T_{CPI}$                                  & -0.1245       & 0.1946        & -0.2324       \\
											   & (0.1663)      & (0.1690)      & (0.2216)      \\
	$D_{CPI}$                                  & -0.1258       & 0.0386        & -0.0400       \\
											   & (0.2747)      & (0.2647)      & (0.4146)      \\
	\multicolumn{2}{l}{Interaction with Personal Experience}   & &  \\
	\cline{1-1} 
	$T_{unemp} \times \text{Great Depression}$ & 0.1370        &                 &                 \\
                                               & (0.3846)      &                 &                 \\
	$D_{unemp} \times \text{Great Depression}$ & -0.0031       &                 &                 \\
											   & (0.4289)      &                 &                 \\
	$T_{CPI} \times \text{Great Depression}$   & 0.4767^{*}    &                 &                 \\
											   & (0.2551)      &                 &                 \\
	$D_{CPI} \times \text{Great Depression}$   & 0.3002        &                 &                 \\
											   & (0.4030)      &                 &                 \\
	$T_{unemp} \times \text{Great Inflation}$  &                 & -0.0127       &                 \\
											   &                 & (0.3484)      &                 \\
	$D_{unemp} \times \text{Great Inflation}$  &                 & 0.4185        &                 \\
											   &                 & (0.3450)      &                 \\
	$T_{CPI} \times \text{Great Inflation}$    &                 & -0.2437       &                 \\
											   &                 & (0.2507)      &                 \\
	$D_{CPI} \times \text{Great Inflation}$    &                 & 0.0881        &                 \\
											   &                 & (0.4030)      &                 \\
	$T_{unemp} \times \text{WWII}$             &                 &                 & -0.0539       \\
											   &                 &                 & (0.3356)      \\
	$D_{unemp} \times \text{WWII}$             &                 &                 & -0.3148       \\
											   &                 &                 & (0.3371)      \\
	$T_{CPI} \times \text{WWII}$               &                 &                 & 0.4902^{*}    \\
											   &                 &                 & (0.2735)      \\
	$D_{CPI} \times \text{WWII}$               &                 &                 & 0.1226        \\
											   &                 &                 & (0.4755)      \\
	\hline
	Fixed Effects  & \multicolumn{1}{c}{Yes} & \multicolumn{1}{c}{Yes} & \multicolumn{1}{c}{Yes} \\
	Log Likelihood & \multicolumn{1}{c}{-328.5406} &\multicolumn{1}{c}{-328.7668} & \multicolumn{1}{c}{-327.3776}\\ 
	$N$ &\multicolumn{1}{c}{2,539} &\multicolumn{1}{c}{2,539} & \multicolumn{1}{c}{2,539} \\ 
	\hline 
	\hline 
	\end{tabular}
	\end{center}
	\begin{tablenotes}
		\small
		\item \emph{Note:} This table examines how members' personal experience may interact with macroecomic situations and affect observed voting results. The sample period spans from the February 1986 FOMC meeting, when Tealbook core CPI data first became available, to the December 2018 meeting.  $^{***}p<0.01$; $^{**}p<0.05$; $^{*}p<0.1$
	\end{tablenotes}
	\label{personal_level_reg_3_appendix}
\end{table}

\clearpage
\newgeometry{left=2cm, right=2cm, top=1.5cm, bottom=1.5cm}  
\begin{table}
\caption{Level of Hidden Dissent and Inverted Yield Curve Periods}
\begin{center}
	\begin{tabular}{@{\extracolsep{5pt}}lD{.}{.}{-4} D{.}{.}{-4} D{.}{.}{-4} D{.}{.}{-4} }
		\hline
		\hline
		& \multicolumn{2}{c}{$HD_i$}  & \multicolumn{2}{c}{$V_i$}\\
		\cline{2-5}
		& (1)              & (2)               & (3)                & (4)    \\
\multicolumn{2}{l}{Macro Factors}   & & & \\
\cline{1-1}
			$T_{unemp}$ & -0.6160^{***} & -0.5908^{***} & 0.0255 & -0.1016 \\ 
			& (0.2358) & (0.2292) & (0.7015) & (0.6811) \\ 
 
			$D_{unemp}$ & -0.1807 & -0.1272 & 0.0449 & -0.1830 \\ 
			& (0.1414) & (0.0889) & (0.4503) & (0.2935) \\ 

			$T_{CPI}$ & 0.7156^{***} & 0.6968^{***} & 1.1985 & 1.2826 \\ 
			& (0.2647) & (0.2623) & (0.8302) & (0.8223) \\ 

			$D_{CPI}$ & 0.2854 & 0.2494 & 0.6612 & 0.7814 \\ 
			& (0.1888) & (0.1752) & (0.6228) & (0.5878) \\ 

			$\text{Diff}_{Y2-Y10}$ & -0.0307 &  & 0.1373 &  \\ 
			& (0.0747) &  & (0.2504) &  \\ 

			$\text{Dummy}_{IYC}$ &  & 0.0251 &  & -0.1612 \\ 
			&  & (0.1171) &  & (0.3820) \\ 
\multicolumn{2}{l}{Member Char.}  & & & \\
\cline{1-1}
			$D_{experience}$ & 0.0144 & 0.0126 & 0.0805 & 0.0877 \\ 
			& (0.0353) & (0.0351) & (0.1164) & (0.1154) \\ 
	
			$D_{age}$ & 0.0398^{*} & 0.0381^{*} & 0.0913 & 0.1001 \\ 
			& (0.0231) & (0.0229) & (0.0800) & (0.0787) \\ 

			$D_{SchoolWealth}$ & 0.0104 & 0.0175 & -0.1999 & -0.2208 \\ 
			& (0.2151) & (0.2147) & (0.7043) & (0.6941) \\ 
	
			$P_{gender}$ & 0.6788 & 0.6077 & 0.6112 & 1.0206 \\ 
			& (0.4477) & (0.4158) & (1.5703) & (1.3909) \\ 
	
			$P_{major}$ & -1.3874^{***} & -1.3653^{***} & -1.3635 & -1.4985 \\ 
			& (0.3589) & (0.3604) & (1.1519) & (1.1508) \\ 
		
			$E_{hometown}$ & -0.2558 & -0.2081 & 0.0019 & -0.2364 \\ 
			& (0.3170) & (0.3116) & (1.0158) & (0.9959) \\ 
		
			$E_{school}$ & 1.0317^{***} & 1.0065^{***} & 1.5067 & 1.5961 \\ 
			& (0.3161) & (0.3166) & (1.0907) & (1.0888) \\ 
		
			$E_{POTUS}$ & 0.2092 & 0.1716 & -0.9115 & -0.7259 \\ 
			& (0.2062) & (0.1995) & (0.6557) & (0.6133) \\ 

			$Party_{Dem}$ & 0.1132 & 0.0928 & 0.4459 & 0.5208 \\ 
			& (0.1092) & (0.1032) & (0.3835) & (0.3691) \\ 

			$P_{depression}$ & -0.1069 & -0.1279 & -0.0682 & -0.0309 \\ 
			& (0.4844) & (0.4811) & (1.6261) & (1.6169) \\ 

			$P_{inflation}$ & 0.1191 & 0.1668 & -0.6957 & -1.1357 \\ 
			& (0.6984) & (0.7225) & (2.4104) & (2.4684) \\ 

			$P_{WWII}$ & 1.1133^{**} & 1.1527^{**} & -0.9135 & -1.1733 \\ 
			& (0.5048) & (0.5189) & (1.6494) & (1.6665) \\ 
			\hline
			Pseudo $R^2$   & \multicolumn{1}{c}{0.3898} & \multicolumn{1}{c}{0.3894} &  \multicolumn{1}{c}{0.4921} &  \multicolumn{1}{c}{0.4918}    \\
			Log Likelihood & \multicolumn{1}{c}{274.7080} & \multicolumn{1}{c}{274.6479} & \multicolumn{1}{c}{-13.4854} & \multicolumn{1}{c}{-13.4924} \\
			$N$                     & 268                  & 268                  & 268          & 268            \\
			\hline
			\hline
		\end{tabular}
	\end{center}
	\begin{tablenotes}
		\small
		\item \emph{Note:} This table examines the hidden dissent level change during the inverted yield curve period. $\text{Diff}_{Y2-Y10}$ denotes the spread between 2-Year and 10-Year Treasury bonds. $\text{Dummy}_{IYC}$ indicates if the inverted yield curve is observed in current period. For other variables, please refer to \autoref{voting_pred_reg1} for their definitions. $^{***}p<0.01$; $^{**}p<0.05$; $^{*}p<0.1$
	\end{tablenotes}
	\label{voting_pred_reg1_inverted_yield_curve}
\end{table}

\clearpage 
\restoregeometry 

\newgeometry{left=2cm, right=2cm, top=1.5cm, bottom=1.5cm}  
\begin{table}
	\caption{Level of Hidden Dissent and Uncertainty Indices}
	\begin{center}
		\begin{tabular}{@{\extracolsep{5pt}}lD{.}{.}{-4} D{.}{.}{-4} D{.}{.}{-4} D{.}{.}{-4} }
			\hline
			\hline
			& \multicolumn{2}{c}{$HD_i$}  & \multicolumn{2}{c}{$V_i$}\\
			\cline{2-5}
			& (1)              & (2)               & (3)                & (4)    \\
			\multicolumn{2}{l}{Macro Factors}   & & & \\
			\cline{1-1}
			$T_{unemp}$ & -0.7007^{**} & -0.6057^{**} & -0.5115 & -0.0713 \\ 
			& (0.2838) & (0.2436) & (0.9517) & (0.7152) \\ 

			$D_{unemp}$ & -0.0897 & -0.1386 & -0.0413 & -0.1474 \\ 
			& (0.0904) & (0.0928) & (0.3241) & (0.3098) \\ 

			$T_{CPI}$ & 0.6378^{**} & 0.6975^{***} & 1.5757 & 1.2533 \\ 
			& (0.3112) & (0.2624) & (1.0192) & (0.8224) \\ 

			$D_{CPI}$ & 0.3160 & 0.2516 & 1.1307 & 0.7646 \\ 
			& (0.1965) & (0.1748) & (0.6887) & (0.5907) \\ 

			$\text{VIX}$ & 0.0050 &  & 0.0095 &  \\ 
			& (0.0054) &  & (0.0175) &  \\ 

			$\text{EPU}$ &  & 0.0002 &  & 0.00002 \\ 
			&  & (0.0013) &  & (0.0038) \\ 
			\multicolumn{2}{l}{Member Char.}  & & & \\
			\cline{1-1}
			$D_{experience}$ & -0.0513 & 0.0125 & 0.0693 & 0.0845 \\ 
			& (0.0415) & (0.0351) & (0.1433) & (0.1156) \\ 

			$D_{age}$ & -0.0040 & 0.0387^{*} & 0.1047 & 0.0983 \\ 
			& (0.0275) & (0.0229) & (0.0989) & (0.0790) \\ 

			$D_{SchoolWealth}$ & -0.0213 & 0.0228 & -0.9370 & -0.2482 \\ 
			& (0.2572) & (0.2133) & (0.8951) & (0.6938) \\ 

			$P_{gender}$ & 0.1148 & 0.6060 & 1.5577 & 0.9971 \\ 
			& (0.4568) & (0.4160) & (1.6067) & (1.3962) \\ 
		
			$P_{major}$ & -1.6489^{***} & -1.3762^{***} & -0.3879 & -1.4270 \\ 
			& (0.4291) & (0.3569) & (1.4509) & (1.1416) \\ 
	
			$E_{hometown}$ & -0.8732^{**} & -0.2291 & -0.0115 & -0.1504 \\ 
			& (0.3699) & (0.3097) & (1.2069) & (0.9841) \\ 

			$E_{school}$ & 0.7530^{**} & 1.0172^{***} & 1.1336 & 1.5519 \\ 
			& (0.3397) & (0.3135) & (1.1935) & (1.0860) \\ 
		
			$E_{POTUS}$ & 0.2128 & 0.1827 & -1.0109 & -0.7704 \\ 
			& (0.2128) & (0.1957) & (0.7200) & (0.6044) \\ 
	
			$Party_{Dem}$ & 0.0905 & 0.0970 & 0.5288 & 0.5027 \\ 
			& (0.1132) & (0.1024) & (0.4404) & (0.3697) \\ 

			$P_{depression}$ & -0.7258 & -0.1407 & -0.0301 & 0.0255 \\ 
			& (0.5420) & (0.4836) & (1.9149) & (1.6232) \\ 
		
			$P_{inflation}$ & 0.5672 & 0.1271 & 0.4175 & -0.8644 \\ 
			& (0.7497) & (0.6984) & (2.7058) & (2.3885) \\ 
	
			$P_{WWII}$ & 1.8389^{***} & 1.1369^{**} & -0.2907 & -1.0183 \\ 
			& (0.5754) & (0.5113) & (1.9666) & (1.6354) \\ 
			\hline
			Pseudo $R^2$    & \multicolumn{1}{c}{0.3187} & \multicolumn{1}{c}{0.3892} & \multicolumn{1}{c}{0.5594} & \multicolumn{1}{c}{0.4914}  \\
			Log Likelihood & \multicolumn{1}{c}{248.9581} & \multicolumn{1}{c}{274.6337} & \multicolumn{1}{c}{-11.6993} & \multicolumn{1}{c}{-13.5026} \\
			$N$                     & 235                 & 268                   & 235          & 268            \\
			\hline
			\hline
		\end{tabular}
	\end{center}
		\begin{tablenotes}
		\small
		\item \emph{Note:} This table examines the correlation between hidden dissent level and uncertainty. VIX denotes the CBOE Volatility Index. EPU is the U.S. Economic Policy Uncertainty Index and we match this monthly data to the nearest FOMC meeting. For other variables, please refer to \autoref{voting_pred_reg1} for their definitions. $^{***}p<0.01$; $^{**}p<0.05$; $^{*}p<0.1$
	\end{tablenotes}
	\label{voting_pred_reg1_uncertainty_indices}
\end{table}

\clearpage 
\restoregeometry 

\clearpage


\renewcommand{\thesection}{C}

\section{Examine Speech Data}
\label{appendix_speech_data}

\noindent Most recent governors' public speeches can be found on the \href{https://www.federalreserve.gov/newsevents/speeches.htm}{Board of Governors of the Federal Reserve} website, and recent presidents' public speeches can be found on the respective regional Federal Reserve Bank website. However, the time coverage varies across different regional Federal Reserve Banks, with many listing speeches from only the two most recent presidents. To access earlier public speeches, we turn to the \href{https://fraser.stlouisfed.org}{FRASER} database maintained by the Federal Reserve Bank of St. Louis.\footnote{In theory, we can download all the public speeches from the FRASER data base. However, Federal Reserve Banks usually provide the HTML version of their presidents' latest speeches which have higher accuracy than the data from the FRASER data base which was converted from pdf files using OCR.} 

In total, we have collected 5,859 speeches from 1970 to 2022. After plotting out the number of speeches per year and the number of labeled transcript data points per year (see \autoref{num_speech}), we notice that prior to 1987, fewer than 50 speeches are successfully retrieved per year. We also find out that the number of public speeches delivered by FOMC members is increasing over time, while the number of labeled transcript observations is relatively stable.

\subsection{Hidden dissent inside FOMC members' speeches}
\label{fomc_member_speeches}

Next we apply the trained deep learning model to analyze the speeches of FOMC members. Following the procedure outlined in \autoref{transcripts_data}, our model assigns a hidden dissent score, $hd_{ij}^{speech}$, to each speech, which reflects the degree of divergence from the chair's stance. A higher $hd_{ij}^{speech}$ score implies a greater deviation from the chair's stance and an increased probability of a dissenting vote (NO) at the end of the meeting. A critical aspect to consider is the selection of an appropriate reference point for the deep learning model, as its effectiveness in measuring hidden dissent depends on identifying a baseline that represents the chair's stance. To capture the hidden dissent demonstrated in members' speeches, we use two distinct reference points: (1) the chair's transcript from the relevant FOMC meeting, with the resulting hidden dissent score denoted as $hd_{ij}^{speech, in}$, (2) the chair's most recent public address preceding the FOMC meeting, and the score is represented as $hd_{ij}^{speech, pre}$.

After obtaining the $hd_{ij}^{speech}$ score, we calculate the annual average to represent the overall level of hidden dissent for speeches and transcripts each year. These values are then compared with the revealed dissent measured by the annual count of NO votes in \autoref{vote_speech_yearly}.

First, we observe that the hidden dissent levels measured in speech data roughly track those derived from transcript. When either measure is high, indicating a strong disagreement among FOMC members, there tend to be a higher number of NO votes recorded in that year. 

Second, compared to the level of hidden dissent derived from speeches (i.e., $hd_{ij}^{speech}$), the values obtained from transcript (i.e., $hd_{ij}$) exhibit a closer alignment with the revealed dissent in the voting records. This finding is not surprising given that we can assign scores to all FOMC meeting members' transcripts. However, not every FOMC member delivered a speech, it inevitably leaves us with a relatively ``sparse" speech dataset. Additionally, certain disparities emerge between the hidden dissent trends and actual voting outcomes. For instance, both measures indicate low hidden dissent level in 1998, but six NO votes were recorded that year. The reason is that five out the six NO votes were cast by President Jerry L. Jordan, making them statistic outlier that are overlooked by the current measure.

Furthermore, the scores measured in speeches tend to be closer to 0.5 compared to those derived from transcripts, particularly after 2000. We interpret it as evidence that FOMC members tend to adopt a more neutral tone in public speeches than in their remarks during FOMC meetings, likely to avoid causing confusion or uncertainty in the financial markets. 

Finally, when using the chair's recent public speech as the reference point, scores derived from members' speeches consistently fall below those obtained from using the chair's meeting transcripts as the reference point. This discrepancy underscores the increased difficulty in identifying divergences between the opinions of FOMC members and their chair through mere examination of their public speeches.

Utilizing the score $hd_{ij}^{speech}$ from members' speeches, we can also construct the variable $HD_i^{speech}$ to measure the hidden dissent level prior to each FOMC meeting. These speeches are linked to meeting $i$ if delivered between meetings $i-1$ and $i$, and $HD_i^{speech}$ is calculated using the same method as $HD_i$.  Summary statistics for these hidden dissent measures are reported in \autoref{disagree_measure_speech}.

\subsection{Do public speeches reveal hidden dissent in advance?}

While FOMC transcripts are released with a 5-year lag, FOMC members deliver speeches regularly. Do these speeches reveal their hidden dissent ahead of the following meeting?

As discussed in \autoref{fomc_member_speeches}, we feed FOMC members' speeches into the deep learning model and generate two hidden dissent scores for each speech, $hd_{ij}^{speech, in}$ and $hd_{ij}^{speech, pre}$, depending on how to chose the reference point for chair. The first measure, $hd_{ij}^{speech, in}$, uses the chair's words from the following meeting as the reference point. Since transcripts are subject to a 5-year delay and are not available in real time, the second measure,  $hd_{ij}^{speech, pre}$, uses the chair's speeches prior to the meeting, which are available in real time, as the reference point instead. For both measures, we compute $HD_i^{speech}$ for each meeting $i$ by averaging the hidden dissent scores for speeches delivered between meetings $i-1$ and $i$. 

Given that not all FOMC members give speeches regularly, $HD_i^{speech}$ is likely a noisier measure than the hidden dissent derived from transcripts $HD_i$.  Nevertheless, as shown in \autoref{vote_speech_yearly}, the two variables are still positively correlated. 

To evaluate the predictive power of speeches, we add the hidden dissent score derived from speeches into the regressions described in \autoref{section4_1} and the results are presented in \autoref{personal_level_reg_2}. Columns (1) and (3) shows a s positive correlation between $hd_{ij}^{speech, in}$ and the hidden dissent score derived from transcripts. This suggests that if we know what the chair said in the subsequent meeting, we can detect hidden dissent in the speeches made by FOMC members. However, when the measure is changed to $hd_{ij}^{speech, pre}$ in columns (2) and (4),  the predictive power of speeches goes down significantly. Knowing only the content of the chair’s speeches during the same period makes it much harder to identify hidden dissent in members’ speeches.

Results at the meeting level are reported in \autoref{voting_pred_reg2_full_fomc_in_meeting} and \autoref{voting_pred_reg2_full_fomc_pre_meeting} and they tell a similar story. The measure $HD_i^{speech, in}$ predicts the subsequent meeting's hidden dissent level $HD_i$, and the predictive power evaporates when we change the measure to $HD_i^{speech, pre}$.  For revealed dissent , as measured by $V_i$, neither of the measures demonstrates any predictive capability. Since not all FOMC members give speeches regularly, we also calculate $HD_i$ considering only the members who delivered speeches before the meeting as a robustness check, and results are similar (see \autoref{voting_pred_reg3_partial_fomc}).

It is well-known that speeches made by FOMC members are carefully crafted and they often do not reveal much about their views on monetary policy. Our findings suggest a more nuanced insight: speeches by FOMC members can be informative only if we have knowledge of what the chair will say in the subsequent meeting. Without that reference point, information in the speeches becomes elusive.

\clearpage


\begin{figure}[hbtp!]
	\begin{center}
		\includegraphics[width=14cm]{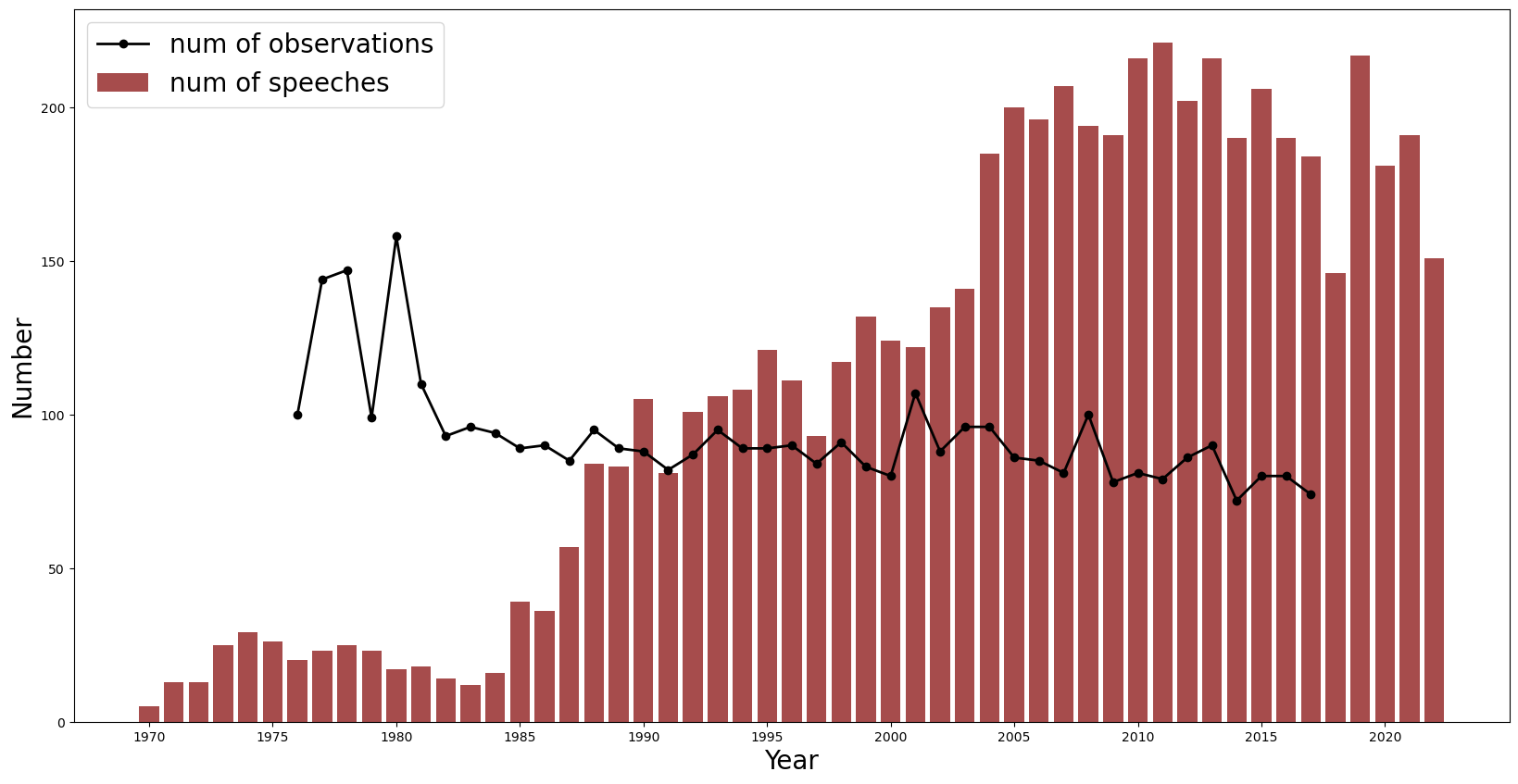}
		\caption{Numbers of Observations and Speeches in Each Year} 
		\label{num_speech}
	\end{center}
	\emph{Note:} The bar chart shows the yearly count of public speeches by FOMC members, with the black dotted line indicating the number of voting members (observations). Most speeches are from recent years, with fewer than 50 speeches retrieved annually before 1987.
\end{figure}

\begin{figure}[hbtp!]
	\begin{center}
		\includegraphics[width=14cm]{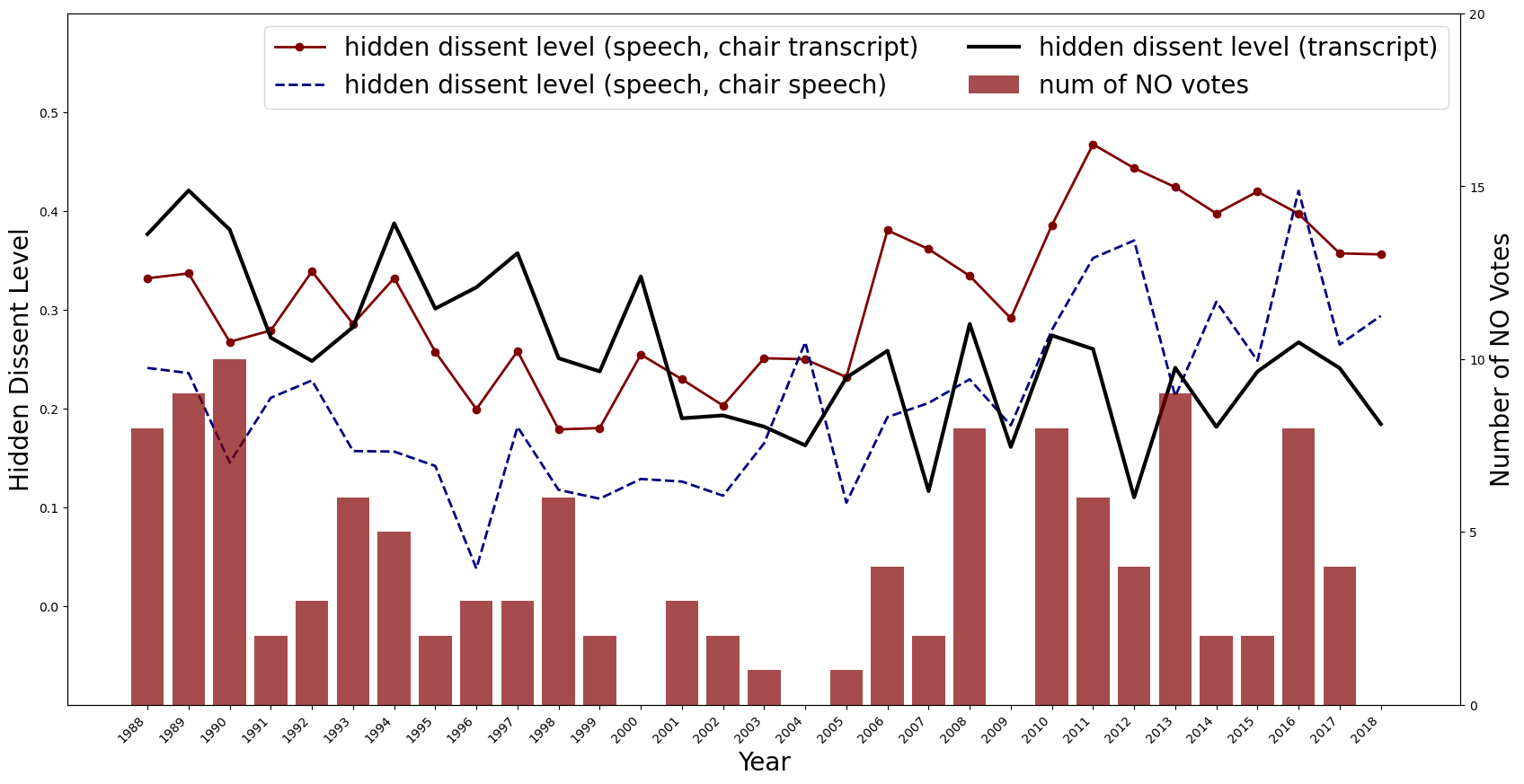}
		\caption{Level of Hidden Dissent in Speeches and Transcripts, NO Votes} 
		\label{vote_speech_yearly}
	\end{center}
	\emph{Note:} The red dotted and blue dashed lines depict hidden dissent levels in speeches, calculated with different reference points and averaged annually. Similarly, the black line shows hidden dissent in transcripts using the same method. The bar chart represents the count of NO votes recorded in each meeting. 
\end{figure}

\clearpage


\begin{table}[!htbp] 
	\begin{center}
		\caption{Summary Statistics for Hidden Dissent Measures} 
		\label{disagree_measure_speech}
		\begin{tabular}{lccccc}
			\hline
			\hline
			Measure            & Mean & S.D. & Min & Max & Count \\
			\hline
			Individual Level & & & & & \\
			\cline{1-1}
			$hd_{ij}^{speech, in}$   & 0.2972 & 0.2764 & 0.0000 & 0.8935 & 1219\\
			$hd_{ij}^{speech, pre}$   & 0.1912 & 0.2344 & 0.0000 & 0.8506 & 1090\\
			[1.2ex] 
			Meeting Level & & & & & \\
			\cline{1-1}
			$HD_{i}^{speech, in}$  & 0.3818 & 0.1954 & 0.0002 & 0.8945 & 345\\
			$HD_{i}^{speech, pre}$  & 0.2101 & 0.1507 & 0.0017 & 0.6399 & 220\\
			\hline
			\hline
		\end{tabular}\\
	\end{center}
	\emph{Note:} This table shows the summary statistics of different hidden dissent measures used in this study. Variable $x_{ij}$ represents hidden dissent directly derived from our deep learning model using transcripts or speeches. Variable $X_{i}$ denotes the hidden dissent level calculated based on $x_{ij}$. Please refer to \autoref{results} for a detailed discussion.  
\end{table}

\begin{table}[!htbp]
	\small 
	\caption{Predicting Hidden Dissent with Speeches (Individual-level)} 
	\begin{center}
		\begin{tabular}{@{\extracolsep{-10pt}}lD{.}{.}{-3} D{.}{.}{-3} D{.}{.}{-3} D{.}{.}{-3}} 
			\hline 
			\hline
			& \multicolumn{2}{c}{$hd_{ij}$ (\textit{Mixed Effects Beta}) } & \multicolumn{2}{c}{$v_{ij}$ (\textit{Mixed Effects Probit})} \\ 
			\cline{2-5}
			& \multicolumn{1}{c}{(1)} & \multicolumn{1}{c}{(2)} & \multicolumn{1}{c}{(3)}& \multicolumn{1}{c}{(4)} \\ 
			$hd_{ij}^{speech, in} $    & 0.4567^{***}  &                 & 0.4276^{**}   &                 \\
			& (0.1145)      &                 & (0.1705)      &                 \\
			$hd_{ij}^{speech, pre}$    &                 & 0.2687^{**}   &                 & 0.2499        \\
			&                 & (0.1368)      &                 & (0.1549)      \\
			\multicolumn{2}{l}{Macro Factors}   & & & \\
			\cline{1-1}
			$T_{unemp}$               & 0.0707        & 0.0743        & 0.0795        & 0.0437        \\
			& (0.1707)      & (0.1772)      & (0.1690)      & (0.1866)      \\
			$D_{unemp}$               & -0.0946       & -0.0896       & 0.3887^{**}   & 0.3094        \\
			& (0.0649)      & (0.0688)      & (0.1944)      & (0.2012)      \\
			$T_{CPI}$                & 0.9795^{***}  & 1.0531^{***}  & 0.1198        & 0.1384        \\
			& (0.2285)      & (0.2365)      & (0.1832)      & (0.1978)      \\
			$D_{CPI}$                & 0.6402^{***}  & 0.6380^{***}  & 0.9726^{***}  & 0.9964^{***}  \\
			& (0.1877)      & (0.1967)      & (0.2801)      & (0.2970)      \\
			\multicolumn{2}{l}{Member Char.}   & & & \\
			\cline{1-1}
			$\text{Econ Major}$       & -0.0608       & -0.0490       & -0.2266       & -0.0419       \\
			& (0.1742)      & (0.1835)      & (0.6350)      & (0.6517)      \\
			$\text{School Northeast}$        & -0.3313       & -0.2925       & -2.3252^{**}  & -2.1137^{**}  \\
			& (0.2586)      & (0.2724)      & (0.9251)      & (0.8669)      \\
			$\text{School South}$     & -0.7587^{***} & -0.7621^{***} & -3.8304^{***} & -4.3953^{***} \\
			& (0.2498)      & (0.2646)      & (1.0934)      & (1.2825)      \\
			$\text{School West}$      & 0.3753        & 0.4161        & -0.5504       & -0.0253       \\
			& (0.3539)      & (0.3721)      & (1.2832)      & (1.1674)      \\
			\hline 
			Control Var. & \multicolumn{1}{c}{Yes} & \multicolumn{1}{c}{Yes} & \multicolumn{1}{c}{Yes} & \multicolumn{1}{c}{Yes} \\
			Log Likelihood & \multicolumn{1}{c}{860.3329} & \multicolumn{1}{c}{776.3906} & \multicolumn{1}{c}{-166.1869} & \multicolumn{1}{c}{-141.9636} \\ 
			$N$ & \multicolumn{1}{c}{1,219} & \multicolumn{1}{c}{1,090} & \multicolumn{1}{c}{1,219}  & \multicolumn{1}{c}{1,090}\\ 
			\hline 
			\hline 
		\end{tabular} 
	\end{center}
	\begin{tablenotes}
		\small
		\item \emph{Note:} This table considers variables that may correlate with hidden dissent and dissent. $hd_{ij}^{speech,in}$ and $hd_{ij}^{speech,pre}$ measure the hidden dissent captured in speeches using chair's meeting transcripts and public speeches as reference points, respectively. Mixed effect regressions is clustered on FOMC member's level. Member characteristics are explained in \autoref{personal_info}. The sample period starts from the FOMC meeting on the February meeting in 1986 (when the Tealbook core CPI first becomes available) and ends on the December meeting in 2018.  $^{***}p<0.01$; $^{**}p<0.05$; $^{*}p<0.1$ 
	\end{tablenotes}
	\label{personal_level_reg_2}
\end{table} 

\newgeometry{left=2cm, right=2cm, top=1.5cm, bottom=1.5cm}  

\begin{table}
	\caption{Predicting Hidden Dissent with Speeches I (Meeting-level)}
	\label{}
	\begin{center}
		\begin{tabular}{@{\extracolsep{-10pt}}lD{.}{.}{-3} D{.}{.}{-3} D{.}{.}{-3} D{.}{.}{-3}}
			\hline
			\hline
			& \multicolumn{2}{c}{$HD_i$ (\textit{Beta})}  &  \multicolumn{2}{c}{$V_i$ (\textit{Fractional Logistic})}\\
			\cline{2-5}
			& (1)              & (2)               & (3)                & (4)            \\
			$HD_i^{speech, in}$ & 0.6770^{***} & 0.4914^{*} & 1.3958^{*} & 1.2482 \\ 
			& (0.2598) & (0.2556) & (0.7996) & (0.8192) \\ 
			\multicolumn{2}{l}{Macro Factors}   & & & \\
			\cline{1-1}
			$T_{unemp}$ &  & -0.5396^{**} &  & 0.0953 \\ 
			&  & (0.2281) &  & (0.6801) \\ 
			
			$D_{unemp}$ &  & -0.1141 &  & -0.1130 \\ 
			&  & (0.0849) &  & (0.2778) \\ 
			
			$T_{CPI}$ &  & 0.6789^{***} &  & 1.1680 \\ 
			&  & (0.2633) &  & (0.8245) \\ 
			
			$D_{CPI}$ &  & 0.2596 &  & 0.6958 \\ 
			&  & (0.1745) &  & (0.5802) \\ 
			\multicolumn{2}{l}{Member Char.}  & & & \\
			\cline{1-1}
			$D_{experience}$ & -0.0464 & -0.0020 & 0.0001 & 0.0665 \\ 
			& (0.0340) & (0.0354) & (0.1050) & (0.1143) \\ 
			
			$D_{age}$ & 0.0206 & 0.0216 & 0.0942 & 0.0934 \\ 
			& (0.0243) & (0.0238) & (0.0797) & (0.0799) \\ 
			
			$D_{SchoolWealth}$ & -0.3072 & -0.0710 & -0.3871 & -0.1747 \\ 
			& (0.2039) & (0.2210) & (0.6280) & (0.7015) \\ 
			
			$P_{gender}$ & 0.6218 & 0.5240 & 0.9499 & 0.8802 \\ 
			& (0.3916) & (0.4128) & (1.2707) & (1.3821) \\ 
			
			$P_{major}$ & -1.2971^{***} & -1.2842^{***} & -1.8010^{**} & -1.3628 \\ 
			& (0.2955) & (0.3592) & (0.8926) & (1.1337) \\ 
			
			$E_{hometown}$ & -0.0683 & -0.4117 & 0.0725 & -0.1057 \\ 
			& (0.3181) & (0.3239) & (0.9395) & (0.9918) \\ 
			
			$E_{school}$ & 0.6363^{**} & 0.8430^{***} & 1.6474 & 1.3021 \\ 
			& (0.3060) & (0.3189) & (1.0351) & (1.1012) \\ 
			
			$E_{POTUS}$ & 0.1288 & 0.1197 & -0.9205 & -0.8107 \\ 
			& (0.1960) & (0.1947) & (0.5720) & (0.5986) \\ 
			
			$Party_{Dem}$ & 0.0808 & 0.0903 & 0.1934 & 0.4686 \\ 
			& (0.0857) & (0.1015) & (0.2825) & (0.3595) \\ 
			
			$P_{depression}$ & -0.4803 & -0.2189 & -0.3122 & -0.1979 \\ 
			& (0.4852) & (0.4762) & (1.5834) & (1.5927) \\ 
			
			$P_{inflation}$ & 0.0477 & 0.3547 & -0.9352 & -0.7398 \\ 
			& (0.6867) & (0.7007) & (2.2780) & (2.3698) \\ 
			
			$P_{WWII}$ & 1.5255^{***} & 1.4318^{***} & -0.0620 & -0.5649 \\ 
			& (0.5025) & (0.5145) & (1.6196) & (1.6521) \\ 
			\hline
			Pseudo $R^{2}$ & \multicolumn{1}{c}{0.3388} & \multicolumn{1}{c}{0.3895} & \multicolumn{1}{c}{0.1095} &  \multicolumn{1}{c}{0.1207}\\ 
			Log Likelihood & \multicolumn{1}{c}{262.8188} & \multicolumn{1}{c}{273.9175} & \multicolumn{1}{c}{-13.4386} & \multicolumn{1}{c}{-13.2695} \\ 
			$N$                        & 265              & 265              & 265                & 265      \\
			\hline
			\hline
		\end{tabular}
	\end{center}
	
	\begin{tablenotes}
		\small
		\item \emph{Note:} This table examines the relationship between two hidden dissent measures constructed respectively from members' speech data and FOMC meeting transcripts. $HD_i^{speech, in}$ measures the level of hidden dissent within speeches delivered right before a certain FOMC meeting, here we use the chair's FOMC meeting transcript as the anchor. For other variables, please refer to \autoref{voting_pred_reg1} for their definitions. $^{***}p<0.01$; $^{**}p<0.05$; $^{*}p<0.1$
	\end{tablenotes}
	\label{voting_pred_reg2_full_fomc_in_meeting}
\end{table}

\clearpage 
\restoregeometry 

\newgeometry{left=2cm, right=2cm, top=1.5cm, bottom=1.5cm}  

\begin{table}
	\caption{Predicting Hidden Dissent with Speeches II (Meeting-level)}
	\label{}
	\begin{center}
		\begin{tabular}{@{\extracolsep{-10pt}}lD{.}{.}{-3} D{.}{.}{-3} D{.}{.}{-3} D{.}{.}{-3}}
			\hline
			\hline
			& \multicolumn{2}{c}{$HD_i$ (\textit{Beta})}  &  \multicolumn{2}{c}{$V_i$ (\textit{Fractional Logistic})}\\
			\cline{2-5}
			& (1)              & (2)               & (3)                & (4)            \\
			$HD_i^{speech, pre}$ & 0.0413 & -0.0538 & -0.7120 & -0.8146 \\ 
			& (0.2615) & (0.2539) & (0.8076) & (0.8274) \\ 
			\multicolumn{2}{l}{Macro Factors}   & & & \\
			\cline{1-1}
			$T_{unemp}$ &  & -0.5393^{**} &  & -0.3595 \\ 
			&  & (0.2490) &  & (0.7886) \\ 
			
			$D_{unemp}$ &  & -0.1494 &  & -0.2295 \\ 
			&  & (0.0956) &  & (0.3278) \\ 
			
			$T_{CPI}$ &  & 0.7396^{**} &  & 1.2022 \\ 
			&  & (0.2913) &  & (0.9283) \\ 
			
			$D_{CPI}$ &  & 0.2493 &  & 0.6392 \\ 
			&  & (0.1943) &  & (0.6731) \\ 
			\multicolumn{2}{l}{Member Char.}  & & & \\
			\cline{1-1}
			$D_{experience}$ & -0.0734^{*} & -0.0157 & -0.0427 & 0.0602 \\ 
			& (0.0400) & (0.0417) & (0.1203) & (0.1364) \\ 
			
			$D_{age}$ & -0.0097 & -0.0045 & 0.1241 & 0.1313 \\ 
			& (0.0297) & (0.0287) & (0.1022) & (0.1027) \\ 
			
			$D_{SchoolWealth}$ & -0.4254^{*} & -0.1764 & -1.0001 & -0.7473 \\ 
			& (0.2259) & (0.2448) & (0.6855) & (0.7772) \\ 
			
			$P_{gender}$ & 0.6776 & 0.6007 & 2.0439 & 1.8688 \\ 
			& (0.4628) & (0.4734) & (1.5047) & (1.6072) \\ 
			
			$P_{major}$ & -1.5921^{***} & -1.6060^{***} & -1.6379 & -1.4599 \\ 
			& (0.3434) & (0.4209) & (1.0479) & (1.3533) \\ 
			
			$E_{hometown}$ & -0.4463 & -0.7999^{**} & 0.5522 & 0.0796 \\ 
			& (0.3841) & (0.3802) & (1.1406) & (1.1967) \\ 
			
			$E_{school}$ & 0.5712^{*} & 0.8263^{**} & 1.9709^{*} & 1.8076 \\ 
			& (0.3393) & (0.3537) & (1.1456) & (1.2466) \\ 
			
			$E_{POTUS}$ & 0.0520 & 0.0299 & -0.9620 & -0.8849 \\ 
			& (0.2140) & (0.2106) & (0.6363) & (0.6710) \\ 
			
			$Party_{Dem}$ & 0.1285 & 0.1502 & 0.1629 & 0.4706 \\ 
			& (0.0970) & (0.1204) & (0.3085) & (0.4529) \\ 
			
			$P_{depression}$ & -0.8023 & -0.4730 & 0.0774 & 0.2133 \\ 
			& (0.5892) & (0.5819) & (1.9334) & (2.0088) \\ 
			
			$P_{inflation}$ & -0.0789 & 0.2312 & 0.0654 & 0.2363 \\ 
			& (0.7835) & (0.7970) & (2.5090) & (2.7398) \\ 
			
			$P_{WWII}$ & 1.5657^{***} & 1.4358^{**} & 0.1359 & -0.3242 \\ 
			& (0.5961) & (0.6045) & (1.8542) & (1.9765) \\ 
			\hline
			Pseudo $R^{2}$ & \multicolumn{1}{c}{0.2972} & \multicolumn{1}{c}{0.3600} &  \multicolumn{1}{c}{0.1168} &  \multicolumn{1}{c}{0.1335} \\ 
			Log Likelihood & \multicolumn{1}{c}{212.7796} & \multicolumn{1}{c}{224.0474} & \multicolumn{1}{c}{-11.0166} & \multicolumn{1}{c}{-10.8088} \\ 
			$N$                        & 218              & 218             & 218                & 218     \\
			\hline
			\hline
		\end{tabular}
	\end{center}
	
	\begin{tablenotes}
		\small
		\item \emph{Note:} This table examines the relationship between two  hidden dissent measures constructed respectively from members' speech data and FOMC meeting transcripts. $HD_i^{speech, pre}$ measures the level of hidden dissent within speeches delivered right before a certain FOMC meeting, here we use the chair's latest public speech before the FOMC meeting as the anchor. For other variables, please refer to \autoref{voting_pred_reg1} for their definitions. $^{***}p<0.01$; $^{**}p<0.05$; $^{*}p<0.1$
	\end{tablenotes}
	\label{voting_pred_reg2_full_fomc_pre_meeting}
\end{table}

\clearpage 
\restoregeometry 

\clearpage
\newgeometry{left=2cm, right=2cm, top=1.5cm, bottom=1.5cm}  
\begin{table}
	\caption{Explanation Power of the Level of Hidden Dissent in Speeches (Selected Members)}
	\label{}
	\begin{center}
		\begin{tabular}{@{\extracolsep{-10pt}}lD{.}{.}{-4} D{.}{.}{-4} D{.}{.}{-4} D{.}{.}{-4} }
			\hline
			\hline
			& \multicolumn{2}{c}{$HD_i$}  & \multicolumn{2}{c}{$V_i$}\\
			\cline{2-5}
			& (1)              & (2)               & (3)                & (4)                \\
			$HD_i^{speech, in}$ & 0.8084^{***} & 0.7161^{**} & 0.1238 & -0.2196 \\ 
			& (0.2964) & (0.2830) & (1.0726) & (1.3850) \\ 
			\multicolumn{2}{l}{Macro Factors}   & & & \\
			\cline{1-1}
			$T_{unemp}$ &  & -0.4559 &  & -1.6798 \\ 
			&  & (0.3036) &  & (1.7819) \\ 
			
			$D_{unemp}$ &  & 0.0491 &  & 0.5951 \\ 
			&  & (0.1049) &  & (0.6184) \\ 
			
			$T_{CPI}$ &  & 1.2017^{***} &  & 3.1763^{*} \\ 
			&  & (0.3337) &  & (1.7795) \\ 
			
			$D_{CPI}$ &  & 1.1683^{***} &  & 5.0066^{***} \\ 
			&  & (0.2580) &  & (1.8020) \\ 
			\multicolumn{2}{l}{Member Char.}  & & & \\
			\cline{1-1}
			$D_{experience}$ & -0.0395 & 0.0036 & -0.1788 & 0.0383 \\ 
			& (0.0376) & (0.0376) & (0.1514) & (0.2095) \\ 
			
			$D_{age}$ & 0.0097 & 0.0152 & 0.0554 & 0.0776 \\ 
			& (0.0288) & (0.0279) & (0.1238) & (0.1698) \\ 
			
			$D_{SchoolWealth}$ & -0.0388 & 0.1481 & -1.0560 & -0.8907 \\ 
			& (0.1865) & (0.1903) & (0.7302) & (1.0819) \\ 
			
			$P_{gender}$ & -0.6698 & -0.7551^{*} & -0.5564 & -0.6630 \\ 
			& (0.4155) & (0.4114) & (1.7403) & (2.3215) \\ 
			
			$P_{major}$ & -0.5989 & -0.0553 & 0.2090 & 1.2691 \\ 
			& (0.3673) & (0.3780) & (1.6012) & (2.2765) \\ 
			
			$E_{hometown}$ & -0.8942^{***} & -0.8791^{***} & -2.6807^{*} & -2.4727 \\ 
			& (0.3317) & (0.3247) & (1.4829) & (2.0934) \\ 
			
			$E_{school}$ & 0.4859^{*} & 0.4356 & 0.3154 & -0.7630 \\ 
			& (0.2787) & (0.2689) & (1.3535) & (1.7440) \\ 
			
			$E_{POTUS}$ & -0.3744^{**} & -0.0054 & -2.0488^{***} & -0.3892 \\ 
			& (0.1807) & (0.1913) & (0.6580) & (1.0240) \\ 
			
			$Party_{Dem}$ & 0.3438^{***} & 0.4860^{***} & 0.7636^{*} & 1.4536^{*} \\ 
			& (0.1040) & (0.1173) & (0.4551) & (0.8305) \\ 
			
			$P_{depression}$ & -2.9320^{***} & -2.7903^{***} & -9.6167^{***} & -10.6846^{***} \\ 
			& (0.5219) & (0.5077) & (2.2679) & (3.1569) \\ 
			
			$P_{inflation}$ & -2.2738^{***} & -1.9453^{***} & -5.0007^{**} & -3.9742 \\ 
			& (0.5883) & (0.5719) & (2.2031) & (2.8402) \\ 
			
			$P_{WWII}$ & 0.6111 & 0.3403 & 0.5340 & -0.3235 \\ 
			& (0.4140) & (0.4057) & (1.7215) & (2.3242) \\ 
			\hline
			Pseudo $R^2$   & \multicolumn{1}{c}{0.2728} & \multicolumn{1}{c}{0.3397} & \multicolumn{1}{c}{0.3763} & \multicolumn{1}{c}{0.4543}\\
			Log Likelihood & \multicolumn{1}{c}{158.7237} & \multicolumn{1}{c}{172.2555} & \multicolumn{1}{c}{-21.0489} & \multicolumn{1}{c}{-18.4155} \\ 
			$N$           & \multicolumn{1}{c}{212} & \multicolumn{1}{c}{212} & \multicolumn{1}{c}{212} & \multicolumn{1}{c}{212} \\ 
			\hline
			\hline
		\end{tabular}
	\end{center}
	
	\begin{tablenotes}
		\small
		\item \emph{Note:} This table examines the relationship between two hidden dissent measures constructed respectively from members' speech data and FOMC meeting transcripts. Data used here only contains FOMC members who gave speeches before meetings. $S_i^{speech,in}$ measures the hidden dissent level within speeches delivered right before a certain FOMC meeting, using chair's  meeting transcripts as the reference point. For other variables, please refer to \autoref{voting_pred_reg1} for their definitions. $^{***}p<0.01$; $^{**}p<0.05$; $^{*}p<0.1$
	\end{tablenotes}
	\label{voting_pred_reg3_partial_fomc}
\end{table}

\clearpage 
\restoregeometry 


\renewcommand{\thesection}{D}

\section{Results for Personal Characteristics}
\label{appendix_personal_detailed}

\begin{table}[!htbp]
	\scriptsize
	\caption{Explaining Hidden vs. Revealed Dissent (Individual-level)} 
	\begin{center}
		\begin{tabular}{@{\extracolsep{-10pt}}lD{.}{.}{-3} D{.}{.}{-3} D{.}{.}{-3} D{.}{.}{-3} D{.}{.}{-3} D{.}{.}{-3}} 
			\hline 
			\hline
			& \multicolumn{3}{c}{$hd_{ij}$ (\textit{Mixed Effects Beta}) } & \multicolumn{3}{c}{$v_{ij}$ (\textit{Mixed Effects Probit})} \\ 
			\cline{2-7}
			& \multicolumn{1}{c}{(1)} & \multicolumn{1}{c}{(2)} & \multicolumn{1}{c}{(3)}& \multicolumn{1}{c}{(4)} & \multicolumn{1}{c}{(5)} & \multicolumn{1}{c}{(6)}\\ 
			$hd_{ij}^{speech, in}$     &                 & 0.4567^{***}  &                 &                 & 0.4276^{**}   &                 \\
			&                 & (0.1145)      &                 &                 & (0.1705)      &                 \\
			$hd_{ij}^{speech, pre}$    &                 &                 & 0.2687^{**}   &                 &                 & 0.2499        \\
			&                 &                 & (0.1368)      &                 &                 & (0.1549)      \\
			\multicolumn{2}{l}{Member Char.}   & & & & & \\
			\cline{1-1}
			$T_{unemp} $             & -0.0136       & 0.0707        & 0.0743        & 0.0584        & 0.0795        & 0.0437        \\
			& (0.1289)      & (0.1707)      & (0.1772)      & (0.1515)      & (0.1690)      & (0.1866)      \\
			$D_{unemp} $              & -0.1056^{**}  & -0.0946       & -0.0896       & 0.1561        & 0.3887^{**}   & 0.3094        \\
			& (0.0462)      & (0.0649)      & (0.0688)      & (0.1345)      & (0.1944)      & (0.2012)      \\
			$T_{CPI} $               & 1.0096^{***}  & 0.9795^{***}  & 1.0531^{***}  & 0.1433        & 0.1198        & 0.1384        \\
			& (0.1436)      & (0.2285)      & (0.2365)      & (0.1200)      & (0.1832)      & (0.1978)      \\
			$D_{CPI} $               & 0.4524^{***}  & 0.6402^{***}  & 0.6380^{***}  & 0.2261        & 0.9726^{***}  & 0.9964^{***}  \\
			& (0.1029)      & (0.1877)      & (0.1967)      & (0.2209)      & (0.2801)      & (0.2970)      \\
			\multicolumn{2}{l}{Macro Factors}   & & & & & \\
			\cline{1-1}
			$\text{Experience}$       & 0.0060        & 0.0105        & 0.0129        & 0.4425^{*}    & 0.7535^{***}  & 0.7254^{***}  \\
			& (0.0116)      & (0.0146)      & (0.0154)      & (0.2302)      & (0.2742)      & (0.2739)      \\
			$\text{Age}$              & -0.0224^{**}  & -0.0195       & -0.0168       & -0.5582       & -0.7545^{*}   & -0.5545       \\
			& (0.0114)      & (0.0131)      & (0.0139)      & (0.3906)      & (0.4208)      & (0.4210)      \\
			$\text{School Wealth}$    & -0.1410       & -0.1463       & -0.1544       & -0.2933       & -0.0956       & -0.2446       \\
			& (0.0946)      & (0.0994)      & (0.1048)      & (0.4140)      & (0.4001)      & (0.3933)      \\
			$\text{Female}$           & 0.0584        & 0.1505        & 0.1268        & 0.0168        & 0.9347        & 1.0798        \\
			& (0.1914)      & (0.2067)      & (0.2180)      & (0.8083)      & (0.7495)      & (0.7074)      \\
			$\text{Econ Major} $      & -0.0755       & -0.0608       & -0.0490       & -0.0955       & -0.2266       & -0.0419       \\
			& (0.1580)      & (0.1742)      & (0.1835)      & (0.6507)      & (0.6350)      & (0.6517)      \\
			$\text{Hometown NE}$     & -0.2422       & -0.2095       & -0.2206       & -1.2368       & -1.3606^{*}   & -1.4149^{*}   \\
			& (0.2095)      & (0.2374)      & (0.2509)      & (0.8505)      & (0.8047)      & (0.7374)      \\
			$\text{Hometown OTH}$     & 0.2916        & 0.1510        & 0.1905        & 0.7758        & -1.2185       & -1.4820       \\
			& (0.3408)      & (0.3772)      & (0.3987)      & (1.3842)      & (1.8136)      & (2.0140)      \\
			$\text{Hometown South}$   & 0.1485        & 0.1483        & 0.1490        & 0.8321        & 0.9830        & 0.8441        \\
			& (0.2292)      & (0.2546)      & (0.2694)      & (0.8781)      & (0.9401)      & (0.8655)      \\
			$\text{Hometown West}$    & 0.2503        & 0.0319        & 0.0359        & 2.4982^{**}   & -0.1155       & -0.4606       \\
			& (0.2977)      & (0.3415)      & (0.3603)      & (1.1656)      & (1.4286)      & (1.3911)      \\
			$\text{School NE}$ & -0.3826       & -0.3313       & -0.2925       & -1.9152^{**}  & -2.3252^{**}  & -2.1137^{**}  \\
			& (0.2335)      & (0.2586)      & (0.2724)      & (0.9074)      & (0.9251)      & (0.8669)      \\
			$\text{School South}$    & -0.7624^{***} & -0.7587^{***} & -0.7621^{***} & -3.8795^{***} & -3.8304^{***} & -4.3953^{***} \\
			& (0.2374)      & (0.2498)      & (0.2646)      & (1.0422)      & (1.0934)      & (1.2825)      \\
			$\text{School West}$     & 0.0669        & 0.3753        & 0.4161        & -1.4407       & -0.5504       & -0.0253       \\
			& (0.3065)      & (0.3539)      & (0.3721)      & (1.2051)      & (1.2832)      & (1.1674)      \\
			$\text{Appt. Dem.}$      & 0.0024        & 0.0689        & 0.0522        & -0.7102       & -0.7950       & -0.8468       \\
			& (0.1469)      & (0.1648)      & (0.1732)      & (0.5885)      & (0.5820)      & (0.5426)      \\
			$\text{Incumbent Dem.}$   & 0.2832^{***}  & 0.3186^{***}  & 0.3520^{***}  & 0.4104        & 0.3213        & 0.5629        \\
			& (0.0495)      & (0.0742)      & (0.0797)      & (0.2866)      & (0.4614)      & (0.4976)      \\
			$\text{Great Depression}$ & 0.6826^{***}  & 0.2938        & 0.3071        & 1.2852        & -1.0949       & -0.8493       \\
			& (0.2438)      & (0.3003)      & (0.3187)      & (1.0124)      & (1.0442)      & (1.0135)      \\
			$\text{Great Inflation}$  & 0.0723        & -0.3957       & -0.3695       & 0.0257        & -2.0905       & -2.0109       \\
			& (0.3269)      & (0.3870)      & (0.4110)      & (1.4051)      & (1.8993)      & (2.2393)      \\
			$\text{WWII}$            & 0.0096        & -0.1732       & -0.1904       & -1.1278       & -2.1481       & -2.6005       \\
			& (0.2662)      & (0.3229)      & (0.3404)      & (1.1719)      & (1.7662)      & (2.0660)      \\
			\hline 
			Log Likelihood &\multicolumn{1}{c}{1642.1933} & \multicolumn{1}{c}{860.3329} & \multicolumn{1}{c}{776.3906} & \multicolumn{1}{c}{-393.9099} & \multicolumn{1}{c}{-166.1869} & \multicolumn{1}{c}{-141.9636} \\ 
			$N$ &\multicolumn{1}{c}{2,528} & \multicolumn{1}{c}{1,219} & \multicolumn{1}{c}{1,090} & \multicolumn{1}{c}{2,528} & \multicolumn{1}{c}{1,219}  & \multicolumn{1}{c}{1,090}\\ 
			\hline 
			\hline 
		\end{tabular} 
	\end{center}
	\begin{tablenotes}
		\small
		\item \emph{Note:} This table considers variables that may correlate with the hidden dissent and revealed dissent. Mixed effect regressions is clustered on FOMC member's level. The variable ``Incumbent Dem.'' equals one if the sitting POTUS during a meeting is a Democrat. The variable ``Incumbent Dem.'' is explained in \autoref{personal_info_summary}. $^{***}p<0.01$; $^{**}p<0.05$; $^{*}p<0.1$
	\end{tablenotes}
	\label{personal_level_reg_detailed}
\end{table} 

\clearpage 

\clearpage

\newgeometry{left=2cm, right=2cm, top=1.5cm, bottom=1.5cm}  
\begin{table}[!htbp]
	\small
	\caption{Explaining Hidden vs. Revealed Dissent (Meeting-level)}
	\begin{center}
		\begin{tabular}{@{\extracolsep{-10pt}}lD{.}{.}{-3} D{.}{.}{-3} D{.}{.}{-3} D{.}{.}{-3}}
			\hline
			\hline
			& \multicolumn{2}{c}{$HD_i$ (\textit{Beta})}  & \multicolumn{2}{c}{$V_i$ (\textit{Fractional Logistic})} \\
			\cline{2-5}
			& (1) & (2) & (3) & (4) \\
			\multicolumn{2}{l}{Macro Factors}   & & & \\
			\cline{1-1}
			$T_{unemp}$ & -0.5488^{**} & -0.5938^{***} & -0.0914 & -0.0703 \\ 
			& (0.2213) & (0.2285) & (0.6175) & (0.6772) \\ 
			
			$D_{unemp}$ & 0.0854 & -0.1334 & 0.4631^{**} & -0.1468 \\ 
			& (0.0642) & (0.0842) & (0.1810) & (0.2812) \\ 
			
			$T_{CPI}$ & 1.1071^{***} & 0.6994^{***} & 1.3953^{**} & 1.2535 \\ 
			& (0.2333) & (0.2621) & (0.6480) & (0.8195) \\ 
			
			$D_{CPI}$ & 0.6876^{***} & 0.2525 & 0.5953^{***} & 0.7648 \\ 
			& (0.0827) & (0.1746) & (0.2255) & (0.5876) \\ 
			\multicolumn{2}{l}{Member Char.}  & & & \\
			\cline{1-1}
			$D_{experience}$ &  & 0.0125 &  & 0.0845 \\ 
			&  & (0.0351) &  & (0.1153) \\ 
			
			$D_{age}$ &  & 0.0384^{*} &  & 0.0982 \\ 
			&  & (0.0228) &  & (0.0787) \\ 
			
			$D_{SchoolWealth}$ &  & 0.0221 &  & -0.2481 \\ 
			&  & (0.2133) &  & (0.6920) \\ 
			
			$P_{gender}$ &  & 0.6089 &  & 0.9974 \\ 
			&  & (0.4158) &  & (1.3916) \\ 
			
			$P_{major}$ &  & -1.3761^{***} &  & -1.4270 \\ 
			&  & (0.3570) &  & (1.1393) \\ 
			
			$E_{hometown}$ &  & -0.2212 &  & -0.1499 \\ 
			&  & (0.3046) &  & (0.9745) \\ 
			
			$E_{school}$ &  & 1.0154^{***} &  & 1.5520 \\ 
			&  & (0.3134) &  & (1.0837) \\ 
			
			$E_{POTUS}$ &  & 0.1806 &  & -0.7706 \\ 
			&  & (0.1954) &  & (0.6017) \\ 
			
			$Party_{Dem}$ &  & 0.0960 &  & 0.5025 \\ 
			&  & (0.1022) &  & (0.3668) \\ 
			
			$P_{depression}$ &  & -0.1337 &  & 0.0263 \\ 
			&  & (0.4806) &  & (1.6106) \\ 
			
			$P_{inflation}$ &  & 0.1253 &  & -0.8639 \\ 
			&  & (0.6982) &  & (2.3806) \\ 
			
			$P_{WWII}$ &  & 1.1260^{**} &  & -1.0188 \\ 
			&  & (0.5046) &  & (1.6272) \\ 
			\hline
			Pseudo $R^{2}$ & \multicolumn{1}{c}{0.2275} & \multicolumn{1}{c}{0.3893} & \multicolumn{1}{c}{0.4547} & \multicolumn{1}{c}{0.4914} \\ 
			Log Likelihood & \multicolumn{1}{c}{241.0162} & \multicolumn{1}{c}{274.6247} & \multicolumn{1}{c}{-14.4783} & \multicolumn{1}{c}{-13.5026} \\ 
			$N$                & 268                  & 268                   & 268          & 268            \\
			\hline
			\hline
		\end{tabular}
	\end{center}
	\begin{tablenotes}
		\small
		\item \emph{Note:} This table examines factors that may correlate with the hidden dissent and revealed dissent. The hidden dissent level, $HD_i$, is quantified as the average of hidden dissent scores, and the revealed dissents, $V_i$, is measured as the percentage of NO votes within a meeting. $T_x$ is the trend of variable $x$, $D_x$ in the macro-variable section means the s.d. of variable $x$ calculated from the past and forecast values in Tealbooks. And in the characteristics section, $D_x$ means the s.d. of the variable $x$ in one meeting. $E_x$ measure the entropy of variable $x$. $P_{gender}$ represents the percentage  of females in the FOMC meeting, $P_{major}$ denotes the percentage  of FOMC members whose highest degree is in econ-related field, and $Party_{Dem}$ denotes the party of the siting president is Democrat. Robust standard errors are reported in all results. $^{***}p<0.01$; $^{**}p<0.05$; $^{*}p<0.1$
	\end{tablenotes}
	\label{voting_pred_reg1_detailed}
\end{table}

\clearpage 
\restoregeometry 

\end{document}